\documentclass[numberedappendix,apj]{emulateapj}
\usepackage{apjfonts}

\input epsf.tex

\begin{document}
\title{RXJ0152.7--1357: Stellar populations in an X-ray luminous galaxy cluster at \lowercase{z}=0.83}
\author{Inger J{\o}rgensen}
\affil{Gemini Observatory, 670 N.\ A`ohoku Pl., Hilo, HI 96720, USA}
\email{ijorgensen@gemini.edu}
\author{Marcel Bergmann}
\affil{NOAO Gemini Science Center, Chile}
\email{mbergmann@ctio.noao.edu} 
\author{Roger Davies, Jordi Barr}
\affil{University of Oxford, United Kingdom}
\email{rld@astro.ox.ac.uk, jmb@astro.ox.ac.uk} 
\author{Marianne Takamiya}
\affil{University of Hawaii, Hilo, Hawaii, USA}
\email{takamiya@hawaii.edu} 
\and
\author{David Crampton}
\affil{HIA, Canada}
\email{david.crampton@nrc.ca} 

\submitted{Accepted for publication in Astronomical Journal, November 30, 2004}

\begin{abstract}
We present a study of the stellar populations of galaxies in the cluster
RXJ0152.7--1357 at a redshift of 0.83. The study is based on new high signal-to-noise
spectroscopy of 29 cluster members covering the wavelength range 5000-10000{\AA} 
as well as $r'i'z'$ photometry of the cluster. 

We use scaling relations between the central velocity dispersions of the galaxies and
their luminosities, Balmer line strengths and various metal line strengths to parameterize
the differences between the members of RXJ0152.7--1357 and our low redshift comparison sample.
The luminosities of the RXJ0152.7--1357 galaxies and the strengths
of the higher order Balmer lines H$\gamma$ and H$\delta$ (for non-emission line galaxies)
appear to be in agreement with pure passive evolution of the stellar populations with a 
formation redshift $z_{\rm form} \approx 4$. 
However, the strengths of the D4000 indices and the metal indices do not support
this interpretation. Compared to our low redshift comparison sample, the metal 
indices (C4668, Fe4383, CN3883, G4300 and CN$_2$) show that at least half of the 
non-emission line galaxies in RXJ0152.7--1357 have an $\alpha$-element abundance 
ratio $\rm [\alpha/Fe]$ of 0.2 dex higher, and about half of the galaxies 
have significantly lower metal content.

X-ray data have previously shown that RXJ0152.7--1357 is in the process of merging from 
two sub-clumps. 
We find that differences in stellar populations of the galaxies are associated with 
the location of the galaxies relative to the X-ray emission. 
The galaxies with weak C4668 and G4300, as well as galaxies for
which weak [\ion{O}{2}] emission indicates a very recent star formation episode
involving about 1 per cent of the mass, are located in areas of low X-ray luminosity, on the 
outskirts of the two sub-clumps.
It is possible that these galaxies are experiencing the effect of the cluster merger 
as (short) episodes of star formation,
while the galaxies in the cores of the sub-clumps are unaffected by the merger.

The spectroscopy of the RXJ0152.7--1357 galaxies shows for the first time 
galaxies in a rich cluster at intermediate 
redshift that cannot evolve passively into the present day galaxy population 
in rich clusters. 
Additional physical processes may be at work and we speculate that 
merging with infalling (disk) galaxies in which stars have formed over an 
extended period might produce the required reduction in $\rm [\alpha/Fe]$.
However, the merging could not be accompanied by star formation involving a
substantial mass fraction.
We note that our conclusions, in part, rely on stellar population models for which
the predictions of the indices in the rest frame blue have not yet been tested
extensively.
\end{abstract}


\section{Introduction}

Studies of nearby galaxies ($z<0.05$) have shown that despite the complex processes involved in the 
formation and evolution of the galaxies (e.g.\ mergers, bursts of star formation, morphological changes), 
the global properties of the galaxies follow very tight empirical scaling relations. 
Spiral galaxies follow the Tully-Fisher (TF) relation (Aaronson et al.\ 1986),
which is a relation between total magnitude and rotational velocity.
Elliptical (E) and lenticular (S0) galaxies follow the Fundamental Plane (FP)
(Dressler et al.\ 1987; Djorgovski \& Davis 1987; J\o rgensen et al.\ 1996),
which is a relation between the effective radius, the mean surface brightness within
that radius and the central velocity dispersion. The FP may be interpreted as a
relation between the masses and the mass-to-light (M/L) ratios of the galaxies.
For E and S0 galaxies the absorption line strengths ($\rm H\beta$, Mg$b$, and $\langle {\rm Fe} \rangle$) as well
as the colors are correlated with the central velocity dispersions of the galaxies 
(Bender et al.\ 1993; J\o rgensen 1997; Colless et al.\ 1999). 

Several authors have used single-stellar population (SSP) models to derive
luminosity weighted mean ages and metal contents from the line strengths. 
Such analysis shows that at a given velocity dispersion metal 
rich cluster E and S0 galaxies are younger than metal poor galaxies, and that many of the galaxies 
have experienced star formation within the last $\approx$ 3 Gyr involving at least 10 per cent 
of the mass (J\o rgensen 1999; Trager et al.\ 2000). 

It is not possible to fully constrain the models for galaxy formation and evolution based on 
observations of one epoch ($z \approx 0$), only.
Therefore, many different groups have studied galaxies at intermediate redshifts, 
typically up to $z\approx 0.8$ 
with a few studies reaching redshift one, in order to constrain the models for galaxy evolution.

The TF relation has been studied by, e.g., Vogt et al.\ (1996), Ziegler et al.\ (2002, 2003), 
Milvang-Jensen et al.\ (2003)
and B\"{o}hm et al.\ (2004).  The results are not all consistent, but in general very small offsets 
relative to the low redshift TF relations are found for cluster galaxies, while field galaxies show
larger offsets.
Ziegler et al.\ (2002) and B\"{o}hm et al.\ (2004) find for field spiral galaxies that the low mass 
galaxies show more evolution between $z=1$ and the present than found for high mass spiral galaxies.
They conclude that the evolution of the M/L ratios depends on the galaxy masses.
All of the studies interpret the offsets relative to the TF relation as an offset in
the luminosity. However, see Kannappan \& Barton (2004) for a discussion of kinematic 
anomalies as the source of some offsets from the TF relation found a high redshifts.

The FP has been used to study the luminosity evolution of E and S0 galaxies as a function 
of redshift 
(e.g., Bender et al.\ 1998; van Dokkum et al.\ 1998; J\o rgensen et al.\ 1999; 
Kelson et al.\ 2000; Ziegler et al.\ 2001; van Dokkum \& Stanford 2003; Wuyts et al.\ 2004). 
The study by van Dokkum \& Stanford is the first
to establish the FP for cluster galaxies at $z > 1$.
The luminosity evolution is usually interpreted within a model that assumes pure passive evolution
of the galaxies. This means the stellar populations of the galaxies are assumed to evolve 
quiescently, with no additional star formation in the redshift interval that is studied. 
In this model, the only differences
between the stellar populations at different redshifts are differences in ages equal to the 
differences in the lookback time for the various redshifts.
All the studies of the FP as a function of redshift conclude that the change in the zero point 
of the FP is consistent with the assumption that E and S0 galaxies evolve passively 
from a high redshift, called the formation redshift, $z_{\rm form}$. 
Most of the authors find that $z_{\rm form} > 2$.

Some attempts have also been made to use other scaling relations for E and S0 galaxies. 
Bender et al.\ (1998) and Ziegler et al.\ (2001) used both the $\rm Mg_2$-$\sigma$ 
relation and the FP, and concluded that both relations were in agreement with pure passive evolution.
Kelson et al.\ (2001) studied the strength of the higher order Balmer lines, H$\delta$ 
and H$\gamma$, as a function
of the galaxy velocity dispersion for four galaxy clusters with redshifts between 0.06 and 0.83.
Kelson et al.\ conclude that the Balmer line strengths are in agreement with pure passive evolution 
and $z_{\rm form} > 2.5$.

Apart from Kelson et al.\ (2000), Ziegler et al.\ (2001), and Wuyts et al.\ (2004),
the rest of the studies have been 
restricted to quite small galaxy samples in each cluster, typically about 10 galaxies 
per cluster, covering a narrow range in luminosities.  
With a narrow coverage in luminosities, and therefore in masses, these studies cannot
address how the evolution may depend on galaxy mass.
Further, the samples are usually selected such that the galaxies are on the red sequence
of the color-magnitude relation and/or their morphologies are early-type (E or S0).
Significant morphological evolution has been found between redshift $\approx 0.5$ and 
the present showing that a substantial number of spiral galaxies must have evolved
into E and S0 galaxies (e.g., Dressler et al.\ 1997). Therefore studying
galaxy evolution by comparing E and S0 galaxies at redshifts between 0.2-1.0 to
those at low redshift, is likely to give a biased impression of the actual
evolution of the galaxies since part of the E and S0 galaxy population at low
redshift originates from galaxies not included in the higher redshift samples.
This ``progenitor bias'' is discussed in detail by van Dokkum \& Franx (2001).
While it is difficult to identify which high redshift galaxies lead to which low redshift
descendants, a safer assumption may be that the whole cluster population at
high redshift leads to the cluster population at low redshift. 
This assumption ignores any cluster infall occurring between redshift $\approx 1$ and
the present that may change the populations of cluster galaxies.
However, to some extent the problem of the ``progenitor bias'' may be addressed 
by studying representative samples of the full population of cluster galaxies.

Two of the major variables that determine the physical processes a galaxy will undergo 
during its evolution are mass and environment. Other major variables like the gas content
may depend on the mass and the early evolution of the galaxy.
Detailed information about how the galaxy evolution 
depends on the mass can be used to constrain the formation scenarios, i.e.\ hierarchical versus 
monolithic collapse (e.g., Kauffmann \& Charlot 1998). 
The scaling relations with low scatter 
represent very powerful tools to address these questions, since we can measure as a 
function of galaxy size, mass, luminosity and central velocity dispersion 
how the galaxies follow or deviate from the 
mean relations and how the relations may change with epoch reflecting the evolutionary 
paths of the galaxies. 
Typical model predictions show changes in the slopes of the scaling relations 
as a function of redshift of $\approx$ 10 per cent between $z=1$ and the present (Ferreras \& Silk 2000a). 
However, empirically larger changes are found, e.g., Ziegler et al.\ (2002) find a 35 per cent change in the 
slope of the TF relation between $z=1$ and the present. 
For nearby galaxies, Concannon et al.\ (2000) find that low mass 
galaxies have a larger age spread than high mass galaxies. 
Their result as well as a recent study of Abell 851 at $z=0.41$ (Ferreras \& Silk 2000b) 
indicate that the evolutionary paths depend significantly on the galaxy mass, and
that these differences should be detectable in samples that reach luminosities
of one to two magnitudes fainter than $L^{\star}$.

From these considerations follow four requirements for a galaxy sample selected for a 
detailed study of galaxy evolution as a function of epoch and mass: (1) Large coverage in 
luminosity, (2) sufficient number of galaxies at each epoch to accurately determine the 
slopes of the scaling relations, (3) sufficient number of epochs to detect the 
possible changes in slopes with redshift, and (4) consistent coverage in distance from
the cluster center at different epochs such that we can differentiate between 
effects due to galaxy mass and due to cluster environment.

In Section 2 we describe the science objectives
and methods of our project, ``The Gemini/HST Galaxy Cluster Project'', which is designed
following the above listed requirements.
The remainder of the paper is based on photometry and spectroscopy of galaxies in the cluster
RXJ0152.7--1357 and addresses a subset of the analysis and questions outlined in Section 2.
Section 3 gives background information about the galaxy cluster RXJ0152.7--1357.
The observational data for the galaxies in RXJ0152.7--1357 are described in Section 4, with
further details given in the Appendix. Our low redshift comparison sample is briefly
described in Section 5.
In Section 6 we describe the stellar population models and the evolutionary 
scenarios that we use for interpreting the data.
Section 7 discusses the cluster sub-structure in RXJ0152.7--1357.
In Section 8 we focus on the properties of the stellar populations in the RXJ0152.7--1357
galaxies. In Section 9 we discuss the stellar populations in the context of 
the cluster sub-structure and in the context of the evolutionary scenarios. 
Section 10 summarizes the conclusions.
Throughout this paper we adopt a $\Lambda$CDM cosmology with 
$\rm H_0 = 70\,km\,s^{-1}\,Mpc^{-1}$, $\Omega_{\rm M}=0.3$, and $\Omega_{\rm \Lambda}=0.7$.

\section{The Gemini/HST Galaxy Cluster Project}

This paper is the first in a series from ``The Gemini/HST Galaxy Cluster Project'',
aimed at studying galaxy evolution during half the age of the Universe.
The science objective of the project is to establish the star formation history
for galaxies in rich clusters as a function of galaxy mass. Among the questions 
that we aim to address are
(1) the role and duration of star formation episodes, (2) the links between morphological 
evolution and the evolution of the stellar populations, (3) the presence of possible
variations in the initial-mass-function (IMF), (4) sub-structure of the clusters and
the location of galaxies containing young stellar populations.

The sample consists of 15 X-ray selected rich galaxy clusters covering 
a redshift interval from $z=0.15$ to $z=1.0$; we adopt a lower limit on the X-ray
luminosity of \mbox{$L_X(0.1-2.4{\rm keV}) = 2 \cdot 10^{44} {\rm ergs\,s^{-1}}$}. 
A full description of the selection of galaxy clusters and sample selection for the
spectroscopic samples in each of the clusters
will be presented in J\o rgensen et al.\ (in preparation).

Our observing strategy is as follows.
We obtain optical photometry in three or four filters with the Gemini Multi-Object Spectrograph 
(GMOS) on either Gemini North or Gemini South.
See Hook et al.\ (2004) for a description of GMOS.
For each cluster we cover approximately the central $\rm 1.7\,Mpc \times 1.7\,Mpc$,
which for the adopted cosmology is about $1\degr \times 1\degr$ at the distance
of the Coma cluster, and $\rm 3.7\,arcmin \times 3.7\,arcmin$ at the distance
of RXJ0152.7--1357.
The photometry is used to select the spectroscopic sample, which
includes all galaxies that are likely to be members of the cluster. 
The selection is based on color-magnitude diagrams as well as color-color diagrams.
Of special importance is that no morphological selection criteria are 
applied to the spectroscopic sample.
For each cluster we obtain high signal-to-noise (S/N) optical spectroscopy of 
30 to 50 cluster members using GMOS.
The S/N of the spectra is typically higher than 25 per {\AA}ngstrom in the
rest frame of the galaxies.
The spectra are used for determining the redshift, the central velocity dispersion
of each galaxy, and line indices for absorption lines (Balmer lines as well as 
several metal lines).
We measure line indices for enough absorption lines to be able to
study differences in ages, metallicities, and $\alpha$-element
abundance ratios $\rm [\alpha/Fe]$.
For galaxies with emission lines, we also determine the equivalent width of these.
Hubble Space Telescope (HST) imaging obtained with either
the Wide Field Planetary Camera 2 or the Advanced Camera for Surveys (ACS)
is used to derive 2-dimensional surface 
photometry of the galaxies. We use available archive data, as well as
data obtained specifically for this project.

We use a two-tiered approach in the analysis of the observational data. 
We establish scaling relations between observable parameters 
(effective radii, surface brightnesses, 
magnitudes, velocity dispersions and absorption line strengths), and 
we use the spectra to derive mean ages, metal content 
and abundance ratios using stellar population models.

The scaling relation zero points track the bulk differences in the
stellar populations. Because most of the scaling relations have either the galaxy mass
or the velocity dispersion as the independent parameter, changes in the slope
of a scaling relation reflect differences in the stellar populations as a
function of galaxy mass or velocity dispersion. 
Finally, the internal scatter of the scaling relations reflect the variations 
in star formation history within a given sample of galaxies.
The most obvious groupings of galaxies are of course groupings with respect to 
redshift. Other groupings make use of information about the cluster environment,
cluster sub-structure, or galaxy morphology.

Using stellar population models like those recently published by Thomas et al.\ (2003, 2004)
we can derive luminosity weighted mean ages, metal content   
and abundance ratios, specifically the $\alpha$-element abundance ratio. 
Most other models do not vary the $\alpha$-element abundance ratio, but assume that it
is solar. This is a significant limitation of the models since data for nearby 
E and S0 galaxies show an $\alpha$-element enhancement of $[\alpha/{\rm Fe}] \approx 0.2$
(e.g., Worthey et al.\ 1992; Davies et al.\ 1993; J\o rgensen 1999; Trager et al.\ 2000).

\begin{deluxetable}{lr}
\tablecaption{Instrumentation \label{tab-inst} }
\tablewidth{230pt}
\tablehead{}
\startdata
Telescope       & Gemini North  \\
Instrument      & GMOS-N       \\
CCDs            & 3 $\times$ EEV 2048$\times$4608 \\
r.o.n.\tablenotemark{a}          & (3.5,3.3,3.0) e$^-$       \\
gain\tablenotemark{a}            & (2.10,2.337,2.30) e$^-$/ADU  \\
Pixel scale     & 0.0727arcsec/pixel \\
Field of view   & $5\farcm5\times5\farcm5$ \\
Imaging filters & $r'$$i'$$z'$ \\
Grating         & R400\_G5305 \\
Spectroscopic filter & OG515\_G0306 \\
Slit width      & 1 arcsec \\
Slit length     & 5 -- 14 arcsec \\
Extraction aperture & 1 arcsec $\times$ 1.15 arcsec \\
$r_{\rm ap}$\tablenotemark{b}   & 0.62 arcsec \\
Spectral resolution\tablenotemark{c}, $\sigma$ & 3.065\AA \\
Wavelength range\tablenotemark{d} & 5000-10000\AA  \\
\enddata
\tablenotetext{a}{Values for the three detectors in the array.}
\tablenotetext{b}{Radius of equivalent circular aperture, see J\o rgensen et al.\ (1995)}
\tablenotetext{c}{Median of the resulting resolutions, each derived as sigma in a Gaussian 
fit to the sky lines in stacked spectra. The resolution is equivalent to 
$\sigma = 116 {\rm km\,s^{-1}}$ at 4300\AA\ in the rest frame of RXJ0152.7--1357.}
\tablenotetext{d}{The exact wavelength range varies from slit-let to slit-let.}
\end{deluxetable}

\begin{figure*}
\epsfxsize 16.5cm
\epsfbox{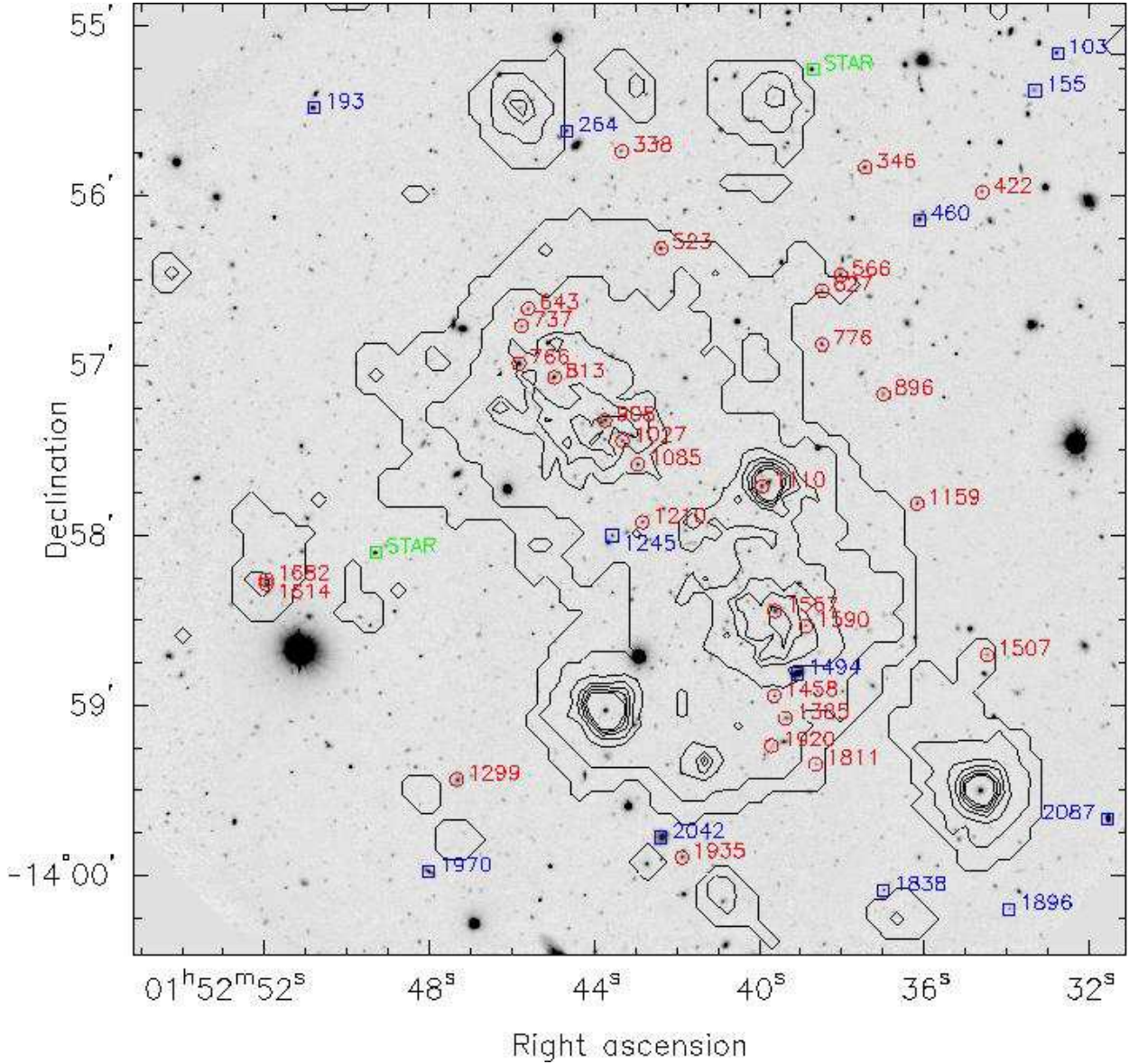}
\caption[]{RXJ0152.7--1357 $i'$-band image with contours of the {\it XMM-Newton} data overlaid.
The image covers approximately 5.5 arcmin $\times$ 5.5 arcmin.
Red circles -- confirmed cluster members labeled with their ID number;
blue boxes -- non-members with spectroscopy labeled with their ID number;
green boxes --the two blue stars included in the mask.
The X-ray image is the sum of the images from the two {\it XMM-Newton} EPIC-MOS cameras.
The X-ray image has been smoothed such that the structure seen is significant
at the 2.5 $\sigma$ level or higher. 
The spacing between the contours is logarithmic with a factor 1.5 between each contour.
\label{fig-grey} }
\end{figure*}

\section{RXJ0152.7--1357: Background information}

The massive cluster of galaxies RXJ0152.7--1357 was discovered from {\it ROSAT}
data by three different surveys: The {\it ROSAT} Deep Cluster Survey (RDCS) and the Wide Angle
{\it ROSAT} Pointed Survey (WARPS) (see Ebeling et al.\ 2000 for the historical account),
as well as the Bright Serendipitous High-Redshift Archival Cluster (SHARC) survey 
(Nichol et al.\ 1999).
Further X-ray observations of the cluster have been carried out with {\it BeppoSAX}
(Della Ceca et al.\ 2000), {\it XMM-Newton} and {\it Chandra}
(Jones et al.\ 2004; Maughan et al.\ 2003).
The data from {\it XMM-Newton} and {\it Chandra} show two sub-clumps and support the view that 
RXJ0152.7--1357 is in the process of merging from two clumps of roughly equal mass, 
see Figure \ref{fig-grey} that shows the {\it XMM-Newton} data together with our $i'$-band image
of the cluster. This figure is discussed further in Section 7.
The total mass of the cluster is estimated to be similar to that of the Coma cluster (Maughan et al.\ 2003). 

Redshifts of six galaxies in RXJ0152.7--1357 have been published by  Ebeling et al.\ (2000),
giving the cluster redshift of 0.833.
Ellis \& Jones (2004) have studied the K-band luminosity function of this
cluster as well as two other massive high redshift clusters, RXJ1226.9+3332 and RXJ1415.1+3612.
They find the luminosity functions to be consistent with passive evolution and a 
formation redshift of $z_{\rm form} \approx 1.5-2.0$.

\section{Observational data}

Imaging and spectroscopy of RXJ0152.7--1357 were obtained with GMOS-N in semester 2002B.
The observations were done in queue in the period from UT 2002 July 18 to UT 2002 September 25.
The data were obtained as part of Gemini programs
GN-2002B-Q-29 (a queue program) and GN-2002B-SV-90 (an engineering program).
Table \ref{tab-inst} summarizes the instrument information, while Tables \ref{tab-imdata} 
and \ref{tab-spdata} summarize the imaging and spectroscopic data, respectively.

\begin{deluxetable}{lrrl}
\tablecaption{GMOS-N Imaging Data \label{tab-imdata} }
\tablewidth{0pc}
\tablehead{
\colhead{Filter} & \colhead{Exposure time} & \colhead{Image quality\tablenotemark{a}} & \colhead{Sky brightness} \\
\colhead{} & \colhead{} & \colhead{arcsec} & \colhead{mag arcsec$^{-2}$} }
\startdata
$r'$     & 12 $\times$ 600sec           & 0.68 & 20.65 \\
$i'$     & 7 $\times$ 450sec (dark sky) & 0.56 & 19.63 \\
         & + 100 $\times$ 120sec (bright sky) & \\
$z'$     & 13 $\times$ 450sec (dark sky) & 0.59 & 19.16 \\
         & + 14 $\times$ 450sec (bright sky) & \\ 
\enddata
\tablenotetext{a}{Average FWHM of 7-10 stars in the field.}
\end{deluxetable}

\begin{deluxetable*}{lrrrr}
\tablecaption{GMOS-N Spectroscopic Data \label{tab-spdata} }
\tablewidth{35pc}
\tablehead{
\colhead{Data set}& \colhead{Exposure time} & \multicolumn{3}{c}{Image~quality\tablenotemark{a}} \\ 
\colhead{}        & \colhead{} & \colhead{at 7000\AA}  & \colhead{at 8000\AA}  & \colhead{at 9000\AA} \\ 
\colhead{}        & \colhead{} & \colhead{(arcsec)} & \colhead{(arcsec)} & \colhead{(arcsec)} }
\startdata
$\lambda_{\rm central} =$ 8050\AA  & 48800 sec (15 exposures) & 0.67 & 0.63 & 0.58 \\
$\lambda_{\rm central} =$ 8150\AA  & 29160 sec (10 exposures) & 0.70 & 0.67 & 0.64 \\ 
Combined                           & 77960 sec (25 exposures) & 0.68 & 0.65 & 0.60 \\
\enddata
\tablenotetext{a}{Average FWHM measured from the two blue stars included in the mask.}
\end{deluxetable*}

\begin{figure*}
\epsfxsize 16.5cm
\epsfbox{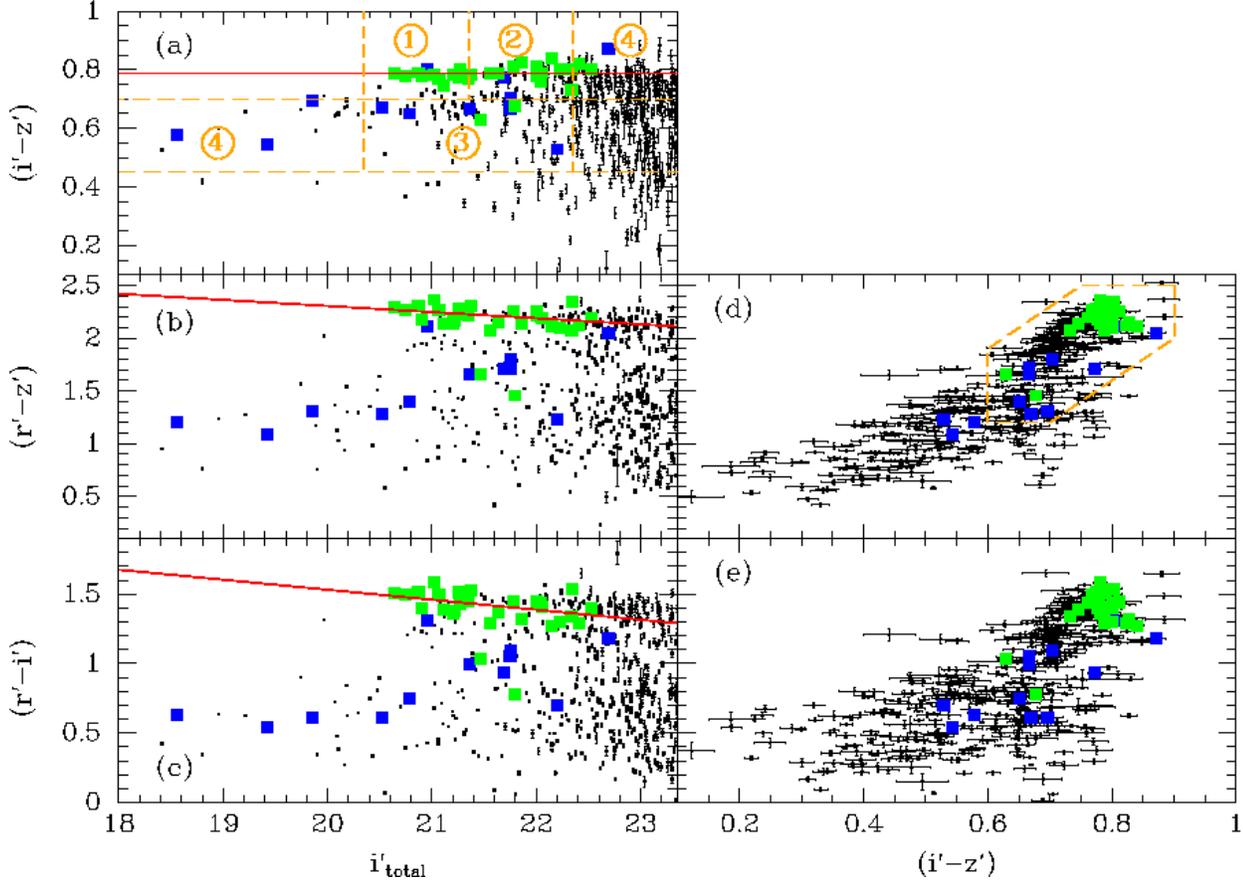}
\caption[]{RXJ0152.7--1357: Color-magnitude and color-color diagrams. Only galaxies
({\it class\_star} $<$ 0.80 in the $i'$ filter) with $i'\le 23.35$ mag are shown.
The magnitudes are the total magnitudes, while all colors are aperture colors.
The photometry has been corrected for the galactic extinction, see Section 4.1.
Green filled boxes -- confirmed cluster members in our spectroscopic sample;
blue filled boxes -- non-members with spectroscopy;
small points with error bars -- galaxies without spectroscopy.
Red lines -- least squares fit to the data for the cluster members, excluding the
two bluest galaxies. The slope for the color-magnitude relation for $(i'-z')$ is
not significantly different from zero, thus the line marks the median color for
the galaxies.  
The orange dashed lines and circled numbers in panel (a) show the object classes, see Section 4.2.
The orange dashed lines on panel (d) outline the sample limits in the colors that would have
been used if photometry in all three passbands had been available at the time of
the sample selection, see Section 4.2 for discussion.
\label{fig-CMCC} }
\end{figure*}

The imaging covers one GMOS-N field, which is approximately 5.5\,arcmin $\times$ 5.5\,arcmin.
Imaging was obtained in three filters.
One GMOS mask was used for the spectroscopy. We used the R400 grating and a slit width of
1 arcsec, resulting in an instrumental resolution of 116 $\rm km\,s^{-1}$ at 4300{\AA} in the rest
frame of RXJ0152.7--1357, see Table \ref{tab-inst}.
The spectroscopic observations were obtained as 25 individual exposures with
exposure time from 2000 to 3600 sec. The total exposure time was $\approx$ 21.7 hours.
Spectroscopy was obtained of 41 galaxies, 29 of which are cluster members.
See Section 7 for the definition of cluster membership.
For the cluster members, the median S/N is 31 per {\AA}ngstrom in the rest frame of the galaxies,
derived in the rest frame wavelength interval 4100-4600 {\AA}. The S/N for the individual
galaxies are listed in Table \ref{tab-speckin} in the appendix. 
Three cluster members have S/N less than 20 per {\AA}ngstrom (ID 896, 1811, and 1920), two of these 
are emission line galaxies.

\subsection{Imaging reductions and derived photometric parameters}

The basic reductions of the data were done using a combination of the Gemini 
IRAF package and custom reduction techniques. 
The Gemini IRAF package is an external package built on core 
IRAF\footnote{IRAF is distributed by National Optical Astronomy Observatories, which
is operated by the Association of Universities for Research in Astronomy, Inc., (AURA),
under cooperative agreement with the National Science Foundation, USA.
The Gemini IRAF package is distributed by Gemini Observatory, which is operated by AURA.}.
The details of the reductions are described in the Appendix. 
The final imaging data products 
are the cleaned and averaged images in each filter, normalized to one of the 
exposures taken in photometric conditions, these images are in the following referred to
as the ``co-added images''. 

The co-added images were processed with SExtractor v.2.1.6 (Bertin \& Arnouts 1996).
The details are described in the Appendix.
We adopt the best magnitudes ({\it mag\_best}) from SExtractor as the total 
magnitudes of the objects. Aperture magnitudes and colors were derived within 
apertures with a diameter of 1.16 arcsec, which is approximately twice the FWHM of the 
point-spread-function of the images. 
From model galaxies with exponential and $r^{1/4}$-profiles, and
with sizes matching our spectroscopic sample, we have found 
that the colors $(r'-i')$ and $(r'-z')$ are affected
by no more than 0.03 due to the small differences in FWHM of the images in the 
different filters. 
The FWHM difference makes the colors systematically too red.
For $(i'-z')$ the effect is less than 0.005. 
In all cases the effect does not affect our analysis.
In the following we use the total magnitudes together with the aperture colors.

For the galaxies in the spectroscopic sample, the typical internal uncertainties on 
the magnitudes and colors due to photon noise only are 0.007 mag and 0.01, respectively. 
The observations in the $i'$-filter were obtained as two independent sets, 
one obtained in dark time and the other in bright time. 
We compare photometry derived from these two data sets and 
use this comparison to estimate the uncertainties on the photometry 
introduced by flat fielding, fringe correction and differences in seeing. 
We compare the total magnitudes as well as the aperture magnitudes
derived from the two data sets. 
All comparisons are done for objects with {\it class\_star}$<$0.80.
From the rms scatter in the comparisons, we find that for
$i' < 21.5$ mag the uncertainties on the total magnitudes and the colors
are 0.035 mag and 0.045, respectively.
For $21.5 < i' \le 22.5$ mag the uncertainties are 0.06 and 0.07.
The standard calibration of the photometry is described in the Appendix.
In general the uncertainties on the calibrations are between 0.04 mag and 0.05 mag
due to the scatter in the relations used for the standard calibration.
Table \ref{tab-photdata} in the Appendix lists the photometry for the spectroscopic sample.

The galactic extinction in the direction of RXJ0152.7--1357 is $A_{\rm B}=0.064$
(Schlegel at al.\ 1998). We use the effective wavelength of the three filters
in which the photometry was obtained, and the calibration from Cardelli et al.\ (1989)
to derive the extinction $A_{\rm r}=0.042$;  $A_{\rm i}=0.033$; $A_{\rm z}=0.027$.
The data in Table \ref{tab-photdata} have not been corrected for galactic extinction.
Two-dimensional surface photometry and the morphologies of the galaxies in the 
cluster will be the topic of a future paper.

\subsection{Spectroscopic sample selection}

The spectroscopic sample was selected based on the photometry.
Stars and galaxies were separated using the SExtractor classification
parameter {\it class\_star} derived from the image in the $i'$-filter.
For the purpose of selecting targets for the spectroscopic observations
we chose a threshold of 0.80, i.e., objects with {\it class\_star}$<$0.80
in the $i'$-image are considered galaxies.
This limit may result in very compact galaxies being excluded from the sample. 
However, it was more important
for the planning of our observations to ensure that all targets for
the spectroscopic observations were indeed galaxies.
Based on the total magnitude in $i'$ and the colors we then define four classes
of objects. 
At the time of the sample selection, only photometry in $i'$ filter and the $z'$ filter
was available. We use a color selection that includes all likely
cluster members. We then define object classes as follows. 
\begin{itemize}
\item 1: $20.35 \le i' \le 21.35 ~ \wedge ~ (i'-z')\ge 0.7$
\item 2: $21.35 < i' \le 22.35 ~ \wedge ~ (i'-z')\ge 0.7$
\item 3: $20.35 \le i' \le 22.35 ~ \wedge ~ 0.45 \le (i'-z') < 0.7$
\item 4: $(i'<20.35 ~ \wedge ~ ~ 0.45 \le (i'-z') < 0.7) \\
\vee ~ (i'>22.35 ~ \wedge ~ (i'-z')\ge 0.7)$
\end{itemize}

Figure \ref{fig-CMCC} summarizes the photometry for the field as color-magnitude
diagrams and color-color diagrams. The spectroscopic sample is marked,
cluster members as solid green boxes. The object classes are visualized on Figure \ref{fig-CMCC}a.
The star-galaxy classification parameter {\it class\_star} in
the $i'$-filter is 0.02-0.04 for all the galaxies in the spectroscopic sample.

Objects in class 1 and 2 are equally important to include in the spectroscopic sample.
We aimed at including roughly the same number of galaxies from each of these two classes. 
Class 3 objects are likely to include blue cluster members. These were observed
when no class 1 or 2 object was available for a given position in the mask. 
Due to the distribution of the class 1, 2 and 3
objects in the field, not all of the available space in the mask could be filled with these.
We therefore included objects from class 4 in order to fill the mask. 
The bright class 4 galaxies are expected to all be foreground galaxies.
The faint class 4 galaxies are expected to include cluster members.

It turned out that most of the observed galaxies from class 3 were in fact foreground 
galaxies. All of the observed bright galaxies from class 4 were indeed foreground galaxies.
Two of the three faint galaxies observed from class 4 were cluster members.

Because only $i'$ and $z'$ band photometry was available at the time of the
sample selection for the spectroscopy, it is relevant to assess if this has significantly
biased the sample selection. 
Figure \ref{fig-CMCC}d shows the outline of the limits
in the color-color diagram that we would have used for the sample selection, had
the $r'$ band photometry been available at that time. These limits are similar to
those used for observations of other clusters in our sample. 
It should be noted that to ensure that we did not exclude blue cluster members, 
the blue limit actually used for the class 3 objects was at a smaller $(i'-z')$ 
than we would have used if the $r'$ band photometry had been available.
There are three galaxies in the 
spectroscopic sample outside the limits shown on Figure \ref{fig-CMCC}d.
Two of these are bright galaxies in class 4. 
These were already expected not to be cluster members, and were only
included in the sample in order to fill the mask. The third galaxy (also 
a foreground galaxy) would probably not have been included in the sample, and 
another target would have been included.
The rest of the sample selection is unaffected by the lack of $r'$ photometry.
We conclude that the sample selection based on $i'$ and $z'$, only, 
was not significantly biased relative to what we would have done using photometry 
in all three passbands.

The spectroscopic sample is marked on  Figure \ref{fig-grey}, cluster members as circles.
Two blue stars were included in the mask in order to obtain a good correction
for the telluric absorption lines. These two stars are also marked on Figure \ref{fig-grey}.

\subsection{Spectroscopic reductions and derived spectroscopic parameters}

The details of the reductions of the spectroscopic data are described in the Appendix. 
The final data products are cleaned and averaged spectra that
have been wavelength calibrated and also calibrated to a relative flux scale.
Both extracted one-dimensional (1D) and the 2-dimensional spectra are
kept after the basic reductions. However, in this paper we use only the 1D spectra.

The co-added 1D spectra were used for deriving the redshifts, velocity dispersions,
absorption line indices, and emission line equivalent widths of the galaxies. 
The details are described in the Appendix. Here we summarize the most important points.

The redshifts and the velocity dispersions were determined by fitting
a mix of three template stars to the spectra. We used software 
made available by Karl Gebhardt. 
The software uses penalized maximum likelihood fitting in pixel space to determine
the velocity dispersion and the redshift, see Gebhardt et al.\ (2000, 2003) for a detailed 
description of the fitting method.  
The template stars were of spectral types K0III, G1V and B8V. Using multiple
template stars limits any systematic effects in the derived velocity dispersions 
due to template mismatch.
The velocity dispersions have been corrected for the aperture size using the technique
from J\o rgensen et al.\ (1995).
Table \ref{tab-speckin} in the Appendix summarizes results from the template fitting.
Measured velocity dispersions as well as aperture corrected velocity dispersions
are listed for the cluster members. For galaxies that are not members of
the cluster, we give the redshift. The detailed data for these galaxies
will be discussed in a future paper.

The Lick/IDS absorption line indices CN$_1$, CN$_2$, G4300, Fe4383, C4668 (Worthey et al.\ 1994),
as well as the higher order Balmer line indices $\rm H\delta _A$ and $\rm H\gamma _A$
(Worthey \& Ottaviani 1997) were derived.
We have also determined the D4000 index (Bruzual 1983; Gorgas et al.\ 1999),
and the blue indices CN3883 and CaHK (Davidge \& Clark 1994).
The indices have been corrected for the aperture size and for the effect of the
velocity dispersions, see Appendix.

\begin{deluxetable}{lrrr}
\tablecaption{Low Redshift Comparison Data \label{tab-compdata} }
\tablewidth{0pc}
\tablehead{
\colhead{Cluster} & \colhead{Redshift} & \colhead{N($\log \sigma$)} &\colhead{N(line indices)} }
\startdata
Perseus & 0.018 & 63 & 51 \\
A0194   & 0.018 & 17 & 14 \\
Coma    & 0.024 & 116 & \nodata \\
\enddata
\end{deluxetable}

\begin{deluxetable*}{llrl}
\tablecaption{Predictions from Single Stellar Population Models \label{tab-models} }
\tablewidth{35pc}
\tablehead{
\multicolumn{2}{l}{Relation} & \colhead{rms} & \colhead{Reference} \\
\multicolumn{2}{l}{(1)} & \colhead{(2)} & \colhead{(3)} }
\startdata
$\rm \log M/L_B $ &$= 0.935 \log {\rm age} + 0.337 {\rm [M/H]} - 0.053 $  & 0.022 & Maraston 2004 \\
$\rm \log M/L_B $ &$= 0.833 \log {\rm age} + 0.409 {\rm [M/H]} + 0.162 $  & 0.013 & Vazdekis-2000 \\
$\rm D4000 $ &$= 0.730 \log {\rm age} + 0.711 {\rm [M/H]} + 1.827$  & 0.052 & Vazdekis-2000 \\
$\rm CN3883 $ &$= 0.173 \log {\rm age} + 0.142 {\rm [M/H]} + 0.086$  & 0.012 & Bruzual \& Charlot 2003 \\
$\rm \log CaHK $ &$= 0.073 \log {\rm age} + 0.061 {\rm [M/H]} + 1.291$  & 0.010 & Bruzual \& Charlot 2003 \\
$\rm (H\delta _A + H\gamma _A)' $ &$= -0.115 \log {\rm age}  -0.095 {\rm [M/H]}  +0.095 {\rm [\alpha/Fe]} + 0.009 $\tablenotemark{a}  &0.008 &  Thomas et al. \\
$\log {\rm H\beta _G} $ &$= -0.221 \log {\rm age} - 0.114 {\rm [M/H]} + 0.055  {\rm [\alpha/Fe]} + 0.500$ & 0.010 & Thomas et al. \\
${\rm CN_2}        $ &$= 0.121 \log {\rm age} + 0.196 {\rm [M/H]} + 0.066 {\rm [\alpha/Fe]} - 0.043$  & 0.025 &  Thomas et al. \\
$\log {\rm G4300}  $ &$= 0.162 \log {\rm age} + 0.163 {\rm [M/H]} + 0.114 {\rm [\alpha/Fe]} + 0.552$  & 0.029 &  Thomas et al. \\
$\log {\rm Fe4383} $ &$= 0.220 \log {\rm age} + 0.342 {\rm [M/H]} - 0.363 {\rm [\alpha/Fe]} + 0.512$  & 0.022 &  Thomas et al. \\
$\log {\rm C4668}  $ &$= 0.145 \log {\rm age} + 0.581 {\rm [M/H]} + 0.023 {\rm [\alpha/Fe]} + 0.529$  & 0.037 &  Thomas et al. \\
$\log {\rm Mg}b    $ &$= 0.173 \log {\rm age} + 0.309 {\rm [M/H]} + 0.210 {\rm [\alpha/Fe]} + 0.354$  & 0.019 &  Thomas et al. \\
$\log \langle {\rm Fe} \rangle   $ &$= 0.113 \log {\rm age} + 0.253 {\rm [M/H]} - 0.278 {\rm [\alpha/Fe]} + 0.343$  & 0.007 &  Thomas et al. \\
\enddata
\tablenotetext{a}{ $({\rm H\delta _A + H\gamma _A})' \equiv -2.5~\log \left ( 1.-({\rm H\delta _A + H\gamma _A})/(43.75+38.75) \right )$, cf.\ Kuntschner (2000).
The rms for the relation translates to an rms of 
$\rm H\delta _A + H\gamma _A$ of $\approx 0.65$ for the typical values of $\rm H\delta _A + H\gamma _A$. 
}
\tablecomments{ (1) Relation established from the published model values. 
${\rm [M/H]}\equiv \log Z/Z_\sun$ is the total metallicity relative to solar.
$[\alpha /\rm{Fe}]$ is the abundance of the $\alpha$-elements relative to iron, and relative to the
solar abundance ratio. The age is in Gyr. The M/L ratios are stellar M/L ratios in solar units. 
(2) Scatter of the model values relative to the relation. (3) Reference for the model values. }
\end{deluxetable*}

Seven of the cluster members have detectable emission lines. 
For these galaxies we determined the equivalent width of the [\ion{O}{2}]$\lambda\lambda$3726,3729 doublet,
in the following referred to as the ``[\ion{O}{2}] line''. 
With an instrumental resolution of $\sigma \approx 3$\,{\AA} (FWHM $\approx 7$\,{\AA}), 
the doublet is not resolved in our spectra.
With our wavelength coverage, the [\ion{O}{2}] line is the only emission 
line that can be measured in the galaxies that are members of RXJ0152.7--1357.
Table \ref{tab-specline} in the Appendix lists the derived line indices and
the measurements of the  [\ion{O}{2}] equivalent widths.

\section{Low redshift comparison data}

The reference sample of galaxies at low redshift used in this paper
consists of 63 galaxies in the Perseus cluster and 17 galaxies
in the cluster Abell 194. Both clusters are at redshift $z=0.018$. 
The Perseus sample covers the central 100 arcmin $\times$ 60 arcmin of 
the cluster. For E and S0 galaxies determination of line indices the 
sample is 96 per cent complete to B=16.05 mag (absolute B-band magnitude of $-18.4$ mag).
The Abell 194 sample is not complete, but covers galaxies of similar
luminosities.
The data for these two clusters will be published and discussed in detail
in a future paper.
We also use the velocity dispersions and photometry for the Coma 
cluster galaxies from J\o rgensen (1999).  This sample contains 116 galaxies
with measurements of the velocity dispersions.
The sample covers the central 64 arcmin $\times$ 70 arcmin of the cluster.
For E and S0 galaxies the sample is 93 per cent complete to B=16.2 mag 
(absolute B-band magnitude of $-18.9$ mag).
All of the galaxies in the low redshift sample are on the red sequence
of the color-magnitude relation and are classified as early-type (E or S0).
Table \ref{tab-compdata} summarizes the low redshift comparison data.

The measurement techniques used for the low redshift comparison data are identical
to those used for our RXJ0152.7--1357, except for the measurements of the 
velocity dispersions of the Coma cluster galaxies. The velocity dispersions
for the Coma cluster galaxies were derived using a Fourier Fitting Technique,
rather than fitting in pixel space. The line-of-sight velocity distribution
for the fits were assumed to be Gaussian. Further, a small fraction of the
Coma cluster velocity dispersions comes from earlier published data.
J\o rgensen (1999) calibrated all the Coma cluster velocity dispersions to
a consistent system. We use the data as given in that paper.
We have tested the consistency of the velocity dispersions of the 
Coma cluster galaxies with those of the Perseus cluster by comparing the 
relation between Mg$b$ and the velocity dispersions
for the two clusters. Under the assumption that this relation is the same
for the two clusters, we find that the velocity dispersions are consistent
within $\Delta \log \sigma = 0.026$. We use this value as a measure of 
the systematic errors that may affect our results due to possible 
inconsistent calibration of the velocity dispersion measurements.
Because of the similar measurement techniques used for all other parameters,
we will assume that the systematic errors in the velocity dispersion
will dominate over possible systematic errors in other parameters.

\section{Stellar population models and evolutionary scenarios}

In this section we first describe the SSP models that 
we have chosen to use. We then discuss the difficulty of using these models
to derive luminosity weighted ages, metal content [M/H], 
and the abundance ratios of the $\alpha$-elements $[\alpha /\rm{Fe}]$.
Finally, we outline the simple evolutionary scenarios that we will reference
in the analysis of the data.

\subsection{Single stellar population models}

In order to interpret the spectroscopic data we use SSP models.
Most SSP models in the literature assume abundance ratios in
agreement with the stars in the solar neighborhood. 
We refer to these as models using ``solar abundance ratios'', though strictly
speaking the abundance ratios may not be solar for the low metallicity
models tied to the abundance ratios of the stars in the solar neighborhood.
Since nearby E and S0 galaxies are known to have non-solar abundance
ratios, specifically the $\alpha$-elements are over-abundant relative to iron
compared to solar abundances, 
we have chosen to use the models from Thomas et al.\ (2003, 2004).

Thomas et al.\ model the Lick/IDS
indices for ages between 1 Gyr and 15 Gyr and total metallicities $\rm [M/H]$ between 
$-2.25$ and 0.67, and they include non-solar abundance ratios for the 
$\alpha$-elements. Models are available for $[\alpha/\rm{Fe}]$= 0.0, 0.2, 0.3 and 0.5.
The M/L ratios for the $[\alpha/\rm{Fe}]$= 0.0 models are published by Maraston (2004).
The models from Thomas et al.\ (2003, 2004) treat the elements
N, O, Mg, Ca, Na, Ne, S, Si, and Ti as $\alpha$-like elements, 
though N is not an $\alpha$-element. $[\alpha/\rm{Fe}]$ from these models
should be interpreted as the abundance of these elements relative to the 
iron-peak elements Cr, Mn, Fe, Co, Ni, Cu, and Zn.
The abundance of carbon is kept fixed in the models.
It is also important to keep in mind that while magnesium and oxygen are primarily 
produced by massive stars and redistributed into the ISM by SNe type II,
elements like carbon and nitrogen primarily originate from intermediate mass
stars and are redistributed into the ISM by these stars during their AGB phase.
Therefore the time scale for the production of carbon and nitrogen is somewhat
longer than that of magnesium and oxygen, though it is shorter than the timescale
for production of the iron-peak elements, which primarily are produced by SNe type Ia. 
The reader is referred to, e.g., 
Chiappini et al.\ (2003), S\'{a}nchez-Bl\'{a}zquez et al.\ (2003)
and Carretero et al.\ (2004) for discussions on the evolution of the 
various element abundances and the implications for the galaxy assembly timescales.
As a further caution, Thomas et al.\ model the effects of
the non-solar $[\alpha/\rm{Fe}]$ on the stellar atmospheres, using the
results from Tripicco \& Bell (1995). The models do not use
evolutionary tracks for non-solar $[\alpha/\rm{Fe}]$. Thus, strictly speaking the 
atmospheres of the models are inconsistent with the stellar interiors of the models. 
However, at the present we find that these models are the most
useful for the interpretation of our data in terms of non-solar abundance ratios.

The indices CN3883, CaHK, and D4000 are not included in the models from Thomas et al.
For D4000 we use models from Vazdekis et al.\ (1996). 
We use the updated 2000-models available on Vazdekis' web-site. 
These models use new isochrones from Girardi et al.\ (2000). 
In the following we refer to the models as the Vazdekis-2000 models. 
The models assume solar abundance ratios, $[\alpha/\rm{Fe}]$=0. 
We use the Vazdekis-2000 models with a Salpeter (1955) IMF, 
in order to match the IMF used by Thomas et al.\ and Maraston.
For CN3883 and CaHK, we use model spectra from Bruzual \& Charlot (2003)
for ages between 1 Gyr and 15 Gyr, and [M/H] of $-0.4$, 0.0, and 0.4, a Salpeter
IMF and the Padova-1994 isochrones.
These models also assume solar abundance ratios. We convolve the model spectra
to the Lick/IDS resolution and derive the line indices using the same method
as used for the observational data. 

\begin{figure*}
\epsfxsize 12.5cm
\epsfbox{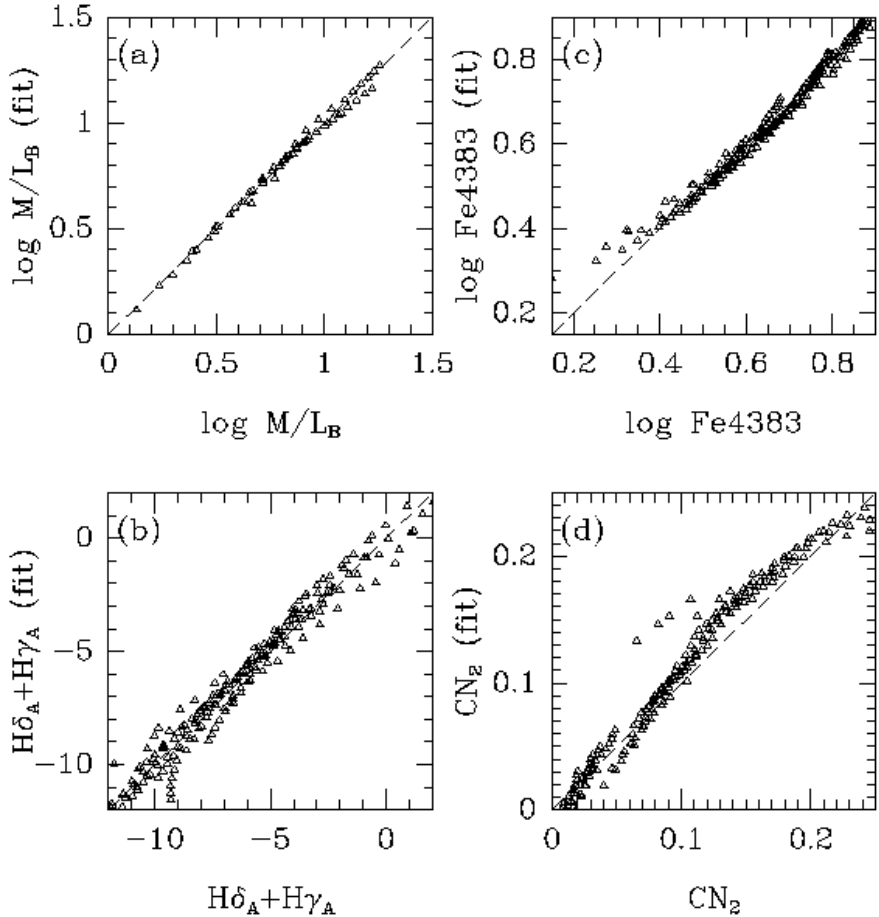}
\caption[]{The model relations listed in Table \ref{tab-models} 
are shown as the values from the fit versus the actual model values.
The lines are the one-to-one relations. If the fits were perfect representations
of the models all the points would be on the one-to-one relations.
Only for CN$_2$ does the fit deviate systematically from the actual
model values, see text for discussion.
\label{fig-models} }
\end{figure*}

When we use the models together, we implicitly assume that the M/L ratios 
and the indices CN3883, CaHK, and D4000 do not depend on $[\alpha/\rm{Fe}]$. 
For the M/L ratios, this may not be a valid assumption. 
Work by Thomas \& Maraston (2003) indicates that the blue luminosity increases with
increasing $[\alpha/\rm{Fe}]$. However, these authors also conclude that the evolutionary
tracks for non-solar $[\alpha/\rm{Fe}]$ do not yet yield realistic ages for elliptical 
galaxies. They find that for tracks with non-solar $[\alpha/\rm{Fe}]$ 
nearby elliptical galaxies become unrealistically old.

In order to interpret the offsets in the scaling relations in terms of differences
in mean ages, metal content, and/or $[\alpha/\rm{Fe}]$, we have used the 
published stellar population model values to establish linear relations between the 
observables and those three parameters.
The relations were derived as least squares fits with the residuals minimized in
the measurable quantities, e.g. log M/L, D4000 etc.
The relations are listed in Table \ref{tab-models}.
For the models from Thomas et al.\ and Maraston, the models were fit for ages 
from 2 Gyr to 15 Gyr, and $\rm [M/H]$ from $-0.33$ to 0.67.
All available $[\alpha/\rm{Fe}]$ values for the models from Thomas et al.\ were included. 
The Vazdekis-2000 models were fit for $\rm [M/H]$ from $-0.68$ to 0.2 and ages of 
2-16 Gyr.  For reference we give the linear relations for the M/L ratios based on 
both the Vazdekis-2000 and the Maraston models.
The line indices derived from the Bruzual \& Charlot models were fit for ages of 2-15 Gyr
and [M/H]=--0.4, 0.0, and 0.4.
The relations listed for the ``visible indices'' Mg$b$, $\langle {\rm Fe} \rangle$, 
and $\rm H\beta _G$ are discussed in Section 6.2.
The model values for $\rm H\beta$ were converted to $\rm H\beta _G$ using the calibration
from J\o rgensen (1997).
Figure \ref{fig-models} shows the relations from Table \ref{tab-models} that are most
important for our conclusions regarding differences in ages and $[\alpha/\rm{Fe}]$.
The fits for the M/L ratio, the higher order Balmer lines, and for Fe4383 show no
systematic effects over the range in observable parameters relevant to our analysis.
The fit for CN$_2$ shows a small systematic effect. The four points deviating most from
the fit are the models for the most metal rich ([M/H]=0.67) and youngest (2 Gyr)
populations. Further, CN$_2$ has a weak non-linear dependency on age for the highest 
metallicity models ([M/H]=0.67). 
The relation will lead to an underestimation of the age dependency for 
stellar populations with very high metallicities.

For the interpretation of the emission line equivalent widths, we use the models from
Magris et al.\ (2003). These authors combine Bruzual \& Charlot model spectra with
emission line spectra from the photoionized gas around massive stars on the main sequence.
We further combine these models with spectra of old stellar populations
using the Bruzual \& Charlot models. The details of this analysis are presented 
in Section 8.3.

\begin{figure*}
\epsfxsize 15.0cm
\epsfbox{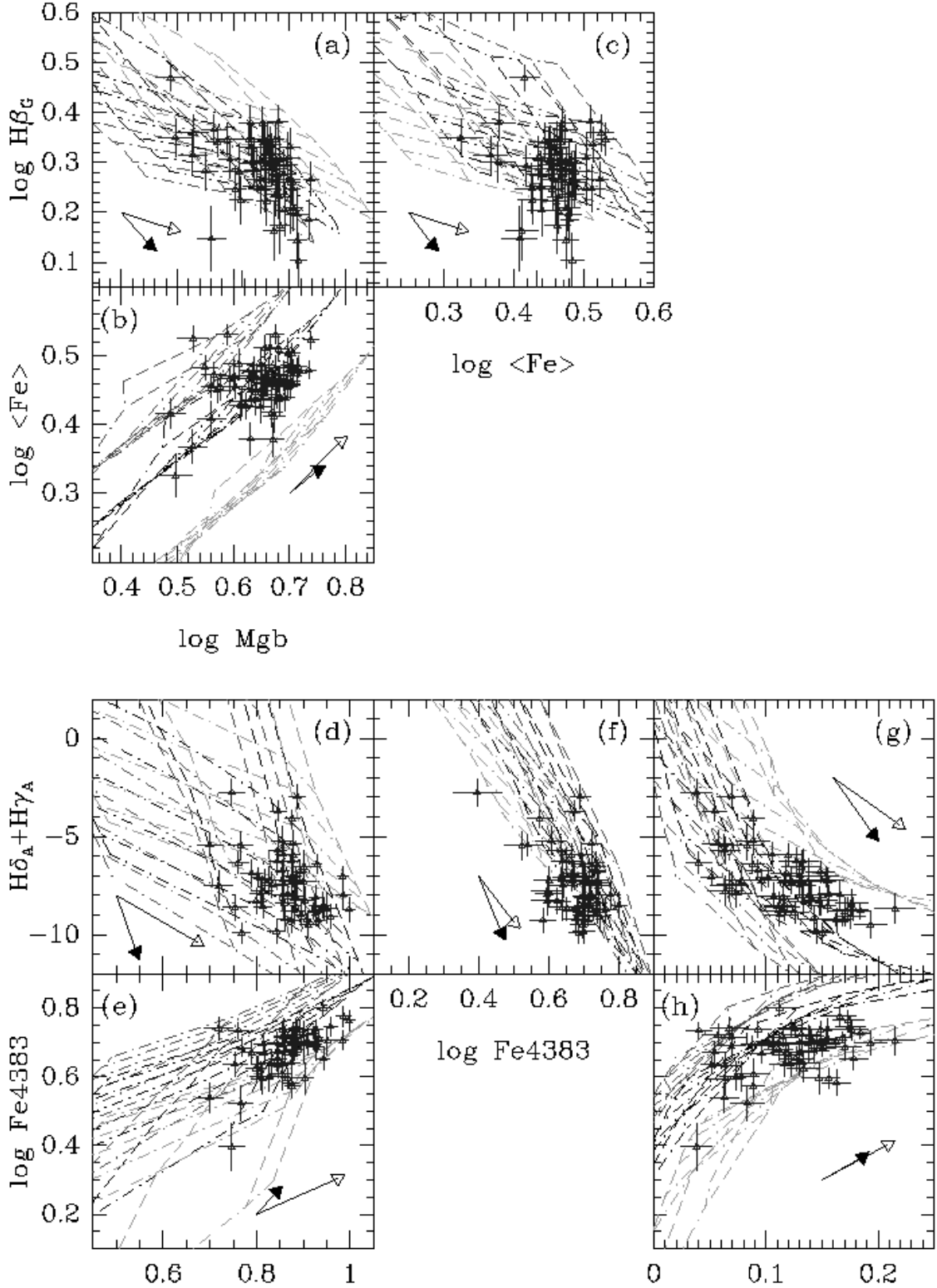}
\caption[]{Line indices plotted versus each other, only the data for the Perseus and Abell194 
sample are shown. Panels (a), (b), and (c) show the indices $\rm H\beta$, Mg$b$, and $\langle {\rm Fe}\rangle$,
which are usually used in the visible wavelength region for investigation of ages, 
metal content and $[\alpha /\rm{Fe}]$.
Panels (d), (e), and (f) show the indices $\rm H\delta _A + H\gamma _A$, C4668, and Fe4383,
which are considered as an alternative to the indices in the visible region.
Panels (g) and (h) show the index CN$_2$ which may be used in place of C4668.
The SSP models from Thomas et al.\ (2003, 2004) are shown as model grids.
We show the models for  $[\alpha /\rm{Fe}] =0.0$, 0.2 and 0.5, as red (dark grey), black and green (light grey) grids, respectively.
Dashed lines -- lines of constant [M/H], for values of -0.33, 0.0, 0.35 and 0.67.
Dot-dashed lines -- lines of constant ages, for values of 1, 2, 3, 5, 8, 11, and 15 Gyr.
The arrows show the approximate changes in the indices for a change of $\Delta \log \rm age = 0.3$
(solid arrow) and $\Delta \rm [M/H]=0.3$ (open arrow).
 \label{fig-metalcomp} }
\end{figure*}

\subsection{Modeling visible and blue line indices}

Ideally, we want to use the SSP models to derive luminosity
weighted mean ages, metal content [M/H], and the $\alpha$-element abundance ratios
$[\alpha /\rm{Fe}]$ of the stellar populations in the galaxies.
In the majority of studies of nearby galaxies, line indices in the visible region
have been used for this purpose. 
Typically the indices $\rm Mg_2$ (or Mg$b$), $\langle {\rm Fe}\rangle$, and H$\beta$ have been used
together with SSP models (e.g., J\o rgensen 1999, Trager et al.\ 2000). 
We call these indices the ``visible indices''.
Figures \ref{fig-metalcomp}a-c show the visible 
indices for our low redshift sample (Perseus and Abell 194) together 
with the SSP models from Thomas et al.\ (2003, 2004).
All three indices depend on both age, [M/H] and $[\alpha /\rm{Fe}]$, but in 
different ways such that it is possible to derive these three parameters from
the measured line indices with the aid of the SSP models. The derived 
values should always be interpreted as the luminosity weighted mean values.

For galaxies at redshift of about 0.6 or larger measuring the visible indices gets
increasingly difficult as the wavelength regions for these indices are 
redshifted into the far red and often affected by the sky subtraction residuals due to
the strong sky lines in this wavelength region. 
For galaxies in RXJ0152.7--1357, none of the visible indices can be measured reliably. 
We are therefore forced to use indices in the rest frame blue. We call these the ``blue
indices''. Thomas et al.\ (2003) discuss various such alternatives to the visible indices. 
Since ages, metallicities
and abundance ratios derived for real galaxies using SSP models are luminosity
weighted quantities, using indices in the rest frame blue instead of at longer
wavelengths will give stronger weight to younger stellar populations that 
dominate the flux at short wavelengths. Thus, it becomes important when tracking
the differences in ages, metallicities and abundance ratios to use a consistent
set of line indices. However, for correct models of the various indices we would
still expect tight correlations between the quantities derived using visible
indices versus using blue indices.

In Figures \ref{fig-metalcomp}d-h
we show the blue indices for our low redshift sample together with the same
SSP models from Thomas et al.\ (2003, 2004) as shown for the visible indices.
The figure shows that the blue indices C4668 (or CN$_2$), Fe4383 and $\rm H\delta _A + H\gamma _A$,
are not simply equivalent to the visible indices Mg$b$, $\langle {\rm Fe}\rangle$, and H$\beta$. 
The blue indices depend in different ways on the age, [M/H] and $[\alpha /\rm{Fe}]$.
The dependencies for both the visible indices and the blue indices are summarized in 
Table \ref{tab-models}.
The Balmer lines primarily depend on the age, but the higher order Balmer lines
have a stronger dependency on $[\alpha /\rm{Fe}]$ than is the case for H$\beta$, as 
described by Thomas et al.
The blue iron index Fe4383 has a strong age dependency. That combined with
the $[\alpha /\rm{Fe}]$ dependency of the higher order Balmer lines make the models in
$\rm H\delta _A + H\gamma _A$ versus Fe4383 almost degenerate.
It is also clear that many of the galaxies have weaker Fe4383 for their Balmer line
strength than predicted by any of the models.
The index C4668 depends very little on $[\alpha /\rm{Fe}]$, compared to its visible
``equivalent'' the Mg$b$ index. Therefore, C4668 versus Fe4383 is less successful in separating
the models according to abundance ratio, than found for Mg$b$ versus $\langle {\rm Fe}\rangle$.
The CN$_2$ appears slightly more useful, but this index is weaker and more difficult to 
measure than C4668.

If we derive age, [M/H] and $[\alpha /\rm{Fe}]$ from \{Mg$b$, $\langle {\rm Fe}\rangle$, H$\beta$\} and from
\{C4668, Fe4383, $\rm H\delta _A + H\gamma _A$\} using the SSP models, it turns out that
the resulting values differ from each other in systematic ways. The general trends
can be seen directly from Figure \ref{fig-metalcomp}. The abundance ratio $[\alpha /\rm{Fe}]$
based on the blue indices is systematically higher than when using the visible indices.
This becomes most obvious when comparing the location of the data points relative
to the model grids in Figures \ref{fig-metalcomp}b and e (or h). The difference is about 0.15 dex.
The metal content [M/H] resulting from the blue indices is slightly lower than when
derived from the visible indices. On Figure \ref{fig-metalcomp}, this can be seen by
comparing panels (a) and (d), keeping the difference in $[\alpha /\rm{Fe}]$ in mind.
The resulting ages span the same range when using the two 
sets of indices, though there is no one-to-one correlation between the ages derived for 
the individual galaxies.

It is beyond the scope of the present paper to resolve these issues, which we believe 
are intrinsic to the SSP models. To our knowledge, there are no other published
studies in which the Thomas et al.\  SSP models have been used to derive 
age, [M/H] and $[\alpha /\rm{Fe}]$ using the blue indices only. 
We have chosen not to present luminosity weighted mean 
ages, [M/H], and $[\alpha /\rm{Fe}]$ derived from
the blue indices. Instead we discuss the measured parameters for RXJ0152.7--1357 
compared to the low redshift sample and to the SSP models.
We primarily discuss the differences between the low redshift sample and 
the  RXJ0152.7--1357 sample, rather than the absolute values of ages, 
[M/H], and $[\alpha /\rm{Fe}]$.

\subsection{Evolutionary scenarios}

In order to simplify the analysis and discussion of our data, we will
refer to some simple evolutionary scenarios
the galaxies may experience between redshift 0.83 and the present. These
scenarios do not represent an exhaustive list of possibilities, but serve as
a framework for our discussion. The scenarios are as follows.

\newcounter{evol}
\begin{list}{(\arabic{evol})}{\usecounter{evol}}
\item Pure passive evolution: In this scenario there is no additional star formation
between $z = 0.83$ and the present. 
The stellar populations present at $z = 0.83$ simply age passively, 
while no other changes take place. 
The difference between the luminosity weighted mean ages of the stellar populations in 
the galaxies at $z = 0.83$ and the present is equal to the lookback time to $z = 0.83$
(7 Gyr with our adopted cosmology).
\item New star formation without galaxy mergers: Like scenario (1), the stellar populations
already present at $z = 0.83$ age passively. In addition new stars are formed. 
Thus, the difference between the luminosity weighted mean ages of the stellar populations
in the galaxies at $z = 0.83$ and the present is smaller than the lookback time to $z = 0.83$.
There may also be differences in [M/H] and $[\alpha /\rm{Fe}]$, 
depending on the details of the star formation.
\item Merging of galaxies and new star formation (perhaps 
limited to the galaxies affected by merging):
The stellar populations already present at $z = 0.83$ age passively.
As for scenario (2), due to the star formation the difference between the luminosity 
weighted mean ages of the stellar populations in the galaxies at $z = 0.83$ and 
the present is smaller than the lookback time to $z = 0.83$. 
There may be differences in [M/H] and $[\alpha /\rm{Fe}]$, 
depending on the details of the star formation and the galaxies entering the mergers.
\item Merging of galaxies without new star formation: In this case, the details of the 
galaxies undergoing merging determine the differences in the
luminosity weighted mean ages, [M/H] and $[\alpha /\rm{Fe}]$
between the stellar populations at $z=0.83$ and the present. 
However, since no new stars are formed we expect that the age difference 
is larger in this scenario than in scenario (3).
\end{list}
For all the scenarios, a potential complication is the sample selection and whether we 
succeed in observing the high redshift progenitors to the galaxies in the low redshift comparison 
sample.
In the following we refer to scenario (1) as ``passive evolution''. Only in this case is it 
always expected that the difference between luminosity weighted mean ages at $z = 0.83$ and 
the present is equal to the lookback time at $z = 0.83$. 
When we state that the data are not in agreement with the passive evolution model, it
means that the differences between the stellar populations of the galaxies 
in RXJ0152.7--1357 and those in our low redshift sample cannot
solely be explained by an age difference equal to the lookback time.

The main question we attempt to address is if we can find a model for the evolution
that makes the stellar populations of the galaxies in RXJ0152.7--1357 evolve into 
stellar populations similar to those in our low redshift comparison sample, 
within the available time which is about 7 Gyr.

\begin{deluxetable*}{rrrrrrrrrr}
\tablecaption{Optical Data for X-ray Point Sources \label{tab-xraysources} }
\tablewidth{0pc}
\tablehead{
\colhead{ID} & \colhead{RA (J2000)} & \colhead{DEC (J2000)\tablenotemark{a}} & 
\colhead{$r'_{\rm total}$} & \colhead{$i'_{\rm total}$} & 
\colhead{$z'_{\rm total}$} & \colhead{$(r'-i)$} & \colhead{$(i'-z')$} & \colhead{FWHM\tablenotemark{b}} & \colhead{class\tablenotemark{c}} }
\startdata
1056 & 1 52 39.78 & -13 57 41.4 & 21.55  & 20.53 & 19.77 & 0.906 & 0.746 & 1.60 & 0.03 \\
1397 & 1 52 43.75 & -13 59 01.8 & 20.83  & 20.58 & 19.99 & 0.077 & 0.519 & 1.21 & 0.41 \\
1824 & 1 52 34.66 & -13 59 30.3 & 20.82  & 20.12 & 19.48 & 0.489 & 0.654 & 1.38 & 0.03 \\
\enddata
\tablenotetext{a}{Positions are consistent with USNO, with an rms scatter of $\approx 0.7$ arcsec.}
\tablenotetext{b}{The FWHM of the object in units of the FWHM of a nearby point source in the
i-band image from GMOS-N.}
\tablenotetext{b}{SExtractor {\it class\_star} from the $i'$-band image.}
\tablecomments{Units of right ascension are hours, minutes, and seconds, and
units of declination are degrees, arcminutes, and arcseconds.}
\end{deluxetable*}

\begin{figure}
\epsfxsize 7.5cm
\epsfbox{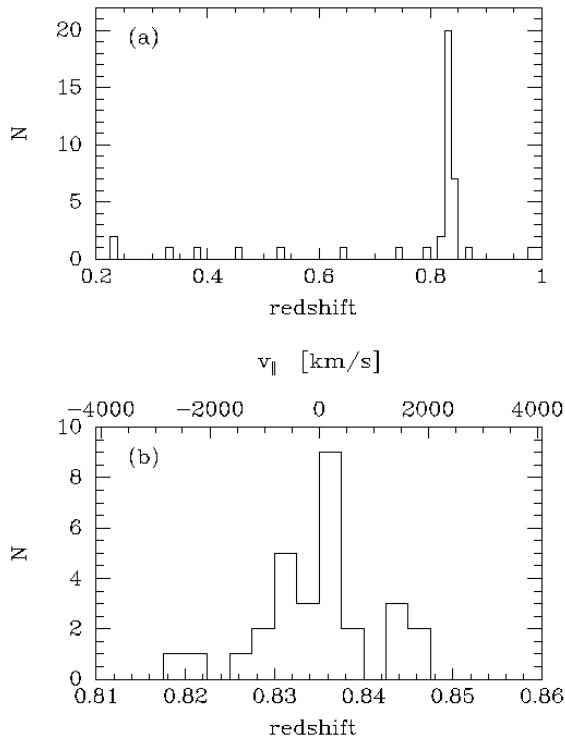}
\caption[]{Redshift distribution of the spectroscopic sample. (a) The full sample. (b) The cluster members.
The top axis on panel (b) gives the radial velocity relative to the cluster redshift 
and in the rest frame of the cluster, $v_{\|} = c (z - z_{\rm cluster}) / (1+z_{\rm cluster})$.
\label{fig-zhist} }
\end{figure}

\section{Cluster velocity dispersion and possible sub-structure}

In this section we establish the criteria for a given galaxy being a cluster
member. We also briefly review the evidence showing that RXJ0152.7--1357 is in the 
process of merging from two sub-clumps. In Section 9, we use
this information in the discussion of the stellar populations 
of the galaxies.

Figure \ref{fig-zhist} shows the redshift distribution of the spectroscopic sample.
In order to derive the median cluster redshift and the the line-of-sight cluster velocity dispersion
we first exclude galaxies more than 3000 km/s from the median redshift.
We then use the biweight method described by Beers et al.\ (1990) to derive
the cluster redshift and velocity dispersion. 
The uncertainties are derived using a bootstrap method.
We note that none of the conclusions would change if we instead had used
the methods described by Danese et al.\ (1980).
We find a cluster redshift of $z_{\rm cluster}=0.8350\pm0.0012$,
and a line-of-sight cluster velocity dispersion, 
$\sigma _{\rm cluster}= (1110 \pm 163) \rm km\,s^{-1}$.
We consider all galaxies within $\pm 3 \sigma _{\rm cluster}$ of the cluster
redshift to be cluster members.
Of the 41 observed galaxies, 29 are cluster members.
Table \ref{tab-speckin} in the Appendix lists which galaxies are considered cluster
members.

Figure \ref{fig-grey} shows the {\it XMM-Newton} data overlaid on our $i'$-band image,
with the cluster members labeled.
The X-ray data for RXJ0152.7--1357 indicates that the cluster is still in the
process of merging from the two sub-clumps that can be seen in the main
X-ray structure (Maughan et al.\ 2003).
The weak lensing analysis presented by Jee et al.\ (2004) also supports
that RXJ0152.7--1357 is an on-going merger.

The three strong X-ray point sources seen on Figure \ref{fig-grey} have obvious 
optical counterparts, all of which are extended. 
Table \ref{tab-xraysources} summarizes the positions and ground based photometry
for these three optical sources.
Two of the sources (ID 1056 and 1397) are included on the HST/ACS imaging
of the cluster, which shows that the sources are galaxies with well-defined spiral arms.
The galaxies were not included in our spectroscopic sample. 
However, Ford et al.\ (2004) have obtained spectroscopy of ID 1056 and 1397, and find
that these are Seyfert galaxies and members of the cluster.
It is not know if the third source ID 1824 is a member of the cluster. 
Maughan et al.\ comment on a fourth X-ray point source. This source is located at
$(\rm{RA}_{\rm J2000},{\rm DEC}_{\rm J2000})=(\rm 1^h52^m41\fs 4,\,-13\degr 59\arcmin 20\farcs 1)$.
We do not find any optical counter part for this source.

From the {\it XMM-Newton} data we estimate that the flux in the three point sources
is about 20 per cent of the total X-ray flux within the central 3.7 arcmin $\times$ 3.7 arcmin
(corresponding to 1.7 Mpc $\times$ 1.7 Mpc).
If we correct the total X-ray luminosity of RXJ0152.7--1357 for these point sources
then the resulting luminosity is about 1.4 times that of the Coma cluster.
The cluster velocity dispersion is only slightly larger than that of
the Coma cluster. The Coma cluster has a line-of-sight velocity dispersion of 
$\sigma _{\rm cluster}= 1010 ^{+51}_{-44} \rm km\,s^{-1}$ (Zabludoff et al.\ 1990).  
Mahdavi \& Geller (2001) established the  $L_X - \sigma$ relation for rich
clusters as $L_X \propto \sigma ^{4.4}$, with a scatter of 0.182 dex in $L_X$.
Thus, if the Coma cluster is on the $L_X - \sigma$ relation (no zero point 
for the relation is given by Mahdavi \& Geller), then RXJ0152.7--1357
is within 0.05 dex in $L_X$ of the relation, and therefore consistent with the relation.
We also note that Maughan et al.\ find excess X-ray emission between
the two sub-clusters, but at a very low level.
The X-ray luminosity of RXJ0152.7--1357 is high enough that it would have
been well above our lower sample limit even without the contribution from the three
X-ray point sources and the excess X-ray emission between the two sub-clusters.

\begin{deluxetable*}{lrrrrrrrrl}
\tablecaption{Scaling Relations \label{tab-relations} }
\tablewidth{0pc}
\tablehead{
\colhead{Relation} & \multicolumn{3}{c}{Low redshift sample} & \multicolumn{3}{c}{RXJ0152.7--1357} & \colhead{$\Delta \gamma$} & 
\colhead{$\sigma _{\rm sys}$} & \colhead{min.coord.} \\
 & \colhead{$\gamma$} & \colhead{$N_{\rm gal}$} & \colhead{rms} &  \colhead{$\gamma$} & \colhead{$N_{\rm gal}$} & \colhead{rms} \\
\colhead{(1)} & \colhead{(2)} & \colhead{(3)} & \colhead{(4)} & \colhead{(5)} & \colhead{(6)} & \colhead{(7)} & \colhead{(8)} & \colhead{(9)} & \colhead{(10)} }
\startdata
$\rm{M_B} = (-8.02 \pm 1.08) \log \sigma + \gamma$  & -2.29 & 116 & 0.81 & -3.16 & 26 & 0.85 & $-0.87\pm 0.18$  & 0.21 & perpendicular\tablenotemark{a} \\
$\rm{H\delta _A + H\gamma _A} = -9.1 \log \sigma + \gamma$         & 13.16 & 65 & 1.53 & 16.64\tablenotemark{c} & 21 & 1.71 & $3.48 \pm 0.42$ & 0.24 & \nodata\tablenotemark{b} \\
$\rm{H\delta _A + H\gamma _A} = (-7.0\pm 1.3) \log \sigma + \gamma$ & 8.39 & 65 & 1.47 & 11.88\tablenotemark{c} & 21 & 1.57 & $3.49 \pm 0.39$ & 0.18 & $\rm H\delta _A + H\gamma _A$\tablenotemark{d} \\
$\rm{H\delta _A + H\gamma _A} = (-5.3\pm 2.3)~{\rm D4000} + \gamma$ & 3.58 & 65 & 1.50 &  7.08 & 22 & 1.44 & $3.50 \pm 0.36$ & \nodata & $\rm H\delta _A + H\gamma _A$\tablenotemark{e} \\
$\rm{D4000} = \gamma$ & 2.10 & 65 & 0.16 & 2.05\tablenotemark{f} & 22 & 0.16 & $-0.05 \pm 0.04$ & \nodata & \nodata \\
${\rm CN3883}      = (0.29\pm 0.04) \log \sigma + \gamma$         & -0.411 & 65 & 0.051 & -0.400\tablenotemark{c} & 20 & 0.046 & $0.011 \pm 0.012$ & 0.008 & perpendicular \\
$\log {\rm CaHK}   = (0.14\pm 0.04) \log \sigma + \gamma$         &  0.997 & 65 & 0.048 &  1.019\tablenotemark{c} & 21 & 0.057 & $0.022 \pm 0.014$ & 0.004 & perpendicular \\
${\rm CN_2}        = (0.22\pm 0.06) \log \sigma + \gamma$         & -0.390 & 65 & 0.034 & -0.416\tablenotemark{c} & 21 & 0.049 & $-0.026 \pm 0.011$ & 0.006 & perpendicular\tablenotemark{d} \\
$\log {\rm G4300}  = (0.14\pm 0.08) \log \sigma + \gamma$         &  0.403 & 65 & 0.051 &  0.303\tablenotemark{c} & 21 & 0.11  & $-0.100 \pm 0.025$ & 0.004 & perpendicular\tablenotemark{d} \\
$\log {\rm Fe4383} = (0.19\pm 0.09) \log \sigma + \gamma$         &  0.263 & 65 & 0.063 &  0.037\tablenotemark{c} & 20 & 0.33  & $-0.226 \pm 0.074$ & 0.005 & perpendicular\tablenotemark{d} \\
$\log {\rm C4668}  = (0.33\pm 0.08) \log \sigma + \gamma$         &  0.107 & 65 & 0.058 &  0.063\tablenotemark{c} & 20 & 0.136 & $-0.044 \pm 0.031$ & 0.009 & perpendicular\tablenotemark{d} \\
%
%
\enddata
\tablecomments{(1) Scaling relation. (2) Zero point for the low redshift sample. (3) Number of galaxies
included from the low redshift sample. (4) rms in the Y-direction of the scaling relation for the low redshift sample.
(5) Zero point for the RXJ0152.7--1357 sample. (6) Number of galaxies
included from the RXJ0152.7--1357 sample. (7) rms in the Y-direction of the scaling relation for the RXJ0152.7--1357 sample.
(8) Zero point differences derived as ``RXJ0152.7--1357''--``low redshift''.
(9) Systematic uncertainties on $\Delta \gamma$, derived as 0.026 times the coefficient for $\log \sigma$.
(10) Coordinate for the minimization when fitting the scaling relation. ''Perpendicular'' means the residuals
are minimized perpendicular to the relation.  }
\tablenotetext{a}{Galaxies with $\log \sigma < 1.8$ excluded, emission line galaxies included.}
\tablenotetext{b}{Slope adopted from Kelson et al.\ (2001).}
\tablenotetext{c}{Emission line galaxies and galaxies with $\log \sigma < 1.8$ excluded.}
\tablenotetext{d}{Slope derived from fit to low redshift sample only}
\tablenotetext{e}{Slope derived from fit to RXJ0152.7--1357 sample only, emission line galaxies excluded}
\tablenotetext{f}{Emission line galaxies excluded.}
\end{deluxetable*}

Two of the galaxies in our sample, ID 1682 and 1614, are associated with
the diffuse X-ray emission to the east of the cluster. 
The mean redshift for these two galaxies is 0.8448, confirming the 
result from Maughan et al.\ that galaxies in this group have a higher redshift
than that of the main cluster.

Maughan et al.\ find that the galaxies associated with the two X-ray sub-clumps have 
slightly different mean redshifts.
We find $z=0.8372 \pm 0.0014$ for the northern sub-cluster (7 galaxies, 
ID 643, 737, 766, 813, 908, 1027 and 1085), and $z=0.8349 \pm 0.0020$
for the southern sub-cluster (6 galaxies, ID 1385, 1458, 1567, 1590, 1811, and 1920). 
This is a somewhat smaller difference than found Maughan et al., and barely 
significant.
We note that both of these sub-clusters of galaxies may have a lower 
line-of-sight velocity dispersion than that of the full sample, though 
of course the uncertainties are large for these small numbers of galaxies. 
We find $\sigma _{\rm northern} = (681 \pm 232) \rm km\,s^{-1}$ and
$\sigma _{\rm southern} = (866 \pm 266) \rm km\,s^{-1}$.
However, we cannot detect any difference between the velocity distribution of all the 
cluster members and a Gaussian distribution. A Kolmogorov-Smirnov test shows that the 
probability that the velocity distribution is drawn from a Gaussian distribution 
is larger than 60 per cent. This is not surprising since we find only a very
small difference between the mean redshifts of the two sub-clusters. If sub-clustering
can be detected from redshift data only, our sample is most likely too small to do so.

\begin{figure*}
\epsfxsize 10.5cm
\epsfbox{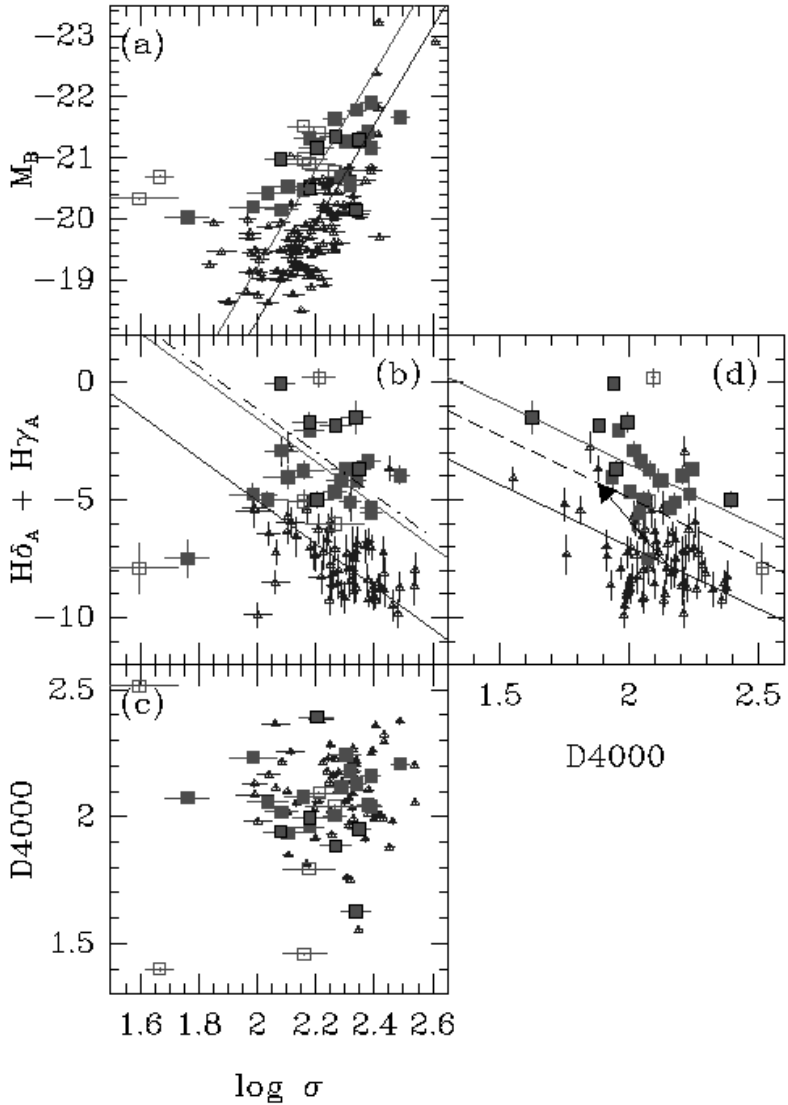}
\caption[]{The absolute B-band magnitudes $\rm M_B$, the Balmer line strengths $\rm H\delta _A + H\gamma _A$,
and the D4000 index shown as a function of the velocity dispersions. The Balmer line strengths are also
shown as a function of the D4000 index. Small blue (black) triangles -- low redshift sample; red (grey) filled squares -- 
galaxies in RXJ0152.7--1357 without emission lines; red (grey) open squares -- emission line galaxies in RXJ0152.7--1357.
The galaxies with low C4668 and G4300 indices discussed in Section 9, are marked with black boxes around the points.
Blue (black) solid lines -- relations for the low redshift sample; red (grey) solid lines -- relations for the RXJ0152.7--1357 
sample. The relations on the panels are listed in Table \ref{tab-relations}. The dot-dashed line on panel (b) 
is the relation from Kelson et al.\ (2001) for the cluster MS1054-03 at redshift 0.83. 
The arrow and the dashed line on panel (d) show how the relation for the low redshift sample 
would be translated in both parameters under assumption of passive evolution and a formation 
redshift of $z_{\rm form} =4.1$, see text for details.
\label{fig-agesfr} }
\end{figure*}

\section{Stellar populations at $z=0.83$}

In this section we characterize the stellar populations in the RXJ0152.7--1357
galaxies by using (1) scaling relations, (2) comparisons with model grids for line indices, and (3)
emission line strengths. The models and their limitations were described in Section 6.

\subsection{Scaling relations}

Table \ref{tab-relations} summarizes the scaling relations that we 
discuss in the following sections.
The scaling relations were fit by minimizing the sum of the absolute residuals.
The zero points are derived as the median of the measurements.
Except for relations involving $\rm H\delta _A + H\gamma _A$, the relations
were fit by minimizing the residuals perpendicular to the relation.
For $\rm H\delta _A + H\gamma _A$ we determine the fit by minimizing the 
residuals in $\rm H\delta _A + H\gamma _A$. 
We use minimization of the sum of the absolute residuals and median zero points
because this technique is very robust to the effect of outliers.
The uncertainties of slopes were derived with a boot-strap method.

Zero point differences between the RXJ0152.7--1357 sample and the low redshift 
sample are also listed in Table \ref{tab-relations}. In all cases the 
differences are derived as ``RXJ0152.7--1357''--``low redshift''.
The random uncertainties on the zero point differences, $\Delta \gamma$, are derived as
\begin{equation}
\sigma _{\Delta \gamma} = \left ( {\rm rms}_{{\rm low}\,z}^2/N_{{\rm low}\,z}
  +  {\rm rms}_{\rm RXJ0152}^2/N_{\rm RXJ0152} \right )^{0.5}
\end{equation}
where subscripts ``low $z$'' and ``RXJ0152'' refer to the low redshift sample
and the RXJ0152.7-1357 sample, respectively.
The systematic uncertainties on the zero point differences are expected to
be dominated by the possible inconsistency in the calibration of the 
velocity dispersions, 0.026 in $\log \sigma$ (cf.\ Section 5). For each scaling relation, we 
estimate the systematic uncertainty in the zero point difference as 
0.026 times the coefficient for $\log \sigma$, see Table \ref{tab-relations}.

\begin{figure*}
\epsfxsize 12.7cm
\epsfbox{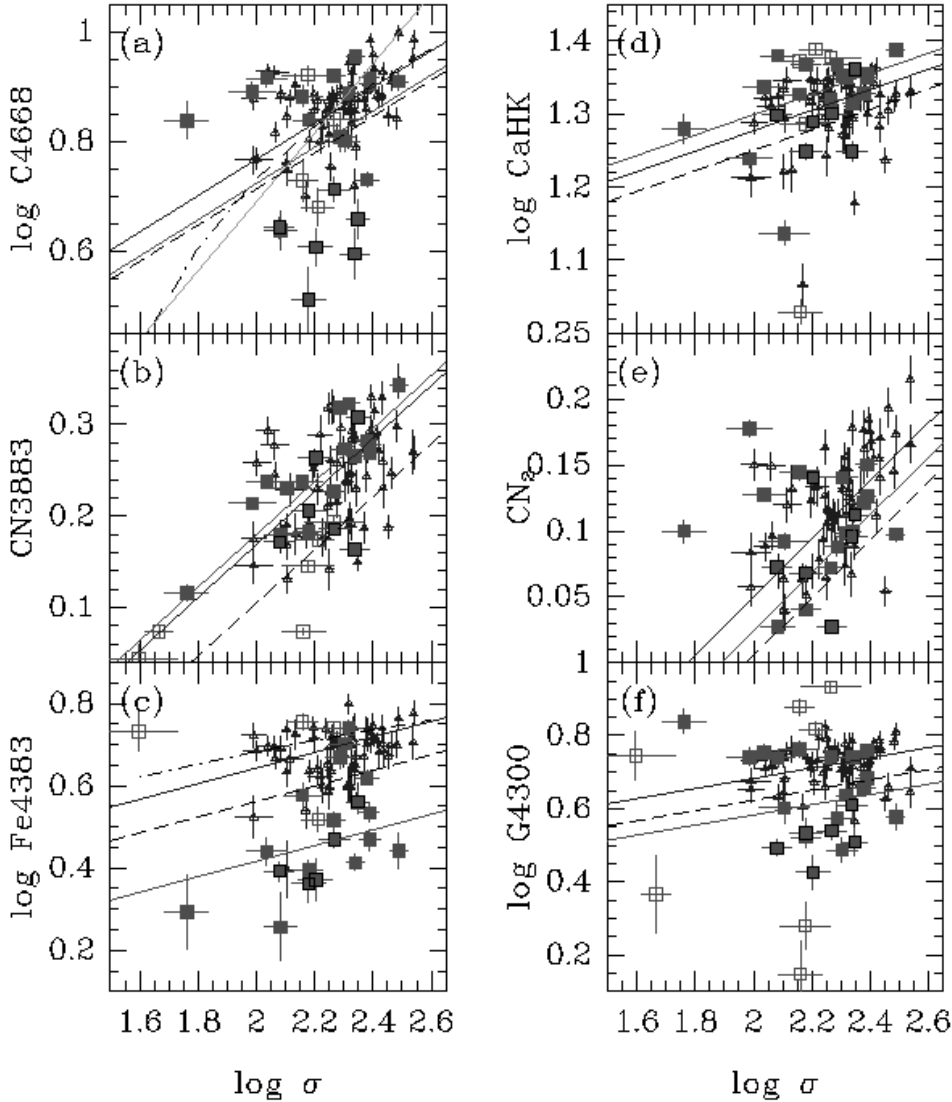}
\caption[]{Scaling relations for the metal indices. Symbols as in Figure \ref{fig-agesfr}. The emission line
galaxies in RXJ0152.7--1357 are not shown on panel (e) because none of the $\rm CN_2$ measurements
for these galaxies are reliable. Blue (black) solid lines -- relations for the low redshift sample; 
red (grey) solid lines -- relations for the RXJ0152.7--1357 galaxies. 
Green (grey steep) line -- relation for C4668 from J\o rgensen (1997) for low redshift galaxies. 
Dot-dashed lines -- relations from Kuntschner (2000) for galaxies in the Fornax cluster.
Dashed lines -- passive evolution predictions for $z_{\rm form}=4.1$, see text.
\label{fig-metal3} }
\end{figure*}

In Sections 8.1.1 and 8.1.2, we discuss the scaling relations for the age indicators
and the metal indices, respectively. The relations are shown on Figures \ref{fig-agesfr}
and \ref{fig-metal3}.

\subsubsection{Scaling relations for the age indicators}

The parameters most sensitive to age differences in the stellar populations
are the absolute rest frame B-band magnitudes, $\rm M_B$, and the strengths of the higher 
order Balmer lines, $\rm H\delta _A + H\gamma _A$. Because the strength of the
4000\AA\ break, D4000, in general is also used to estimate the ages of
stellar populations we include this index in the discussion of age indicators.
We use the scaling relations of these parameters to test if the data for 
RXJ0152.7--1357 are consistent with the hypothesis of pure passive evolution.

Figure \ref{fig-agesfr} shows the scaling relations for the age indicators.
The total magnitude $\rm M_B$ versus the velocity dispersion is the Faber-Jackson (1976) relation,
which is a projection of the FP.
The zero point difference between the RXJ0152.7--1357 sample and the
low redshift comparison sample 
is $\Delta \rm M_B = -0.87 \pm 0.18$, see Table \ref{tab-relations}. 
The uncertainty includes only the random uncertainties. The systematic uncertainty 
is estimated to be 0.21 based on the possible systematic uncertainties affecting
the measurements of the velocity dispersions. Adopting a slope of $-9.1$ or 
$-6.94$ instead of $-8.02$ does not change the zero point difference significantly.
We find $-0.89\pm0.20$ and $-0.92\pm0.15$ in the two cases, respectively (the uncertainties 
are the random uncertainties). 
The values $-9.1$ and $-6.94$ are the best fitting slope minus and plus, respectively,
the 1\,$\sigma$ uncertainty on the slope.
Thus, the accurate slope is not critical for the derived zero point.
Further, we have tested whether the distribution of the velocity dispersions
for the RXJ0152.7--1357 sample is significantly different from that of the low
redshift sample. A Kolmogorov-Smirnov test gives a probability of 24 per cent that
the two samples are drawn from the same parent sample. Thus, the two samples are 
not significantly different.
In summary, the zero point difference between the RXJ0152.7--1357 sample and the
low redshift comparison sample is $\Delta \rm M_B = -0.87 \pm 0.39$ including the 
systematic uncertainties.

For $\rm H\delta _A + H\gamma _A$ versus $\log \sigma$, we adopt the slope of $-9.1$ from
Kelson et al.\ (2001). The relation from Kelson et al.\ is shown on Figure \ref{fig-agesfr}b
with the zero point that these authors find for the cluster MS1054-03 at redshift 0.83.
Our data are in agreement with the data from Kelson et al.\ at redshift 0.83.
With the slope from Kelson et al., we find $\Delta (\rm H\delta _A + H\gamma _A) = 3.48 \pm 0.42$
(random uncertainty).
The adopted slope is not critical for the determination of the zero point difference.
The zero point difference does not change significantly if we instead use
the slope derived from fitting our low redshift sample and the RXJ0152.7--1357 sample,
cf.\ Table \ref{tab-relations}.
The systematic uncertainty on the zero point difference is 
$\approx 0.2$, see Table \ref{tab-relations}.

Using models from Maraston for the luminosities and models from Thomas et al.\ (2004) for the 
Balmer lines, we can translate these zero point differences to differences in the logarithm of the age in Gyr
under the assumption that [M/H] and $\rm [\alpha/\rm{Fe}]$ do not change, i.e., pure passive evolution.
For both parameters, we find $\Delta \log {\rm age}=-0.37$.
Including the systematic errors, the uncertainties are 
$\pm 0.17$ and $\pm 0.065$ on the measurement based on $\rm M_B$
and on $\rm H\delta _A + H\gamma _A$, respectively. 
Thus, $\rm H\delta _A + H\gamma _A$ provide the tighest constraint.
For the assumed cosmology, the look-back times at redshifts 0.02 and 0.83 
are 0.3 Gyr and 7.0 Gyr, respectively.
From the offset in $\rm H\delta _A + H\gamma _A$ we derive
the mean ages of the galaxies in RXJ0152.7--1357 must be $5.0^{+1.5}_{-1.0}$ Gyr,
and therefore that the lookback time to the epoch of formation must be 
$12.0^{+1.5}_{-1.0}$ Gyr.
This gives a formation redshift formation redshift $z_{\rm form} =4.1$, with a lower 
limit of 2.7 (one sigma uncertainty).
The 95 per cent confidence limit is $z_{\rm form} > 1.9$.
This is consistent with previous results based on either luminosity offsets 
(e.g., J\o rgensen et al.\ 1999; Kelson et al.\ 2000; Ziegler et al.\ 2001) 
or Balmer line strengths (Kelson et al. 2001).
Specifically, based on $\rm H\delta _A + H\gamma _A$
measurements of galaxies in four clusters,
Kelson et al.\ (2001) find 95 per cent confidence 
limits on $z_{\rm form}$ that are very similar to our result.

Next we examine if the D4000 indices are consistent with pure passive evolution and 
$z_{\rm form} =4.1$.
For passive evolution D4000 gets stronger with the age of the stellar population,
cf.\ Table \ref{tab-models}, see also Barbaro \& Poggianti (1997) for a more
detailed discussion of how the D4000 index depends on current and past star formation
rates.

Figure \ref{fig-agesfr}c and d show the D4000 index versus the velocity dispersion,
and  $\rm H\delta _A + H\gamma _A$ versus D4000. 
D4000 does not correlate with the velocity dispersion. A Kendall's $\tau$ 
correlation test (suitable for the small RXJ0152.7--1357 sample as
well as the larger low redshift sample) gives a probability of 60 per cent
or larger that there is no correlation between the two parameters.
Further, the distribution of the D4000 measurements for the low redshift sample
and for the RXJ0152.7--1357 are not significantly different.
A Kolmogorov-Smirnov test shows that there is a probability of about 60 per cent that
two samples are drawn from the same parent distribution. 
In Table \ref{tab-relations} we list the median values and rms scatter for the two samples. 
There is no significant difference in the median values of D4000, 
we find $\Delta \rm{D4000}=-0.05\pm0.04$. 

Pure passive evolution with $z_{\rm form} =4.1$ predicts an
offset in D4000 of $-0.28$, the index is expected to be weaker in the
RXJ0152.7--1357 galaxies.
On Figure \ref{fig-agesfr}d we illustrate this offset as the translation of
the low redshift relation between D4000 and $\rm H\delta _A + H\gamma _A$
that is predicted by passive evolution with $z_{\rm form} =4.1$.

\subsubsection{Scaling relations for the metal indices}

Figure \ref{fig-metal3} shows the indices for the metal lines, C4668, CN3883, Fe4383, CaHK, 
CN$_2$, and G4300 versus the velocity dispersions. Table \ref{tab-relations} summarizes the 
relations shown on the figure, the zero point differences, as well as the 
information about how the fitting was done.

All the metal indices depend on both the metallicity and the age of the 
stellar population. Thus, the scaling relations for the metal indices are 
expected to evolve for a pure passive evolution model.
Under the assumption that there is an age difference of $\Delta \log {\rm age} = -0.37$ 
between the RXJ0152.7--1357 sample and the low redshift sample, as derived 
from the B magnitudes and the higher order Balmer lines, we can derive the predicted 
offsets in the metal indices using the relations given in Table \ref{tab-models}. 
We assume that the velocity dispersions do not evolve, and then
use these offset to predict the location of the relations for the RXJ0152.7--1357 sample
by shifting the relations for the low redshift sample.
The predicted relations are shown on Figure \ref{fig-metal3}.
These predicted relations can be seen as the 
predictions for the pure passive evolution model with $z_{\rm form}=4.1$.
For C4668, $\rm CN_2$, G4300, and Fe4383, the predictions are based on 
models from Thomas et al.\ (2003). For CN3883 and CaHK, the predictions are
based on the Bruzual \& Charlot (2003) models.

Formally, the zero point differences for $\rm CN_2$, G4300, and Fe4383 are in marginal
disagreement with the pure passive evolution model.
If the stellar populations in both samples are very metal rich, the offset
predicted for pure passive evolution would be larger than shown for $\rm CN_2$
on Figure \ref{fig-metal3}e
due to the non-linear dependency on the age discussed in Section 6.1.
This does not significantly affect the conclusions we draw based on the data.
The differences between the model predictions and the data for CN3883 and CaHK are 
significant at the 5.4$\sigma$ and 3.5$\sigma$ level, respectively.

For C4668, G4300 and Fe4383 the scatter of the RXJ0152.7--1357 sample relative to
the adopted relations is more than twice the scatter of the low redshift sample.
The increased scatter is unlikely to be due to higher measurements uncertainties; in fact
the measurement uncertainties for the RXJ0152.7--1357 sample are in general lower
than those for the low redshift sample, see Figure \ref{fig-metal3}.
For CN$_2$ the scatter  of the RXJ0152.7--1357 sample is about 1.6 times that of the
low redshift sample, while for CN3883 and CaHK the scatter for the two samples is very similar.
The higher scatter of the RXJ0152.7--1357 sample is not an effect of more blue 
(star-forming) galaxies being included in that sample compared to the low redshift
sample. Except for the two blue emission-line galaxies in the RXJ0152.7--1357 sample,
all the galaxies in that sample are on the red sequence of the color-magnitude
diagram, as is the case for the low redshift sample. The emission line galaxies
in the RXJ0152.7--1357 sample are excluded from the determination of the 
scatter for all the relations.

\begin{figure*}
\epsfxsize 15.0cm
\epsfbox{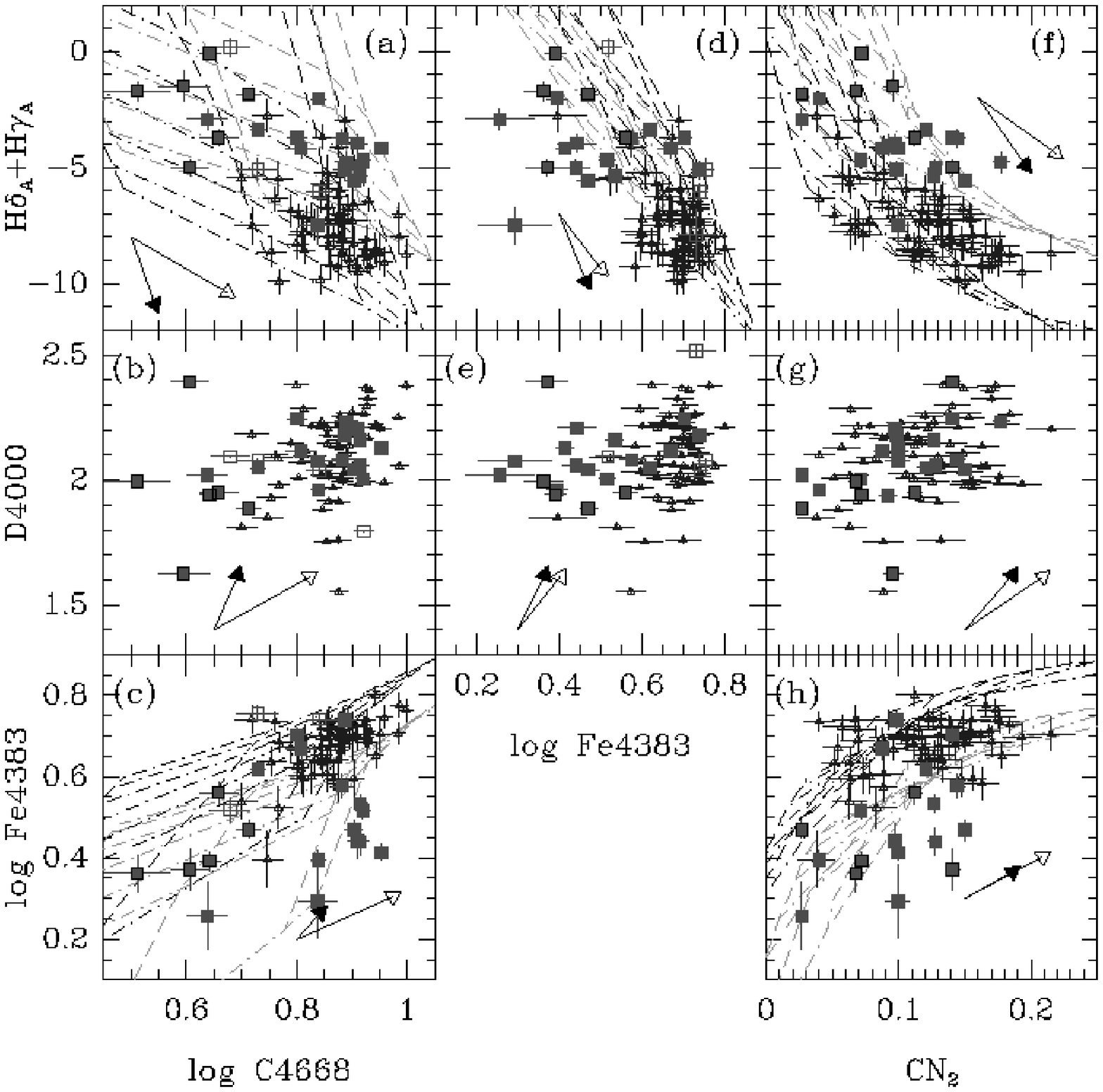}
\caption[]{The primary blue line indices versus each other. Data symbols as in Figure \ref{fig-agesfr}.
The grids on panels (a), (c), and (d) are SSP models from Thomas et al.\ (2003, 2004).
The black grid and green (grey) grid have $[\alpha /\rm{Fe}] =0.2$ and 0.5, respectively.
Dashed lines -- lines of constant [M/H], for values of -0.33, 0.0, 0.35 and 0.67.
Dot-dashed lines -- lines of constant ages, for values of 1, 2, 3, 5, 8, 11, and 15 Gyr.
The arrows show the approximate changes in the indices for a change of $\Delta \log \rm age = 0.3$
(solid arrow) and $\Delta \rm [M/H]=0.3$ (open arrow).
 \label{fig-metal1} }
\end{figure*}

\begin{figure*}
\epsfxsize 15.0cm
\epsfbox{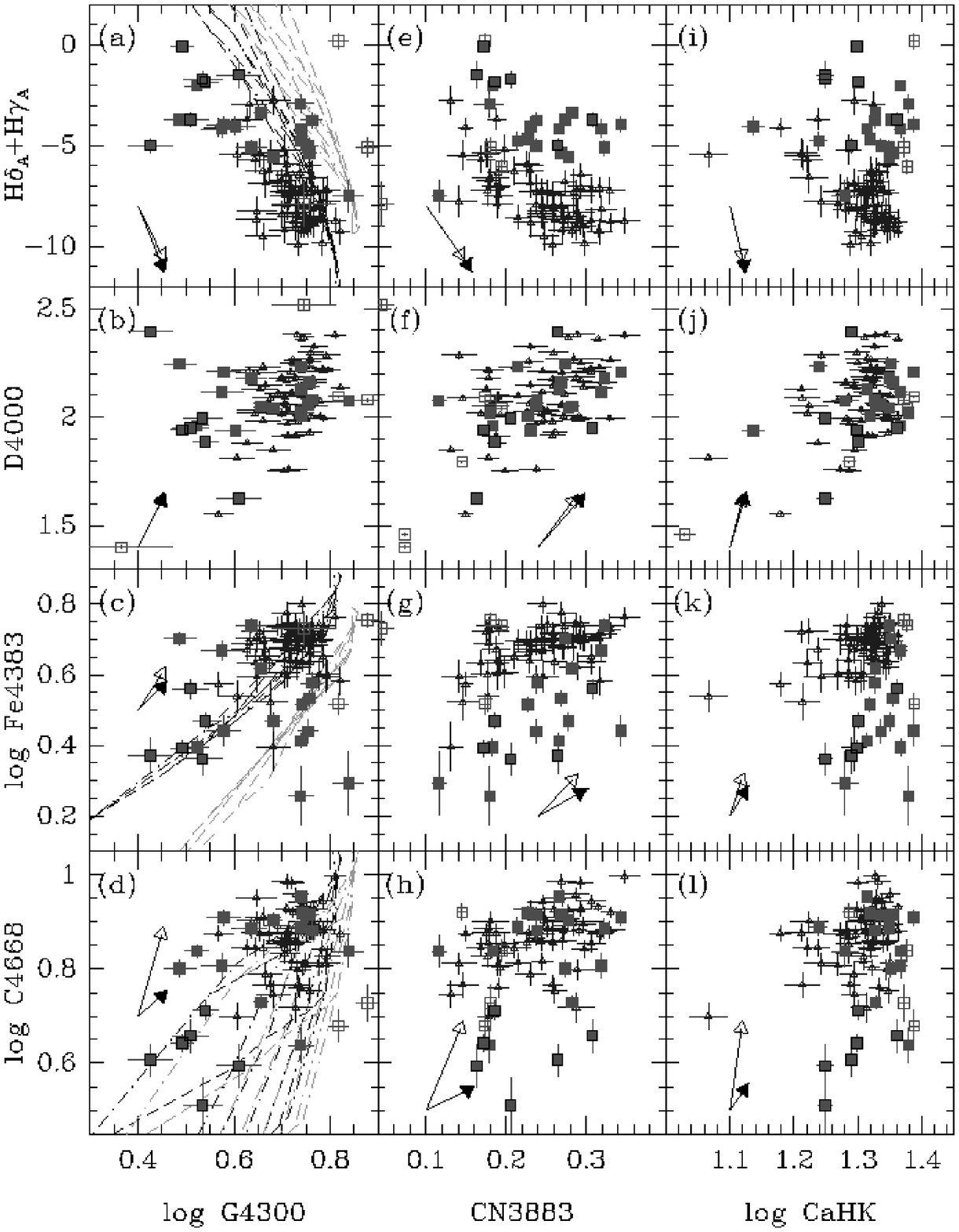}
\caption[]{The primary blue line indices C4668, Fe4383, D4000, and $\rm H\delta _A + H\gamma _A$ versus
the secondary indices G4300, CN3883, and CaHK. Data symbols as in Figure \ref{fig-agesfr}.
The grids on panels (a), (c), and (d) are SSP models from Thomas et al.\ (2003, 2004). The
same models are shown in these panels as those included on Figure \ref{fig-metal1}.
The arrows show the approximate changes in the indices for a change of $\Delta \log \rm age = 0.3$
(solid arrow) and $\Delta \rm [M/H]=0.3$ (open arrow).
\label{fig-metal2} }
\end{figure*}

\subsection{Evidence for differences in metallicities and abundance ratios}

Using the line indices together, and comparing these to the SSP models offer
the possibility of disentangling differences in ages from differences in metallicities
and abundance ratios. As described in Section 6.2, we use the models only to 
quantify relative differences between the RXJ0152.7--1357 sample and the low redshift
comparison sample, rather than actually derive luminosity weighted mean
ages, [M/H], and $[\alpha /\rm{Fe}]$ for the individual galaxies.

Figure \ref{fig-metal1} shows what we call the primary indices C4668 (and $\rm CN_2$), Fe4383, D4000, 
and $\rm H\delta _A + H\gamma _A$ versus each other. 
Model grids from Thomas et al.\ (2003, 2004) are overlaid. D4000 is not included in these models.

The location of the RXJ0152.7--1357 data points on Figures \ref{fig-metal1}c, f, and h 
cannot be an effect of age differences, only, between the RXJ0152.7--1357 sample and the low
redshift sample. We estimate based on these figures that
at least half the galaxies in the RXJ0152.7--1357 sample have $[\alpha /\rm{Fe}]$ approximately
0.2 dex higher than the low redshift sample. 
Except for a few of the  RXJ0152.7--1357 galaxies, they could in fact all have 
$[\alpha /\rm{Fe}]$ of 0.2 dex higher than the low redshift sample.
Figure \ref{fig-metal1}d indicates that the effect may be even stronger, with many of
the RXJ0152.7--1357 galaxies having unusually weak Fe4383 indices.
The location of the RXJ0152.7--1357 data points on Figures \ref{fig-metal1}a and c, further
indicates that about half of the galaxies 
have metal content significantly below that of the low redshift sample.
This conclusion is based primarily on the C4668 measurements, but is supported
by the $\rm CN_2$ measurements, see Figures \ref{fig-metal1}f and h.

While we have no models for D4000 that include $[\alpha /\rm{Fe}]$, it is striking that
D4000 is strongly correlated with $\rm CN_2$, and that the correlation is the same for the
low redshift sample and for the RXJ0152.7--1357 sample. $\rm CN_2$ increases with 
increasing $[\alpha /\rm{Fe}]$ though the dependence is relatively weak, see Table \ref{tab-models}. 
The correlation between D4000 and $\rm CN_2$ may indicate that D4000 also increases
with increasing $[\alpha /\rm{Fe}]$. We return to this issue in Section 9.

Figure \ref{fig-metal2} shows C4668, Fe4383, D4000, and $\rm H\delta _A + H\gamma _A$ versus
what we call the secondary indices, G4300, CN3883, and CaHK.
Model grids from Thomas et al.\ (2003, 2004) are overlaid for G4300.
D4000, CN3883 and CaHK are not included in the models.
We use these plots to check whether the secondary indices give results that are consistent 
with the primary indices, and we discuss if there are better choices of metal
indices than the primary indices.

Figure \ref{fig-metal2}d (G4300 versus C4668) supports the lower metal content of 
part of the RXJ0152.7--1357 sample compared to the low redshift sample.
The plots of G4300 versus $\rm H\delta _A + H\gamma _A$ and Fe4383 (Figure \ref{fig-metal2}a and c)
indicate that about a third of the galaxies in the RXJ0152.7--1357 sample have $[\alpha /\rm{Fe}]$ 
roughly 0.2 dex higher than the low redshift sample. 
However, closer inspection of these plots and comparison with Figure \ref{fig-metal1}f and h
reveals that the high $[\alpha /\rm{Fe}]$ galaxies are not the same galaxies 
on the two sets of plots. This can be seen by looking at the location of the six weak-lined
galaxies (discussed in Section 9), which are marked with black boxes around the points.
This inconsistency may be due to the modeling of G4300. Thomas et al.\ (2003) comment
that this index is very poorly calibrated.
Even though the models for different $[\alpha /\rm{Fe}]$ are well-separated in G4300 versus Fe4383
(Figure \ref{fig-metal2}c), we therefore conclude that G4300 with the present modeling is not
a convincing blue alternative to Mg$b$.

The remainder of the indices shown on Figure \ref{fig-metal2} have not been modeled by
Thomas et al. However, D4000 and CN3883 show a similar strong correlation as seen for
D4000 and $\rm CN_2$, and the correlation is the same for the RXJ0152.7--1357 sample
and the low redshift sample. 

Fe4383 versus CN3883 (Figure \ref{fig-metal2}g)
as well as $\rm H\delta _A + H\gamma _A$ versus CN3883 (Figure \ref{fig-metal2}e)
show a clear separation of the two samples, similar to separation seen when using $\rm CN_2$
instead of CN3883. 
If we assume that Fe4383 primarily probes the iron abundance, while CN3883 probes the 
carbon and nitrogen abundance, then the distribution of data points on
Figure \ref{fig-metal2}g indicates that the low metallicity galaxies in RXJ0152.7--1357
have significantly higher [CN/Fe] than seen in the low redshift sample.

The index CN3883 is stronger than $\rm CN_2$, in the sense that for a given S/N of the
spectra the relative uncertainties on CN3883 is smaller than that of $\rm CN_2$. 
Further, for galaxies in RXJ0152.7--1357 and in general for galaxies for which 
the blue continuum passband is redshifted to wavelengths redwards of 4000{\AA} 
the CN3883 index is also very easy to measure. 
Thus, this index is potentially very useful for studies of galaxies at redshifts above
about 0.1. It would be valuable to have the dependence of $[\alpha /\rm{Fe}]$
and [CN/Fe] modeled for CN3883.

The CaHK index adds very little information. The separation of the low redshift sample
and the RXJ0152.7--1357 sample on Figure \ref{fig-metal2}i, k, and l is due to the indices 
plotted versus CaHK, and is not caused by CaHK itself.

\subsection{Recent star formation -- the emission line galaxies}

There are seven galaxies in the RXJ0152.7--1357 sample that have significant 
emission lines. Five of these follow the red sequence in color-magnitude
diagram, Figure \ref{fig-CMCC}, though one of them (ID 1920) has a quite low
S/N spectrum.
The main questions of interest are the mass fraction involved in the star formation
and the duration of the star formation episode.

In order to convert the equivalent width of the [\ion{O}{2}] to star formation 
rates we assume that the equivalent width measured within the aperture we use 
is representative of the global value.
We then derive the observed luminosity of the [\ion{O}{2}] line as
\begin{equation}
L({\rm [O II]})_{\rm obs} = 1.4 \cdot 10^{29} \frac{L_{\rm B}} {L_{\rm B\sun}} {\rm EW([O II])} ({\rm ergs\,s^{-1}})
\end{equation}
with $\frac{L_{\rm B}}{L_{\rm B\sun}}=10^{0.4(5.48-M_{\rm B})}$, 
see Kennicutt (1992) and Balogh et al.\ (1997).
We convert the observed $L(\rm{[O II]})_{\rm obs}$ to the star formation rate (SFR) 
in $M_{\sun}\, {\rm yr^{-1}}$ using the calibration from Kewley et al.\ (2004)
\begin{equation}
{\rm SFR} = 6.58 \cdot 10^{-42} \, L({\rm [O II]})
\end{equation}
with
\begin{equation}
\label{eq-red}
L({\rm [O II]}) = 3.11 \cdot 10^{-20} \, L({\rm [O II]})_{\rm obs}^{1.495}
\end{equation}
Equation \ref{eq-red} takes into account the intrinsic reddening. 
Since we do not know the oxygen abundances,
we have chosen to use the average calibration established by Kewley et al.

We find that the mean SFR for the four red emission line galaxies is $1.3 M_{\sun}\, {\rm yr^{-1}}$.
If we estimate the mean mass of these four galaxies based on their absolute B magnitude,
the median luminosity difference between the RXJ0152.7--1357 sample and the low 
redshift sample, and the FP for the low redshift sample, we get a mean mass of
about $1.5 \cdot 10^{11} M_{\sun}$.
Thus, a 1 Gyr burst of star formation with the mean SFR found from the 
[\ion{O}{2}] line would have involved about 1 per cent of the mass.

We attempted to further constrain the mass fraction and duration of 
the star formation episode
by creating ``toy'' models, which consist of a mix of two stellar populations.
We assumed that some mass fraction was involved in the star formation episode, while 
the remainder of the mass, the underlying ``old'' stellar population,
is not forming stars. We used models from Magris et al.\ (2003) for the star
forming population and models from Bruzual \& Charlot (2003) for the 
``old'' stellar populations. However, none of our modeling resulted in 
further constraints on the duration or the mass involved in the star formation
episode. The emission lines are either caused by quite small mass fraction (1 per cent)
involved in a very recent star formation burst (less the 0.3 Gyr prior),
or if a larger mass fraction is involved the episode must have started earlier.

\section{Discussion}

Using the results presented in Sections 7 and 8,
we now attempt to determine if there is an evolutionary scenario for the 
galaxies in the RXJ0152.7--1357 sample that will make these galaxies evolve
into galaxies similar to those in our low redshift sample, within the 
available time which is about 7 Gyr. 

We assume that the galaxies in the RXJ0152.7--1357 sample can be considered the
progenitors for the galaxies in the low redshift sample. There is no guarantee that this
assumption is correct. However, we note that all the non-emission galaxies in
the RXJ0152.7--1357 sample are on the red sequence.
Based on the HST/ACS archive data available for RXJ0152.7--1357 
we find that none of these galaxies show any 
obvious spiral structure, see Figure \ref{fig-stamps} in the Appendix. 
Thus, these galaxies are morphologically similar to the low redshift sample.
The four red emission line galaxies are on the red sequence and they 
show weak spiral structure (see Figure \ref{fig-stamps}). 
We do not know whether their morphologies will 
evolve such that they will resemble our low redshift sample after $\approx$ 7 Gyr, 
but we can still discuss whether there is an evolutionary scenario for their
stellar populations that will lead to stellar populations similar to those
of the low redshift sample.
The two blue emission line galaxies in the RXJ0152.7--1357 sample are not 
considered in this discussion.

The simplest evolutionary scenario is pure passive evolution (scenario 1
in Section 6.3). 
No new stars are formed and the model prediction is that the only difference 
between the stellar populations in the RXJ0152.7--1357 sample and those in 
the low redshift sample should be an 
age difference equal to the difference in the lookback time.
In Section 8.1.1 we found that this is in agreement with the B-band luminosities
and the strength of the higher order Balmer lines, and that the model
implies a formation redshift of $z_{\rm form} \approx 4$.
However, once we take the other spectral indices into account, the picture
is no longer this simple. First, the SSP models predict that for a pure
passive evolution model we should have found a difference in the D4000
strengths between the RXJ0152.7--1357 sample and the low redshift sample.
The difference between the predicted offset in D4000 and the measured 
(insignificant) offset is larger than five times the uncertainty.
Further, both the scatter and the zero point differences for several of the 
scaling relations for the metal indices are in contradiction with the pure
passive evolution model, cf.\ Table \ref{tab-relations}.

If pure passive evolution is the correct evolutionary scenario, then either the 
data or the models (or both) must be incorrect for many of the indices.
The SSP models from Thomas et al.\ (2003, 2004) are ambitious in the sense
that they include non-solar $[\alpha /\rm{Fe}]$. However, the age dependencies
of the various parameters predicted by these models are in general agreement
with other models, e.g., the Vazdekis-2000 models. Even if the details
of the models are incorrect, we find it very unlikely that the significantly
higher scatter in some of the scaling relations for the RXJ0152.7--1357 sample compared
to those of the low redshift sample can be due to age differences only.
The most troubling is perhaps that the D4000 indices for the RXJ0152.7--1357 sample 
are very similar to those of the low redshift sample. This index is strong 
and has low measurement uncertainties. We therefore consider it unlikely that
the data are grossly incorrect. Thus, if pure passive evolution is the
correct evolutionary scenario, then the models predicting a significant
age dependency for D4000 must be incorrect.

Because of the many contradictions between the data and predictions for 
the pure passive evolution scenario, we consider this evolutionary scenario unlikely. 
Next we discuss the 
evidence for differences in metal content [M/H] and abundance ratios $[\alpha /\rm{Fe}]$
between the RXJ0152.7--1357 sample and the low redshift sample, and use these
differences to discuss evolutionary scenarios that involve star formation and/or
merging, in addition to passive evolution of the already existing stellar population.

First we examine in more detail the possible cause of the higher scatter for C4668, G4300 and Fe4383
scaling relations found for the RXJ0152.7--1357 sample (see Section 8.1.2 and Table \ref{tab-relations}).
For C4668 and G4300 it appears that there is a group of galaxies roughly following the relations
established for the low redshift sample, while another group of galaxies have significantly 
weaker metal indices, see Figure \ref{fig-metal3}.
The majority of the RXJ0152.7--1357 galaxies has significantly weaker Fe4383 indices
than found for the low redshift sample. 
We isolate the galaxies with C4668, G4300 and Fe4383 weaker than the relations established 
for the low redshift sample by more than twice the rms of that sample.
There are six galaxies that meet that criterion for all three indices, none of them are emission line
galaxies. In fact Fe4383 does not limit this sample more than just requiring C4668 and G4300
to be weak. The six galaxies are ID 737, 776, 1507, 1614, 1682 and 1935.
Five of these lie, in projection, in areas of the cluster that have low X-ray flux 
(see Figure \ref{fig-grey}). 
Two of them (ID 1614 and 1682) are in the diffuse X-ray emission east of the main
X-ray structure.
If we exclude C4668 as a selection criteria and then select the group of galaxies
with weak G4300 as before, this group then contains 10 non-emission line galaxies.
Eight of these are in areas of low X-ray flux.
If we on the other hand exclude G4300 as a selection criteria, then there are eight non-emission
line galaxies with weak C4668, seven of which are in areas of low X-ray flux.
For reference, 14 of the 29 cluster members are located in areas of low X-ray flux.
The four red emission line galaxies in the RXJ0152.7--1357 sample, which have
good S/N spectra (ID 566, 643, 1159 and 1299), also appear to be located 
in areas of low X-ray luminosity. 
Further,
Ford et al.\ (2004) find in their study of this cluster that the star-forming spiral galaxies 
in the cluster are located in areas of low X-ray luminosity away from the cores of the two 
sub-clusters.

We speculate that the unusually weak C4668, G4300 (and Fe4383), as well as the
weak emission lines in the red emission line galaxies, are the effect of short bursts
of star formation in these galaxies, maybe triggered as a result of their first 
infall into the cluster. Only the galaxies with the weak emission lines still have
small amounts of on-going star formation, while in the galaxies without emission
lines the star formation has stopped.
Using the scaling relation between the higher order Balmer lines $\rm H\delta _A + H\gamma _A$ 
and the velocity dispersions (Figure \ref{fig-agesfr}b) we find that the 
six weak-lined galaxies are offset with $\Delta (\rm H\delta _A + H\gamma _A)= 2.26 \pm 0.77$
from the rest of the non-emission line galaxies in the RXJ0152.7--1357 sample.
A similar offset is not seen in the absolute B magnitudes where the difference
is insignificant $\Delta \rm M_B = -0.07 \pm 0.45$.
It is therefore possible that these galaxies are offset in $\rm H\delta _A + H\gamma _A$
primarily because of a higher $[\alpha /\rm{Fe}]$. The offset in $\rm H\delta _A + H\gamma _A$
is equivalent to an offset in $[\alpha /\rm{Fe}]$ of 0.3 dex, assuming no differences in
the mean ages and the mean [M/H] between these six galaxies and the 
remainder of the RXJ0152.7--1357 galaxies. If these six galaxies do contain younger 
stellar populations than the remainder of the RXJ0152.7--1357 galaxies
this offset in $[\alpha /\rm{Fe}]$ should be seen as an upper limit.


Based on our comparison of $\rm H\delta _A + H\gamma _A$, Fe4383 and $\rm CN_2$
with the model grids from Thomas et al.\ (2003, 2004) (see Figure \ref{fig-metal1})
we found that a large fraction of the galaxies in the RXJ0152.7--1357 sample
may have $[\alpha /\rm{Fe}]$ about 0.2 dex higher than seen in the low redshift sample.
If we assume that the median difference in $[\alpha /\rm{Fe}]$ between 
the RXJ0152.7--1357 sample and the low redshift sample is 0.2 dex, then this
explains about half of the offset in $\rm H\delta _A + H\gamma _A$ between
the two samples.
This in turn means that the difference required in the mean ages to explain
the remainder of the offset is only $\Delta \log {\rm age} \approx -0.2$. 
Formally this leads to a formation redshift higher than found in Section 8.1.1, 
but the pure passive evolution scenario is
of course no longer valid because of the difference in the abundance ratios.
We also found that because of the empirical correlation between D4000 and $\rm CN_2$
that D4000 would have to increase with $[\alpha /\rm{Fe}]$, if $\rm CN_2$
increases with $[\alpha /\rm{Fe}]$ as the models indicate. It is therefore possible
that the D4000 indices for the RXJ0152.7--1357 sample can be explained by a combination
of younger mean ages (leading to weaker D4000 indices) combined with higher
$[\alpha /\rm{Fe}]$ (leading to stronger D4000 indices), giving the net result
that the D4000 indices for the RXJ0152.7--1357 sample are similar to those of
the low redshift sample. Until models are available that predict how D4000
depends on $[\alpha /\rm{Fe}]$, this remains a speculation.
Models with non-solar $[\alpha /\rm{Fe}]$
are also needed for the M/L ratios, since current models cannot be used to
test if the differences in both ages and $[\alpha /\rm{Fe}]$ would lead to 
inconsistencies with the measured luminosities.

It is difficult (if not impossible) to envision an evolutionary path that 
within the available time (about 7 Gyr) will lead from the stellar 
populations in the RXJ0152.7--1357 sample to those in the low redshift sample. 
The hardest challenge would be to identify processes that would reduce 
$\rm [\alpha/Fe]$ with 0.2 dex for a large number of the galaxies, 
without causing the resulting mean ages of the stellar populations in 
the galaxies to be much too young.
One possibility would be mergers between these high $\rm [\alpha/Fe]$
galaxies and galaxies containing stellar populations with much lower $\rm [\alpha/Fe]$,
i.e.\ (disk) galaxies whose stellar populations formed over a much longer time scale.
Thus, this may be either scenario (3) or (4) as described in Section 6.3,
depending on whether there is star formation involved.
In any case, in order to 
maintain a mix of stellar populations with fairly high mean ages
such mergers can involve only a small amount of new star formation.
Further, one would need a source of low $\rm [\alpha/Fe]$ galaxies that are
dominated by ``old'' (5 Gyr) stellar populations. Our RXJ0152.7--1357 sample is
by no means complete, but it is difficult to envision the
presence of a large populations of $\rm [\alpha/Fe]=0$ galaxies in the cluster --
only three out of the 29 galaxies in our sample have line indices in agreement
with $\rm [\alpha/Fe]=0$.

Perhaps an easier task would be to determine how the galaxies in the RXJ0152.7--1357
sample would appear if they were to evolve passively for 7 Gyr, without merging.
After 7 Gyr of passive evolution, the red emission line galaxies will 
no longer have emission lines. Further, since the mass involved in the star formation
is around 1 per cent of the total mass, the properties of the galaxies
will be dominated by the underlying ``old'' stellar population.
Thus, the signature of the short episode of star formation will no longer be 
detectable. We can only speculate as to whether these galaxies retain their 
weak spiral structure.
The galaxies in RXJ0152.7--1357 that have higher $\rm [\alpha/Fe]$
than the low redshift sample will still have higher $\rm [\alpha/Fe]$ after 7 Gyr of passive 
evolution. The galaxies will also still cause the scatter in the 
scaling relations for C4668, G4300 and Fe4383 to be higher than
found for the low redshift sample. Thus, pure passive evolution of the
RXJ0152.7--1357 galaxies will result in stellar populations
different from those in the low redshift sample.

\section{Conclusions}

We have studied the stellar populations in galaxies in the rich galaxy cluster
RXJ0152.7--1357 at a redshift of 0.83. Our conclusions are based on high signal-to-noise
optical spectroscopy and ground based photometry in three filters. 
From the spectroscopy we have measured redshifts and central velocity dispersions.
Of the 41 galaxies observed, 29 are cluster members.
For these, we present absorption line indices for the Lick/IDS
lines in the wavelength region from 4000{\AA} to 4700{\AA} in the rest frame.
We have also measured the indices CN3883, CaHK and D4000 located at shorter
wavelengths and not part of the Lick/IDS system.
For the seven emission line galaxies in our RXJ0152.7--1357 sample, we measure
the equivalent width of the [\ion{O}{2}] line.

We have established scaling relations between the absolute total B magnitude
and the central velocity dispersions of the galaxies (the Faber-Jackson relation)
as well as relations between absorption line indices and
the central velocity dispersions of the galaxies. Comparison of these relations
with those for our low redshift comparison sample shows that the B magnitudes
and the strengths of the high order Balmer lines, $\rm H\delta _A + H\gamma _A$,
are in agreement with pure passive evolution and a formation redshift $z_{\rm form} \approx 4$.
However, the measurements of D4000 do not support this conclusion. 
The strengths of the line indices for the metal lines, as well as the scatter
in the scaling relations for C4668, Fe4383 and G4300 also indicate that
pure passive evolution will not result in the stellar populations of the galaxies 
in the RXJ0152.7--1357 sample evolving into stellar populations similar to those in
our low redshift comparison sample, within the available time.

We find that six galaxies with weak G4300 and C4668 lines for their velocity 
dispersion, as well as four red galaxies with emission lines appear to be located
in areas of low X-ray luminosity. It is likely that these galaxies are experiencing
the effect of the merger of the two sub-clusters that make up the  RXJ0152.7--1357 cluster
as (short) episodes of star formation triggered by the cluster merger.

Comparing the strengths of C4668, Fe4383, CN3883, G4300 and CN$_2$ to
predictions from the stellar population models from Thomas et al.\ (2003, 2004)
we find that at least half of the galaxies in the cluster have
an $\alpha$-element abundance ratio $\rm [\alpha/Fe]$ of 0.2 dex higher than the
galaxies in our low redshift comparison sample. 
This also indicates that the evolution of these galaxies may have involved more
short bursts of star formation than has been the case for the low redshift sample.
For about half of the  non-emission line galaxies in the cluster, the metal
content is (still) significantly below that of the low redshift comparison sample.

The study of RXJ0152.7--1357 presented in this paper contains the first published
study of metal line indices at this redshift. Previous studies have concentrated
on luminosities, either using the Faber-Jackson relation or the Fundamental Plane, 
and Balmer lines.  
Our results are in agreement with these previous studies in the sense that if
we had limited our analysis to the Faber-Jackson relation and the Balmer lines we 
would have concluded that the stellar populations of the galaxies in the 
RXJ0152.7--1357 sample by pure passive evolution can evolve into stellar 
populations similar to those seen in our low redshift sample, and stellar 
populations have a formation redshift $z_{\rm form} > 2$. 
Only because we also have measured the indices for the metal lines and D4000, do
we find that pure passive evolution no longer appears to be a viable scenario.
An evolutionary path that makes the galaxies in the RXJ0152.7--1357 sample 
evolve into galaxies similar to our low redshift sample has to involve
star formation and/or merging.

\vspace{0.5cm}
Acknowledgments:
Claudia Maraston is thanked for making M/L ratio information available for her
SSP models prior to publication.
Karl Gebhardt is thanked for making his kinematics software available.
We thank the anonymous referee for constructive suggestions that helped improve
this paper.
Based on observations obtained at the Gemini Observatory, which is operated by the
Association of Universities for Research in Astronomy, Inc., under a cooperative agreement
with the NSF on behalf of the Gemini partnership: the National Science Foundation (United
States), the Particle Physics and Astronomy Research Council (United Kingdom), the
National Research Council (Canada), CONICYT (Chile), the Australian Research Council
(Australia), CNPq (Brazil) and CONICET (Argentina).
The data presented in this paper originate from the following Gemini programs:
GN-2002B-Q-29 (a queue program) and GN-2002B-SV-90 (an engineering program).
Observations have been used that were obtained with {\it XMM-Newton}, an ESA science mission 
funded by ESA member states and NASA.
In part, based on observations made with the NASA/ESA Hubble Space Telescope, 
obtained from the data archive at the Space Telescope Science Institute. 
STScI is operated by the Association of Universities for Research in Astronomy, 
Inc. under NASA contract NAS 5-26555.

\appendix

\section{Reduction of GMOS data}

The data were reduced using the Gemini IRAF package.  
In order to make the description of the reductions useful to a wider audience,
we give the names of the tasks in the Gemini IRAF package used 
for each of the reduction steps. 
It is also noted when the reduction techniques differ from the techniques 
implemented in the Gemini IRAF package.

GMOS-N has three detectors placed side by side. The raw
images are multi-extension FITS (MEF) images with three 
image extensions, one for each of the detectors.
The data format for GMOS data is described in detail on
the Gemini web pages (http://www.gemini.edu). 

\subsection{Imaging data}

\subsubsection{Basic reduction steps handled by the Gemini IRAF package}

Mean bias frames were created from all the available
bias frames taken during each of the GMOS-N observing runs (task {\tt gbias}).
The overscan level was not subtracted, since it is known
that the overscan region on the GMOS-N detectors contain
residual signal that depends on the signal level of
the images. Further, the bias level for the GMOS-N detectors
is very stable, varying less than 0.2ADU from run to run.
Twilight flat fields were created from all the available
twilight flats taken during each of the GMOS-N observing runs ({\tt giflat}).
The overscan section of the images were trimmed off, and
the images were bias subtracted and flat fielded using the 
matching calibration frames ({\tt gireduce}).
The three image extensions of each image were then mosaiced
to form one image containing the data from all three GMOS-N detectors ({\tt gmosaic}).

\subsubsection{Fringing and scattered light}

The detectors in GMOS-N have fringing at long wavelengths. 
The $i'$ images have fringes which are about one per cent 
of the dark sky background peak-to-peak, while the 
fringing in the $z'$ filter is about 5 per cent peak-to-peak.
Example fringe frames are available on the Gemini web pages.
The $r'$ images have a scattered light at a low level.

Fringe frames were derived for the $i'$ and $z'$ images; a scattered 
light frame was derived for the $r'$ images. 
The technique is the same for all three filters.
We describe the technique actually used, but note that from the Gemini IRAF package
v1.7 the construction of the fringe frames can be done with the task {\tt gifringe}.
The starting point is the mosaiced images from {\tt gmosaic}. 
The goal is to produce frames that are cleaned of signal from objects 
and have a mean level of zero, such that they contain only the contribution from the 
fringes and/or the scattered light.
Mask files for the objects in the images were produced using the NIRI task {\tt nisky}.
The median sky level was subtracted from each image, and the
images were median combined, omitting pixels that were flagged
in the masks as containing signal from objects. The task
{\tt gemcombine} in the Gemini IRAF package was used. This task
is just a wrapper for {\tt imcombine} that enables processing of MEF files.
The resulting frames for the $i'$ and $r'$ filters  were then median filtered with a 
box of 3.6 arcsec by 3.6 arcsec (25 by 25 pixels for the binned data). 
The $z'$ fringe frame cannot be median filtered, as the fringing occurs at a spatial 
scale of a few pixels.

The fringe and scattered light frames were subtracted from the individual mosaiced images. 
The level of fringing in the $z'$ images varies significantly. The best scaling of the $z'$
fringe frame was determined by visual inspection of the results from scalings between 
0.9 and 1.1 in steps of 0.05.
After subtraction of the fringe and scatter light frames the variations in the 
background are less than 0.1 per cent of the background level.

\begin{deluxetable}{lrrr}
\tablecaption{GMOS-N Photometric Zero Points \label{tab-mzero} }
\tablewidth{20pc}
\tablehead{
\colhead{Filter} & \colhead{$m_{\rm zero}$} & \colhead{$N_{\rm obs}$\tablenotemark{a}} 
  & \colhead{$k$\tablenotemark{b}}
}
\startdata
$r'$     & $28.05\pm 0.015$ & 7            & 0.11 \\
$i'$     & $27.78\pm 0.023$ & 7            & 0.10 \\
$z'$     & $26.63\pm 0.014$ & 8            & 0.05 \\
\enddata
\tablenotetext{a}{Number of individual measurements}
\tablenotetext{b}{Median atmospheric extinction for Mauna Kea}
\end{deluxetable}

\begin{deluxetable*}{lrrrrr}
\tablecaption{GMOS-N Color Terms: Linear Relations \label{tab-color} }
\tablewidth{0pc}
\tablehead{
\colhead{$\Delta {\rm m}$} & \colhead{rms} & \colhead{Color term fit} & \colhead{rms(fit)} & \colhead{N} & Color interval \\
\colhead{(1)} & \colhead{(2)} & \colhead{(3)} & \colhead{(4)} & \colhead{(5)} & \colhead{(6)}
}
\startdata
$\Delta r_{\rm zero}$   & 0.045\tablenotemark{a} & $(0.042\pm 0.004) (r'-i') - (0.011\pm 0.002)$ & 0.043 & 1084 & $-0.35\le (r'-i')\le 2.2$\\
$\Delta i_{\rm zero}$   & 0.054\tablenotemark{a} & $(0.113\pm 0.008) (i'-z') - (0.010\pm 0.002)$ & 0.049 & 1081 & $-0.3\le (i'-z')\le 0.85$ \\
$\Delta z_{\rm zero}$   & 0.055\tablenotemark{b} & $(0.125\pm 0.012) (i'-z') - (0.014\pm 0.003)$ & 0.050 & 492 & $-0.3\le (i'-z')\le 0.68$ \\
       &       & $(1.929\pm 0.177) (i'-z') - (1.188\pm 0.127)$ & 0.043 & 12 & $0.6\le (i'-z')\le 0.85$\\
\enddata
\tablenotetext{a}{$-0.7\le (r'-i')\le 2.5$}
\tablenotetext{b}{$-0.7\le (r'-z')\le 2.25$}
\tablecomments{(1) Residual zero point, (2) rms of $\Delta m$, equivalent to the
expected uncertainty on the standard calibration if the color terms are ignored, (3) linear fits to the color terms,
(4) rms of the linear fits, (5) number of individual measurements included in the fits, (6) color interval
within which the linear fit applies.}
\end{deluxetable*}

\subsubsection{Co-addition of the images}

The images were registered and co-added using the task
{\tt imcoadd}, which is part of the {\tt gemtools} package in the Gemini 
IRAF package. 
The resulting image from {\tt imcoadd} is the average of all 
the good pixels. The input images are scaled such that
images taken under non-photometric conditions and/or at
different airmasses can be correctly combined.
The photometric zero point of the co-added image is 
identical to the first image in the stack. Therefore we
always use images taken under photometric conditions
as the first images in the stacks.
The total exposure times, sky background and image qualities
of the co-added images are listed in Table \ref{tab-imdata}.

\subsubsection{Standard calibration of the photometry}

The imaging have been calibrated using the exposures taken under photometric
conditions on UT 2002 September 14 and using standard star observations 
obtained the same night. Four exposures in each filter were obtained of the 
Landolt (1992) standard star field PG0231+051.
We use the transformation from Smith et al.\ (2002) to convert the standard magnitudes 
from the system used by Landolt to the Sloan-Digital-Sky-Survey (SDSS) photometric system.
Then we establish the median zero points for each filter, see Table \ref{tab-mzero}.
Further, we have used all available standard star observations obtained with GMOS-N in the period
between UT 2001 August 20 and UT 2003 December 26 to establish the color terms
for the transformations. Full information about this calibration will be 
included in our paper that describes all the GMOS photometry for this project
(J\o rgensen et al., in preparation). Here we give only the relations used for the
observations of RXJ0152.7--1357, see Table \ref{tab-color}.
The standard magnitudes are then derived as 
\begin{equation}
m_{\rm std} = m_{\rm zero} + \Delta m_{\rm zero} - 2.5 \log (N/t) - k\,(airmass-1)
\end{equation}
where $t$ is the exposure time, $N$ is the number of electrons
above the sky level, $airmass$ is the mean airmass for the exposure,
$k$ is the median atmospheric extinction at Mauna Kea, and $\Delta m_{\rm zero}$ is derived 
using one of the relations listed in Table \ref{tab-color}.
In order to make the information in the table more generally useful, we have also
given the scatter relative to the nightly zero points in the case where the
color terms are ignored (column (2) in Table \ref{tab-color}). This is the average
uncertainty to be expected on standard calibrated magnitudes if the color term is ignored,
for objects with colors in the intervals listed in the notes to the table.
The color terms are fairly small for $r'$ and $i'$, while for $z'$
the color term for RXJ0152.7--1357 cluster members with $(i'-z') \approx 0.8$ is about 0.35 mag.

\begin{deluxetable*}{rrrrrrrr}
\tablecaption{Photometric Data for the Spectroscopic Sample \label{tab-photdata} }
\tablewidth{0pc}
\tablehead{
\colhead{ID} & \colhead{RA (J2000)} & \colhead{DEC (J2000)\tablenotemark{a}} & 
\colhead{$r'_{\rm total}$} & \colhead{$i'_{\rm total}$} & 
\colhead{$z'_{\rm total}$} & \colhead{$(r'-i)$} & \colhead{$(i'-z')$} }
\startdata
        103&   1 52 32.77&  -13 55 09.8&  23.02&  21.78&  21.10&  1.107&  0.710\\
        155&   1 52 33.31&  -13 55 23.0&  22.76&  21.72&  21.03&  0.942&  0.779\\
        193&   1 52 50.80&  -13 55 28.9&  21.46&  20.82&  20.17&  0.753&  0.656\\
        264&   1 52 44.66&  -13 55 37.3&  22.30&  21.40&  20.68&  1.001&  0.672\\
        338&   1 52 43.33&  -13 55 44.4&  23.67&  22.37&  21.61&  1.349&  0.738\\
        346&   1 52 37.42&  -13 55 50.1&  22.68&  21.24&  20.56&  1.390&  0.791\\
        422&   1 52 34.59&  -13 55 58.8&  23.59&  22.18&  21.43&  1.284&  0.847\\
        460&   1 52 36.11&  -13 56 08.5&  22.26&  20.99&  20.24&  1.320&  0.810\\
        523&   1 52 42.38&  -13 56 18.7&  22.37&  21.10&  20.36&  1.508&  0.773\\
        566&   1 52 38.03&  -13 56 28.1&  22.33&  21.14&  20.40&  1.399&  0.752\\
        627&   1 52 38.48&  -13 56 33.6&  23.23&  21.89&  21.11&  1.326&  0.833\\
        643&   1 52 45.60&  -13 56 40.0&  23.15&  21.82&  21.03&  1.457&  0.817\\
        737&   1 52 45.77&  -13 56 46.1&  23.25&  22.09&  21.29&  1.414&  0.796\\
        766&   1 52 45.83&  -13 56 59.2&  22.10&  20.67&  19.88&  1.515&  0.795\\
        776&   1 52 38.48&  -13 56 52.5&  22.86&  21.59&  20.84&  1.298&  0.793\\
        813&   1 52 44.97&  -13 57 04.2&  22.17&  20.78&  19.99&  1.503&  0.783\\
        896&   1 52 36.99&  -13 57 10.1&  22.69&  21.83&  21.23&  0.788&  0.683\\
        908&   1 52 43.74&  -13 57 19.4&  22.13&  20.93&  20.14&  1.410&  0.784\\
       1027&   1 52 43.32&  -13 57 26.7&  23.24&  22.03&  21.23&  1.448&  0.817\\
       1085&   1 52 42.94&  -13 57 35.0&  22.91&  21.40&  20.61&  1.534&  0.791\\
       1110&   1 52 39.93&  -13 57 42.6&  22.59&  21.29&  20.50&  1.521&  0.781\\
       1159&   1 52 36.18&  -13 57 48.8&  23.02&  21.66&  20.93&  1.374&  0.794\\
       1210&   1 52 42.83&  -13 57 55.3&  23.40&  22.04&  21.31&  1.453&  0.785\\
       1245&   1 52 43.57&  -13 58 00.0&  22.72&  21.78&  21.07&  1.059&  0.672\\
       1299&   1 52 47.34&  -13 59 26.1&  22.59&  21.05&  20.30&  1.593&  0.787\\
       1385&   1 52 39.36&  -13 59 04.5&  22.58&  21.50&  20.87&  1.040&  0.635\\
       1458&   1 52 39.64&  -13 58 56.6&  23.59&  22.08&  21.35&  1.444&  0.767\\
       1494&   1 52 39.08&  -13 58 48.8&  19.22&  18.59&  18.00&  0.635&  0.584\\
       1507&   1 52 34.48&  -13 58 42.2&  23.86&  22.44&  21.74&  1.305&  0.829\\
       1567&   1 52 39.62&  -13 58 26.7&  22.10&  20.90&  20.09&  1.526&  0.796\\
       1590&   1 52 38.87&  -13 58 32.0&  24.00&  22.37&  21.58&  1.547&  0.808\\
       1614&   1 52 51.96&  -13 58 17.1&  22.52&  21.30&  20.55&  1.434&  0.806\\
       1682&   1 52 51.96&  -13 58 15.6&  22.39&  21.23&  20.45&  1.369&  0.788\\
       1811&   1 52 38.63&  -13 59 20.8&  23.90&  22.56&  21.89&  1.406&  0.807\\
       1838&   1 52 37.00&  -14 00 05.3&  22.93&  22.23&  21.57&  0.710&  0.535\\
       1896&   1 52 33.95&  -14 00 11.9&  23.99&  22.72&  21.94&  1.187&  0.878\\
       1920&   1 52 39.70&  -13 59 14.3&  23.24&  22.25&  21.53&  1.312&  0.807\\
       1935&   1 52 41.88&  -13 59 53.6&  22.88&  21.38&  20.61&  1.459&  0.777\\
       1970&   1 52 48.03&  -13 59 58.6&  21.15&  20.56&  19.94&  0.621&  0.676\\
       2042&   1 52 42.38&  -13 59 46.6&  19.91&  19.45&  18.91&  0.551&  0.550\\
       2087&   1 52 31.55&  -13 59 40.1&  20.53&  19.88&  19.28&  0.618&  0.702\\
\enddata
\tablenotetext{a}{Positions are consistent with USNO, with an rms scatter of $\approx 0.7$ arcsec.}
\tablecomments{Units of right ascension are hours, minutes, and seconds, and
units of declination are degrees, arcminutes, and arcseconds.}
\end{deluxetable*}

\subsubsection{Derived photometric parameters}

We used SExtractor (v2.1.6, Bertin \& Arnouts 1996) in 2-image mode, with the 
images pre-registered to each other. 
The image in the $i'$-filter was used for detections, while
the images in the other two filters were used only for photometry.
The background mesh size was adjusted to avoid systematic effects from the
galaxies with the largest angular size. We use a background mesh size of 256 pixels, 
with a filter size of 5 pixels.
Since the field is quite crowded, we used 64 sub-thresholds for the deblending
of objects and a minimum contrast for the deblending of only 0.0005 (the 
default is 0.005). The results were visually inspected to ensure that all
potential targets for spectroscopic observations were correctly deblended.
Other adjustments to the SExtractor parameters were trivial adjustments
for the magnitude zero points, the effective gains in the co-added images,
and the image quality.

We adopt the best magnitudes ({\it mag\_best}) from SExtractor as the total 
magnitudes of the objects. Aperture magnitudes and colors were derived within 
apertures with a diameter of 1.16 arcsec, which is approximately twice the FWHM of the 
point-spread-function of the images. The typical uncertainties on the 
magnitudes and colors are discussed in Section 4.1.
Table \ref{tab-photdata} summarizes the total magnitudes and aperture colors
for the galaxies included in the spectroscopic sample.
The data in the table have not been corrected for the galactic extinction.

\subsubsection{Calibration of the photometry to the rest frame}

We have derived total magnitudes in the rest frame B-band from the observed $i'$ magnitudes
and colors $(i'-z')$. The calibration was established from Bruzual \& Charlot (2003) stellar population
models spanning the observed color range.
These models are used to give the rest frame B magnitudes, as well as the observed
$i'$ and $(i'-z')$ at a given redshift.
For each redshift, in steps of 0.025, we then fit the rest frame B magnitudes with the observed $i'$ 
magnitude plus a second order polynomial in $(i'-z')$.
When the calibration is used, the fitting coefficients are then interpolated to the exact 
redshift of the galaxy.  More details will appear in our paper
that describes all the GMOS photometry for the project (J\o rgensen et al., in preparation).
For reference, we here give the resulting equation for the cluster median redshift of 0.835,
\begin{equation}
\label{eq-restframe}
B_{\rm rest} = i' + 0.8026 - 0.4268 (i'-z') - 0.0941 (i'-z')^2
\end{equation}
The magnitude $i'$ and the color $(i'-z')$ are corrected for galactic extinction before using
Equation \ref{eq-restframe} to determine $B_{\rm rest}$.
The absolute B-band magnitude, $M_{\rm B}$, is then derived as
\begin{equation}
M_{\rm B} = B_{\rm rest} - DM(z) + 2.5 \log (1+z)
\end{equation}
where $z=0.835$ is the median cluster redshift and $DM(z)= 43.60$ is the distance modulus for 
RXJ0152.7--1357 for our adopted cosmology.
See also Blanton et al.\ (2003) on techniques for how to calibrate to a ``fixed-frame'' system,
like the rest frame B.

\subsection{Spectroscopic data}

\subsubsection{Bias subtraction and flat field correction}

The bias subtraction for the spectroscopic data is done in the same way as for the imaging data.

Quartz-halogen flat fields were taken during the nights, bracketing the 
science exposures. For each science exposure, the combination
of the two flat fields taken closest in time was used
to construct the normalized flat field.
The flat fields were bias subtracted and trimmed the same
way as for the imaging data, except the task {\tt gsreduce} was
used enabling the mask definition file (MDF) to be attached as an extension to the 
resulting MEF file. 
The MDF contains the information about the location of the slit-lets on the detector array.
The flats were combined in sets
corresponding to the science frames, and the flat fields were normalized
extension by extension.  To do so, a cubic spline was fit to the 
data, rejecting pixels that deviated more than $\pm 2.5 \sigma$
from the fit. This technique is now part of {\tt gsflat}.

The flat fields have emission lines at 6926 {\AA} and 6940 {\AA}.
These emission lines originate from a surface inside the Gemini calibration unit.
The pixels affected in the flat fields were set to 1.0,
and these pixels cannot be flat field corrected.

The normalized flat fields were then applied to the science
exposures, extension by extension ({\tt gsreduce}).
The resulting science exposures still have three extensions
as the raw exposures, but have now been trimmed, bias
subtracted and flat fielded.

\subsubsection{Sky subtraction and fringe frames}

The techniques implemented in Gemini IRAF package v1.7 for 
the sky subtraction of MOS data are different from the ones used for the current dataset.
Because of the strong sky lines in the red, we developed 
a technique to subtract off the sky before any interpolation
was done on the data. We also construct fringe frames
and subtract these off prior to any interpolation of the data.
For the spectroscopic data, the fringing is larger than $\pm 2$ per cent for wavelengths 
longer than 700 nm.

The three extensions in the science exposures were temporarily pasted 
together to one image, using {\tt gmosaic} with the paste flag set.
This has the effect that no interpolation is done at this point, which
is essential for the sky subtraction technique. The fact that
no correction is done for the accurate positions of the three chips 
relative to each other is not important at this stage.
The images were then cut into pieces corresponding to the 
spectra from the slit-lets. The first iteration of this was done
with {\tt gscut}, but the sections had to be manually adjusted to avoid
the very edges of the slit-lets. The resulting images are MEF
files with one image extension per slit-let.

The sky subtraction was done by fitting each slit-let
in the spatial direction with a 2nd order
polynomial rejecting pixels that deviate with more than
$\pm 3 \sigma$. An aperture of 1.45 arcsec  at the object position
was omitted from the fit. For the brightest cluster galaxy
an aperture of 2.2 arcsec was used. Further, the fitting was iterated 
five times and the rejected pixels grown with two pixels to 
effectively eliminate signal from the objects and from cosmic ray hits.

For each of the two wavelength settings a fringe frame was constructed from
the sky subtracted images by median combining these.
Apertures around the objects were masked and not included in the combined images. 
The aperture sizes were between 1.67 and 3.13 arcsec, depending on the 
angular size of the objects.
Because the observations are dithered in the spatial direction
the resulting fringe frames have full spatial coverage.
The result is two fringe frames, one for the 
central wavelength setting of 805 nm and one for 815 nm.

The fringe frames were subtracted from each of the 
individual images. The level of fringing varies and 
the best scaling of the fringe frames was determined 
by inspecting the results from scaling between 0.60 and
2.00, in steps of 0.05.

\begin{figure*}
\epsfxsize 16.5cm
\epsfbox{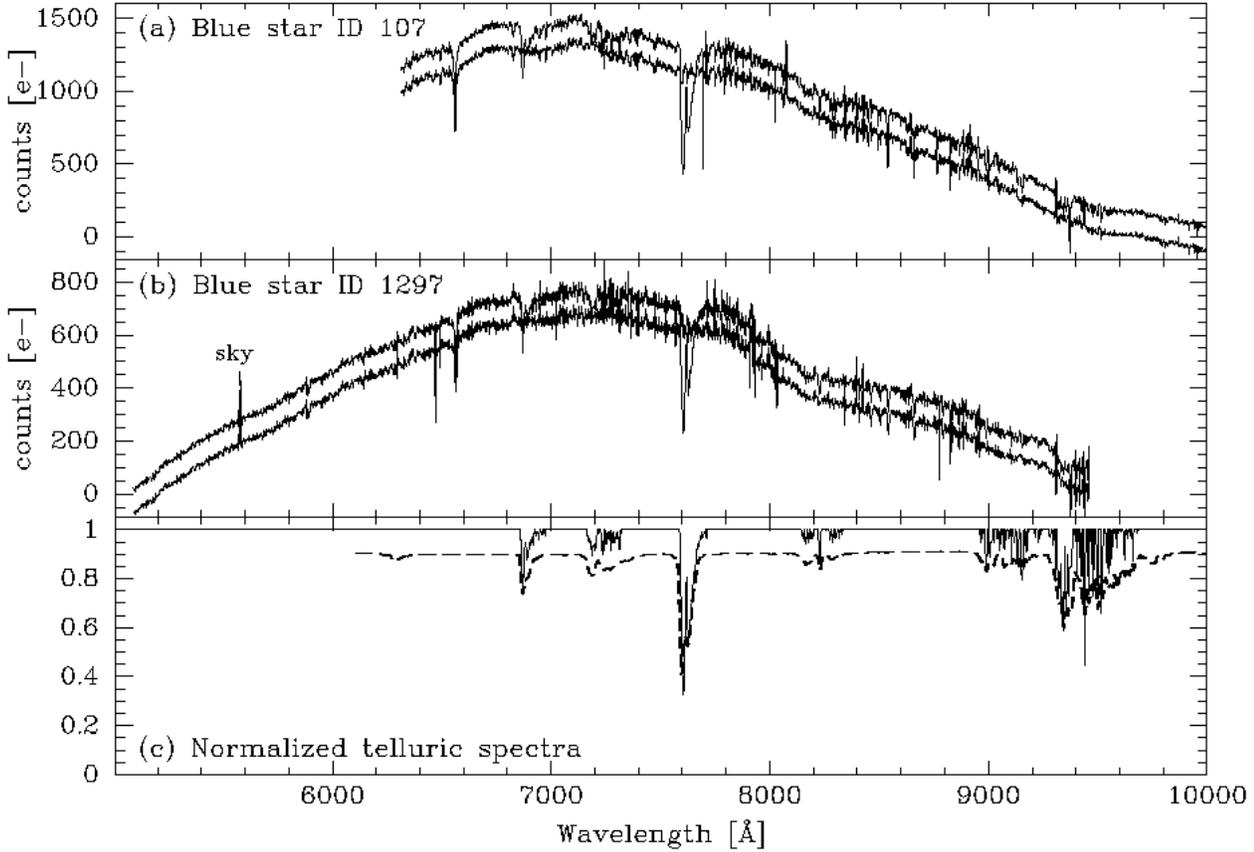}
\caption[]{(a) \& (b) Spectra of the two blue stars used to construct the
telluric absorption line spectrum. Top line --
the spectra before applying the telluric absorption correction;
bottom line --  the spectra after applying the telluric absorption correction,
for clarity offset with 10 per cent of the peak signal.
(c) Thin solid line -- normalized telluric absorption spectrum;
thick dashed line -- for reference a lower resolution normalized telluric absorption spectrum
from the Gemini Integration Time Calculator, for clarity offset with 10 per cent.
\label{fig-telluric} }
\end{figure*}

\subsubsection{Combination of the individual exposures}

The images taken at each wavelength setting were then combined.
The images were shifted in the spatial direction to compensate
for the dithering along the slit.
The combined images were put back into the original three
extensions and then mosaiced using {\tt gmosaic} with the full 
transformations enabled. Thus, in this step the task is used
to correct for the relative position of the three chips. 
The slit-lets were then cut out and put in individual extensions,
using the task {\tt gscut}.

\subsubsection{Wavelength calibration and extraction}

The wavelength calibrations were established from CuAr spectra ({\tt gswavelength}). 
The dispersion function was fit with a 6th order polynomial. 
The rms of the fits were typically 0.25 {\AA} (0.4 pixels).
The wavelength calibrations were applied to the combined 
spectra for each wavelength setting ({\tt gstransform}).
The gaps between the CCDs in GMOS result in gaps in the 
wavelength coverage for each of the wavelength settings. 

The wavelength calibrated spectra from the two wavelength settings were
combined using {\tt scombine}. 
The gaps were masked by setting the pixel values to $-1000$ ADU 
and the pixels were excluded from the combined images by applying
a threshold in {\tt scombine}.
With the release of IRAF 2.12.2a, this step can be accomplished
with the task {\tt lscombine}.

The spectra were then traced and extracted using {\tt gsextract},
which is a wrapper for {\tt apall} enabling easy processing of MEF files. 
The spectra were extracted in apertures that are 1.15 arcsec wide.
For some of the slits, the wavelength range goes redwards of
10300 {\AA}ngstrom, where 2nd order contamination is significant.
This part of the spectrum was deleted from the trace as it 
is offset from the 1st order trace because of the atmospheric
differential refraction.

\subsubsection{Correction for telluric lines}

Slits for two blue stars were included in the mask design in order to enable
correction for the telluric lines. Together the two blue stars
provide complete wavelength coverage for all the spectra of the science targets.
The spectra of the two blue stars were combined using {\tt scombine}.
The resulting spectrum was normalized by fitting it with a
27-piece cubic spline and dividing it with the fit.
The wavelength ranges unaffected by telluric lines were set to one.
The science spectra were then divided by the telluric line spectrum.
Figure \ref{fig-telluric} shows the spectra of the blue stars before and
after removal of the telluric lines as well as the resulting 
normalized telluric absorption spectrum that was applied to all 
the galaxy spectra. The stellar spectra shown on this figure have not been flux calibrated.

\subsubsection{Relative flux calibration}

Because MOS spectra cover different wavelength ranges depending
on where in the field of view the slit-lets are located, the
spectra cannot be flux calibrated using a single observation
of a standard star. Instead several observations were used
spanning a range in central wavelength, such that when these
were combined the resulting sensitivity function covers all
wavelength ranges covered by the MOS spectra.
The extracted spectra were flux calibrated with this
sensitivity function. The calibration is relative since the
slit width used for the standard star observations is 1 arcsec
and some of the observations of RXJ0152.7--1357 were taken through thin cirrus.

\begin{figure*}
\epsfxsize 12cm
\epsfbox{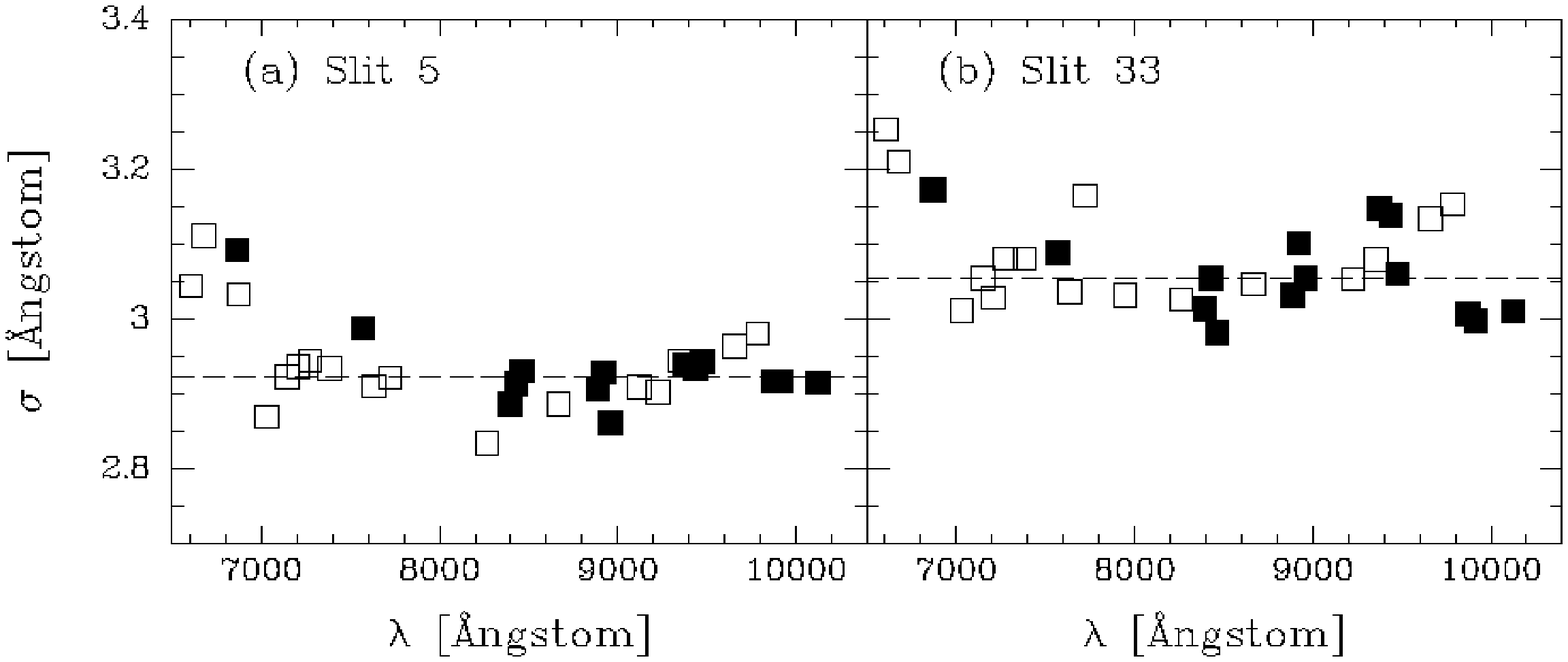}
\caption[]{Instrumental resolution as a function of wavelength for two of the 
slit-lets. Open symbols -- instrumental resolution derived from CuAr arc spectra;
filled symbols -- instrumental resolution derived from single sky frame.
\label{fig-instres_indext} }
\end{figure*}

\begin{figure*}
\epsfxsize 12cm
\epsfbox{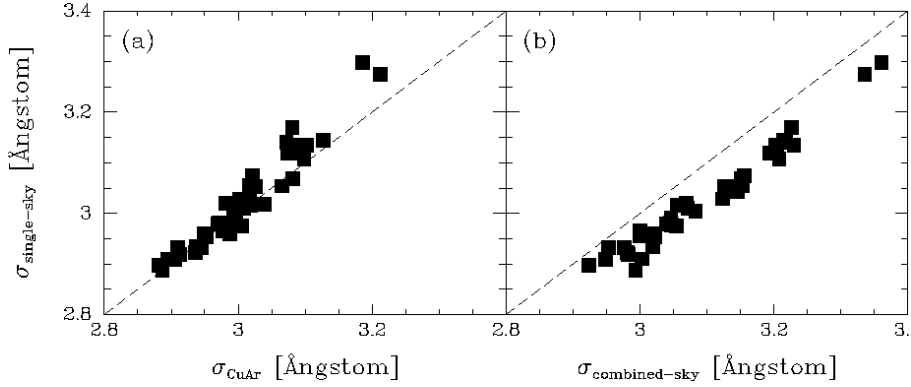}
\caption[]{Median instrumental resolution derived from a single sky frame versus
(a) the instrumental resolution derived from CuAr arc spectra, and (b)
the instrumental resolution derived from the sky spectra combined the same
way as the science spectra. The dashed lines on both panels are the one-to-one
relations. Combining the science spectra degrades the instrumental resolution with 0.065\AA.
\label{fig-instres_arcsky} }
\end{figure*}

\begin{figure*}
\epsfxsize 12cm
\epsfbox{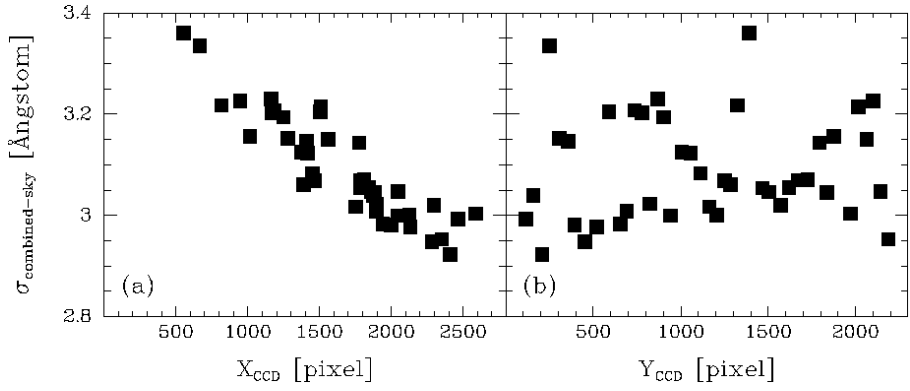}
\caption[]{Median instrumental resolution derived from the combined sky frame versus
position on the detector. The instrumental resolution depends on the X-position
of the slit.
\label{fig-instres_ccdpos} }
\end{figure*}

\begin{figure*}
\epsfxsize 6.6cm
\epsfbox{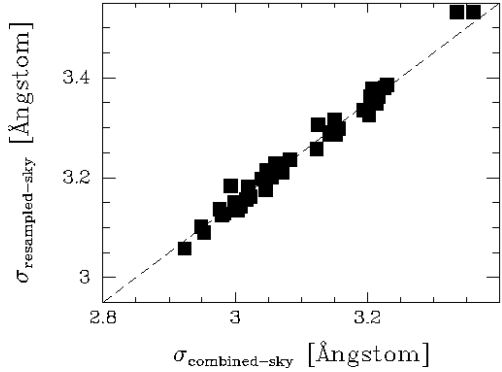}
\caption[]{Median instrumental resolution derived from the combined sky frame 
median filtered and resampled to the critical resolution versus
the median instrumental resolution before median filtering and resampling.
The median filtering and resampling degrades the instrumental resolution with 0.15{\AA}.
The dashed line is offset with 0.15{\AA} from the one-to-one relation.
\label{fig-instres_sesky} }
\end{figure*}

\subsubsection{Instrumental resolution and resampling of the spectra}

Determination of the instrumental resolution is needed in order
to derive the internal velocity dispersions of the galaxies.
We derived the instrumental resolution both from gaussian fits
to the isolated lines in the CuAr arc spectra and from gaussian fits to 
skylines. Since the final science spectra were extracted from the
combination of many individual frames, the analysis of the sky lines
was done on both a single frame and on a combined frame processed
in the same way as the science spectra. By comparing the results
from the single sky frame and the combined frame we can assess to
what extent the process of combining the individual frames degrades
the instrumental resolution.

The skylines are significantly more crowded than the CuAr lines, 
due to the wavelength interval covered by the spectra.
Therefore, the continuum of the sky was first subtracted off by
fitting a low order polynomial to the sky spectra.
The signal from the sky emission lines were omitted by using
iterative rejection of positive residuals from the fit.
Then the sky lines were fit with gaussian profiles, assuming that the background was zero.

Figure \ref{fig-instres_indext} shows the instrumental resolution as a function
of wavelength for two of the slit-lets. Both the CuAr spectra and 
the sky spectra show that the instrumental resolution does not
depend significantly on the wavelength. 
Figure \ref{fig-instres_arcsky}a shows the median values of the instrumental resolution 
derived from the single sky frame versus the results from the CuAr arc spectra for 
all slit-lets. The resolutions derived from the CuAr arc spectra and the single 
sky frame are consistent. Figure \ref{fig-instres_arcsky}b shows the results from
the single sky frame versus the combined sky frame. The instrumental resolution
of the combined sky frame is on average 0.065 \AA\ worse than the resolution of 
the single sky frame. This small degradation is caused by combining the 25 individual 
frames.

The instrumental resolution varies from slit to slit. In Figure \ref{fig-instres_ccdpos}
we show the instrumental resolution as a function of the slit's position on the
detector array. The instrumental resolution depends on the X-position. While the 
cause of this is not clear, the important conclusion for the
present use of the data is that we will use the individual values for the 
instrumental resolution when determining the internal velocity dispersions of 
the galaxies.

The original sampling of the spectra oversamples the spectral resolution. 
We median filtered the spectra with 5 pixels in the 
spectral direction, and then resampled them to have 2.75{\AA} per pixel, which is just
above the critical sampling for the instrumental resolution. This process
degrades the resulting instrumental resolution slightly, see 
Figure \ref{fig-instres_sesky}. This degradation is taken into account in our
determination of the internal velocity dispersions of the galaxies. 

\subsubsection{Determination of redshifts and velocity dispersions}

For the determination of the redshifts and the velocity dispersions, we 
used observations of three template stars obtained with GMOS-N using the
B600 grating and a slit width of either 0.75 arcsec or 0.50 arcsec.
Information about the template spectra is listed in Table \ref{tab-templates}.
The template stars were observed with the star stationary in the slit. 
Thus, the effective instrumental resolution is partly determined by the
image quality of the observation. Therefore,
the effective instrumental resolution for each of the template spectra 
was derived by comparing the spectra to other template stars observed with 
GMOS-N using a technique that fills the slit with signal. Those other
template star observations have shorter and more redwards wavelength coverage 
and are therefore not optimal for use for the spectra of RXJ0152.7--1357.
The effective instrumental resolution for each of the template spectra is
given in Table \ref{tab-templates}. 

\begin{deluxetable}{lrrrr}
\tablecaption{GMOS-N Template Spectra \label{tab-templates} }
\tablewidth{30pc}
\tablehead{
\colhead{Star} & \colhead{Spectral type} & \colhead{$V_{\rm hel}$} 
& \colhead{Wavelength coverage} & \colhead{$\sigma_{\rm inst}$\tablenotemark{a}} \\
\colhead{} & \colhead{} & \colhead{$\rm km\,s^{-1}$} & \colhead{({\AA})} }
\startdata
HD172958 & B8V   & --16.0 & 3674.8\AA\ -- 6539.2\AA\ & 1.253 \\
HD030649 & G1V   &   26.0 & 3746.7\AA\ -- 6630.8\AA\ & 0.940 \\
HD172401 & K0III &   23.6 & 3674.3\AA\ -- 6539.1\AA\ & 1.243 \\
\enddata
\tablenotetext{a}{Effective instrumental resolution, see text for details}
\end{deluxetable}

\begin{deluxetable*}{rrrrrrrrrr}
\tablecaption{Results from Template Fitting \label{tab-speckin} }
\tablewidth{0pc}
\tablehead{
\colhead{ID} & \colhead{Redshift} & \colhead{Member\tablenotemark{a}} &  
\colhead{$\log \sigma$} & \colhead{$\log \sigma _{\rm cor}$\tablenotemark{b}} & 
\colhead{$\sigma _{\log \sigma}$} & \multicolumn{3}{c}{Template fractions} & \colhead{S/N\tablenotemark{c}} \\
\colhead{}&\colhead{} &\colhead{} &\colhead{} &\colhead{} &\colhead{} & \colhead{B8V} & \colhead{G1V} & \colhead{K0III} & }
\startdata
        103& 0.6406&  0&  \nodata & \nodata & \nodata & \nodata & \nodata & \nodata & 34.8\\
        155& 0.9955&  0&  \nodata & \nodata & \nodata & \nodata & \nodata & \nodata & 22.1 \\
        193& 0.4562&  0&  \nodata & \nodata & \nodata & \nodata & \nodata & \nodata & 34.2 \\
        264& 0.5341&  0&  \nodata & \nodata & \nodata & \nodata & \nodata & \nodata & 26.5 \\
        338& 0.8193&  1&  2.057&  2.084&  0.057&   0.19&   0.54&   0.28 & 29.8 \\
        346& 0.8367&  1&  2.152&  2.179&  0.050&   0.08&   0.69&   0.23 & 52.4 \\
        422& 0.8342&  1&  2.009&  2.036&  0.070&   0.00&   0.65&   0.35 & 35.0 \\
        460& 0.8649&  0&  \nodata & \nodata & \nodata & \nodata & \nodata & \nodata & 80.0 \\
        523& 0.8206&  1&  2.352&  2.379&  0.038&   0.00&   0.65&   0.35 & 62.1 \\
        566& 0.8369&  1&  2.184&  2.211&  0.055&   0.29&   0.46&   0.25 & 31.4 \\
        627& 0.8324&  1&  2.259&  2.286&  0.048&   0.00&   0.55&   0.45 & 25.4 \\
        643& 0.8445&  1&  2.238&  2.265&  0.102&   0.08&   0.68&   0.24 & 30.8 \\
        737& 0.8384&  1&  2.152&  2.179&  0.056&   0.00&   0.65&   0.35 & 27.2 \\
        766& 0.8346&  1&  2.362&  2.389&  0.036&   0.00&   0.51&   0.49 & 45.7 \\
        776& 0.8325&  1&  2.052&  2.079&  0.049&   0.19&   0.48&   0.32 & 57.5 \\
        813& 0.8351&  1&  2.312&  2.339&  0.031&   0.00&   0.56&   0.44 & 64.9 \\
        896& 0.8458&  1&  1.639&  1.666&  0.049&   0.65&   0.35&   0.00 & 16.1 \\
        908& 0.8393&  1&  2.237&  2.264&  0.041&   0.00&   0.69&   0.31 & 45.8 \\
       1027& 0.8357&  1&  2.290&  2.317&  0.034&   0.00&   0.47&   0.53 & 29.6 \\
       1085& 0.8325&  1&  2.361&  2.388&  0.032&   0.00&   0.47&   0.53 & 49.0 \\
       1110& 0.8322&  1&  2.276&  2.303&  0.051&   0.00&   0.48&   0.52 & 39.8 \\
       1159& 0.8357&  1&  2.150&  2.177&  0.087&   0.29&   0.49&   0.22 & 32.3 \\
       1210& 0.8372&  1&  2.079&  2.106&  0.085&   0.00&   0.51&   0.49 & 24.8 \\
       1245& 0.7875&  0&  \nodata & \nodata & \nodata & \nodata & \nodata & \nodata & 27.8 \\
       1299& 0.8374&  1&  2.130&  2.157&  0.052&   0.30&   0.45&   0.26 & 24.8 \\
       1385& 0.8368&  1&  2.134&  2.161&  0.075&   0.63&   0.37&   0.00 & 26.6 \\
       1458& 0.8324&  1&  2.132&  2.159&  0.066&   0.00&   0.66&   0.34 & 30.9 \\
       1494& 0.2374&  0&  \nodata & \nodata & \nodata & \nodata & \nodata & \nodata & 380. \\
       1507& 0.8289&  1&  2.311&  2.338&  0.051&   0.19&   0.61&   0.20 & 22.9 \\
       1567& 0.8291&  1&  2.461&  2.488&  0.030&   0.10&   0.41&   0.49 & 32.7 \\
       1590& 0.8317&  1&  1.959&  1.986&  0.078&   0.21&   0.19&   0.60 & 28.0 \\
       1614& 0.8433&  1&  2.321&  2.348&  0.041&   0.06&   0.66&   0.28 & 29.8 \\
       1682& 0.8463&  1&  2.240&  2.268&  0.050&   0.11&   0.57&   0.33 & 43.2 \\
       1811& 0.8351&  1&  1.735&  1.762&  0.076&   0.00&   0.55&   0.45 & 18.0 \\
       1838& 0.7450&  0&  \nodata & \nodata & \nodata & \nodata & \nodata & \nodata & 28.2 \\
       1896& 0.9810&  0&  \nodata & \nodata & \nodata & \nodata & \nodata & \nodata & 29.0 \\
       1920& 0.8442&  1&  1.569&  1.596&  0.134&   0.06&   0.34&   0.60 & 10.0 \\
       1935& 0.8252&  1&  2.177&  2.204&  0.058&   0.00&   0.71&   0.29 & 34.4 \\
       1970& 0.3775&  0&  \nodata & \nodata & \nodata & \nodata & \nodata & \nodata & 21.7 \\
       2042& 0.2362&  0&  \nodata & \nodata & \nodata & \nodata & \nodata & \nodata & 42.5 \\
       2087& 0.3320&  0&  \nodata & \nodata & \nodata & \nodata & \nodata & \nodata & 102. \\
\enddata
\tablenotetext{a}{Adopted membership: 1 -- galaxy is a member of RXJ0152.7--1357; 
0 -- galaxy is not a member of RXJ0152.7--1357}
\tablenotetext{b}{Velocity dispersions corrected to a standard size aperture equivalent to a
circular aperture with diameter of 3.4 arcsec at the distance of the Coma cluster.}
\tablenotetext{c}{S/N per {\AA}ngstrom in the rest frame of the galaxy. The S/N was derived 
in the rest frame wavelength interval 4100-4600 {\AA}, 
except for ID 1494, 2042, and 2087 for which the wavelength interval 5500-6000 {\AA} was used.}
\end{deluxetable*}

Initial redshifts were derived by cross-correlating the spectra with 
a spectrum of the K0III star HD172401. Several template spectra
were constructed by offsetting the template spectrum to redshifts 
between 0.3 and 0.9. Then the galaxy spectra were cross-correlated with
the template closest in redshift, using the task {\tt xcor}, which is part
of STSDAS\footnote{STSDAS is a product of Space Telescope Science Institute,
which is operated by AURA for NASA.}.
For galaxies with obvious emission lines, the emission lines were used
for the initial determination of the redshift of the galaxy.

Once the redshift had been determined to an accuracy of about 200 $\rm km\,s^{-1}$,
velocity dispersions as well as more accurate redshifts were derived using 
software made available by Karl Gebhardt.
The software uses penalized maximum likelihood fitting in pixel space to determine
the velocity dispersion and the redshift, see Gebhardt et al.\ (2000, 2003) for a detailed 
description of the fitting method.

\begin{figure*}
\epsfxsize 12cm
\epsfbox{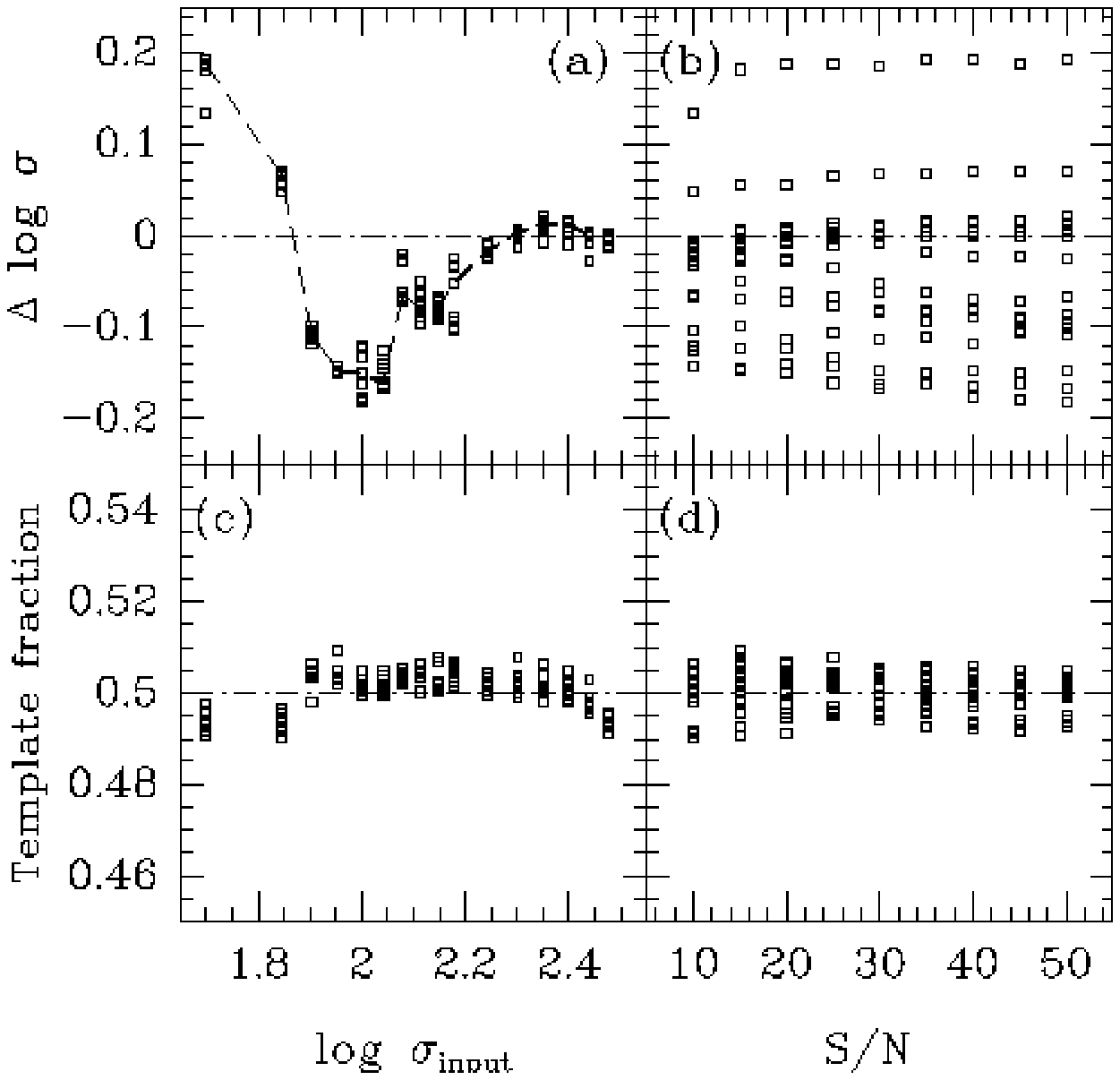}
\caption[]{Results from simulations. The systematic error on $\log \sigma$,
$\Delta \log \sigma = \log \sigma _{\rm output} - \log \sigma _{\rm input}$,
(a) as a function of $\log \sigma _{\rm input}$, and (b) as a function of S/N per {\AA}ngstrom.
The dashed line on panel (a) marks the median values of $\Delta \log \sigma$.
The output template fractions (c) as a function of $\log \sigma _{\rm input}$, 
and (d) as a function of S/N per {\AA}ngstrom. See text for discussion.
\label{fig-simulate} }
\end{figure*}

We used the three template stars simultaneously in order to limit the 
systematic errors introduced by template mis-match. 
Further, wavelength ranges affected by emission lines or strong residuals 
from the sky subtraction were masked and not used for the fitting. 
The blue limit for the fitted wavelength ranges was in all cases 3750\AA\ in
the rest frame of the galaxy such
that the [\ion{O}{2}] emission line was excluded from all fits.
The red limit was chosen to be as far red as the S/N would allow.
For the cluster members we typically used a limit of 5050\AA\ in the rest 
frame of the galaxy. 

Gebhardt's software is written primarily for processing of spatially
resolved spectra of nearby galaxies. We therefore wrote an interface to
the program that takes care of shifting both input galaxy spectra and
template stars to the rest frame, after which all fitting is done in the
rest frame. The interface limits the wavelength range fitted to
the wavelength range in common after the shifts to the rest frame.
The interface also convolves the template spectra to the instrumental resolution
for each of the galaxy spectra.
The results from the fitting are a refined redshift and the best fitting
velocity dispersion. Uncertainties are determined from Monte-Carlo simulations.
Further, the output contains information about the fractional composition
of the best fitting template spectrum made from the three input template
spectra. This can be used as a crude spectral classification of the 
galaxy spectra.

The derived velocity dispersions were corrected for the size of the aperture
using the technique described in J\o rgensen et al.\ (1995). 
We correct the velocity dispersions to a standard aperture size with
a radius $r_{\rm norm}$, which is equivalent to 1.7 arcsec at the distance of
the Coma cluster.
For members of RXJ0152.7--1357 the correction to the velocity 
dispersions is $\approx$ 6 per cent.

Table \ref{tab-speckin} summarizes results from the template fitting.
Measured velocity dispersions as well as aperture corrected velocity dispersions
are listed for the cluster members. For galaxies that are not members of
the cluster, we give the redshift. The detailed data for these galaxies
will be discussed in a future paper.

\subsubsection{The instrumental resolution -- effects on the derived velocity dispersions}

In order to determine at which velocity dispersion and S/N the determination
of the velocity dispersions becomes affected by systematics we have made
model galaxy spectra from our three template stars. The model galaxy spectra
cover velocity dispersions from 50 $\rm km\,s^{-1}$ to 300 $\rm km\,s^{-1}$,
and average S/N from 10 to 50 per {\AA}ngstrom in the rest frame.
The noise is modeled from a noise spectrum that contains the average variation
with wavelength, due to sky lines and system throughput, that applies to the 
RXJ0152.7--1357 cluster members.
The model spectra match the instrumental resolution and the sampling 
of the real data.
Three sets of model spectra were made with mixes of the templates \{B8V, G1V, K0III\}
of \{0.25, 0.5, 0.25\}, \{0., 0.5, 0.5\}, and \{0., 0., 1.\}, respectively.
For each model spectrum we derive the velocity dispersion as for the 
real data, using all three template stars. 
We made 100 realizations of each combination of velocity dispersion and S/N.
We then derived the median values of $\log \sigma$ for the 100 realizations.

Figure \ref{fig-simulate}a and b shows the systematic errors in $\log \sigma$
as the difference between the input $\log \sigma$ and median value from the simulation,
as a function of the S/N of the model and the input $\log \sigma$.
The template mix turned out to have very little effect on the systematic errors. Thus,
Figure \ref{fig-simulate} shows only the result from the 50/50 mix of the
G1V and K0III templates.
Further, the S/N of the spectra does not significantly affect the 
systematic error, see Figure \ref{fig-simulate}b. 
This is different from what has been found for the 
Fourier Fitting Technique where low S/N leads to larger systematic errors 
(e.g., J\o rgensen et al.\ 1995).
Velocity dispersions below the instrumental resolution ($\log \sigma = 2.06$) 
may be subject to systematic errors as large as $\pm 0.15$ in $\log \sigma$;
for velocity dispersions smaller than 0.5 times the instrumental resolution
the systematic effects are larger.
According to the models, the systematic errors lead to both underestimated 
and over estimated velocity dispersions depending on the value of the input
velocity dispersion. For input velocity dispersions of 0.5-1.0 times 
the instrumental resolution the velocity dispersions get underestimated,
while for smaller input velocity dispersions the systematic errors
lead to overestimated velocity dispersions.
The galaxies ID 896, 1811, and 1920 have velocity dispersions
of less than $\approx 0.5$ times the instrumental resolution 
and the measurements therefore are expected to be affected by systematic effects.
The galaxies ID 422 and 1590 have velocity dispersions slightly below
the instrumental resolution. According to the models, the measurements 
for these galaxies may be subject to systematic errors of the order 
0.1 on $\log \sigma$. This is similar to
the size of the random uncertainties for these galaxies.
In the analysis of scaling relations involving $\log \sigma$ we exclude
the three low dispersion galaxies from the fits. Further, we use median
zero points for our analysis in order to limit the effect of the systmatic errors
in $\log \sigma$ on our results.

Figure \ref{fig-simulate}c and d show the output template fractions
as a function of the input velocity dispersion and the S/N. Independent of the
velocity dispersion and the S/N, the output template fractions are
within 0.01 of the input template fractions.

\begin{figure*}
\epsfxsize 17.5cm
\epsfbox{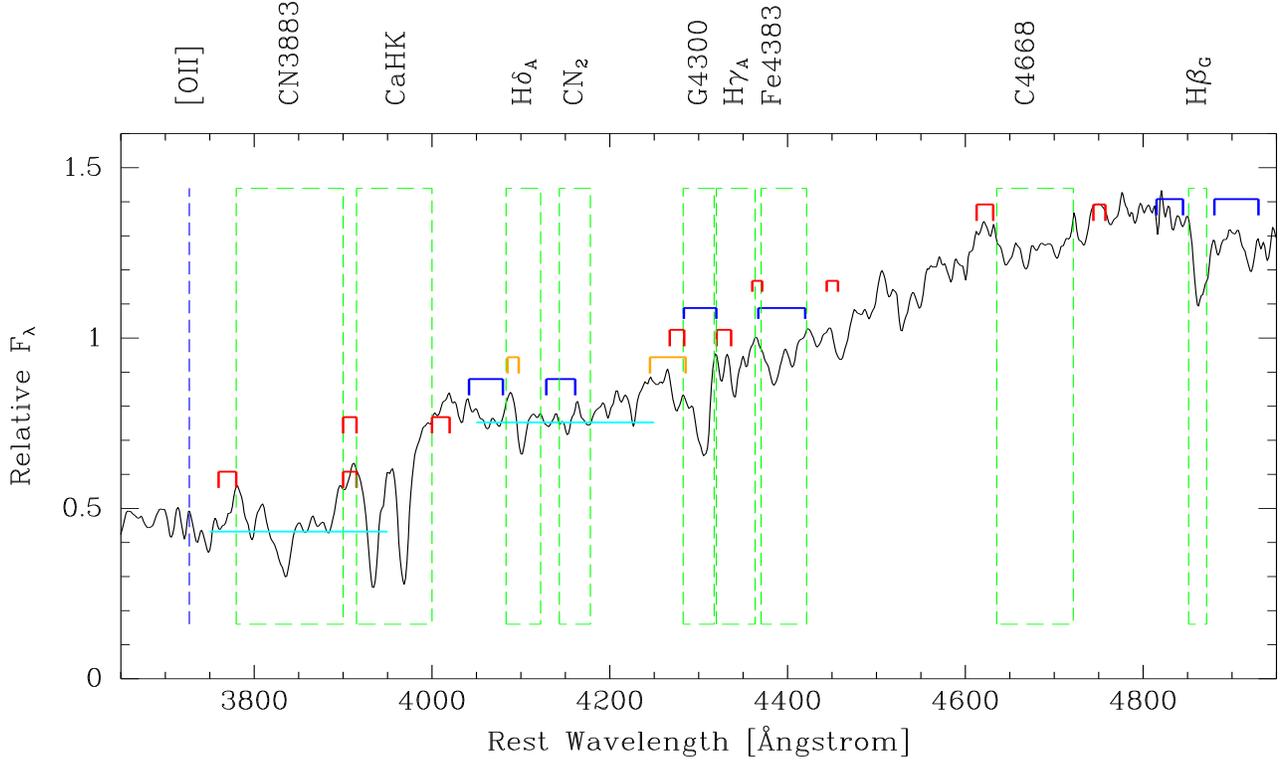}
\caption[]{Passbands for the line indices. The spectrum is the composite of the
22 cluster members that do not have emission lines. The on-line passband for each index is 
marked as the dashed green box. The continuum passbands are marked in blue (for the 
Balmer lines), orange (for CN$_2$) or red (all other metal line indices). 
Except for CN$_2$, the continuum passbands are very close the the on-line passbands. 
The D4000 passbands are marked as horizontal cyan lines. The location of the [\ion{O}{2}]
emission line is marked, though the line is not present in  the spectrum.
The passbands for $\rm H\beta _G$ are marked, though this index cannot be measured
reliably from the spectra of individual RXJ0152.7--1357 galaxies.
\label{fig-bands_comp} }
\end{figure*}

\subsubsection{Determination of line indices}

Absorption line indices were derived from the spectra. For indices in the Lick/IDS system,
the definitions from Worthey et al.\ (1994) were used. In addition we determined the indices
for H$\gamma$ and H$\delta$ defined by Worthey \& Ottaviani (1997), the
H$\beta _{\rm G}$ index defined by J\o rgensen (1997),
the D4000 index defined by Bruzual (1983), see also Gorgas et al.\ (1999), and
the blue indices CN3883 and CaHK defined by Davidge \& Clark (1994).
The passbands for the indices are shown on Figure \ref{fig-bands_comp}.
Indices redwards of C4668 are for galaxies in RXJ0152.7--1357 severely affected by
sky subtraction residuals. Thus, these indices were not measured.
Figure \ref{fig-bands_comp} shows the passbands for the indices measured for the
RXJ0152.7--1357 galaxies.

The spectra were convolved to the instrumental resolution of the Lick/IDS spectra,
in the rest frame of the galaxies. We adopted the instrumental resolution of the Lick/IDS
spectra presented in Worthey \& Ottaviani (1997). The indices and uncertainties, based on the
noise spectra, were then derived. 

The raw values of the indices were corrected 
for the velocity dispersion of the galaxy, following the technique described by,
e.g., Davies et al.\ (1993). 
We used the spectra of three template stars used for the determination
of the velocity dispersions to derive the corrections for the velocity dispersion.
The indices are of three different types, equivalent widths (EW), ratios or magnitudes,
see Table \ref{tab-apcor}.
For indices derived as EW or ratios the corrected indices, $\rm index_0$ were derived as
\begin{equation}
\label{eq-velcor1}
{\rm index_0} = C(\sigma ) \cdot {\rm index}
\end{equation}
where {\rm index} is the value derived for the galaxy, $C(\sigma )$ is the correction
factor, and ${\rm index_0}$ is the corrected index for a zero velocity dispersion.
For indices derived in magnitudes, we used
\begin{equation}
\label{eq-velcor2}
{\rm index_0} = C(\sigma ) + {\rm index}
\end{equation}
For each galaxy, a composite template was constructed using the template
fractions found from the template fitting, see Table \ref{tab-speckin}.
The composite template was convolved to the Lick/IDS resolution and the 
line indices derived. Then the composite template was convolved with the
velocity dispersion found for the galaxy, and the line indices derived.
The corrections $C(\sigma )$ for the galaxy were derived as the ratio or 
difference, as specified in Eqs.\ \ref{eq-velcor1} and \ref{eq-velcor2}.

The velocity dispersion correction factor for a composite template of 
$0.5 {\rm G1V} + 0.5 {\rm K0III}$ and a velocity dispersion of 230 $\rm km\,s^{-1}$ is 
listed in Table \ref{tab-apcor}. 
The corrected indices are in general stronger than the uncorrected indices.
For the indices UVcont, NH3360, and BL3580 we cannot determine the velocity dispersion
correction since the template star spectra do not cover these short wavelengths.
However, since all the passbands are quite broad for these indices we expect the
corrections to be insignificant.

\begin{deluxetable}{lrrrrrrrrrrrrrrrrrrr}
\tablecaption{Corrections to the Line Indices\label{tab-apcor} }
\tablewidth{14pc}
\tablehead{
\colhead{Index} & \colhead{Type} & \colhead{$C(\sigma )$\tablenotemark{a}} & 
\colhead{$\alpha$} & \colhead{Reference\tablenotemark{b}} }
\startdata
UVcont & Ratio & \nodata & 0.035 & 1 \\
NH3360 & EW    & \nodata & 0.05  & 1 \\
BL3580 & EW    & \nodata & 0.05  & 1 \\
CN3883 & mag   & 0.003  & 0.05  & 1 \\
CaHK   & EW    & 1.009  & 0.025 & 2  \\
D4000  & Ratio & 1.000  & 0.03  & 4  \\
H$\delta_{\rm A}$ & EW & 1.114 & 0.0 & \\
H$\delta_{\rm F}$ & EW & 1.158 & 0.0 & \\
CN1    & mag   & 0.002  & 0.04  & 2  \\
CN2    & mag   & 0.006  & 0.04  & 2  \\
Ca4227 & EW    & 1.297  & 0.05  & 3  \\
G4300  & EW    & 1.035  & 0.015 & 2  \\
H$\gamma_{\rm A}$ & EW & 1.009  & 0.0 &  \\
H$\gamma_{\rm F}$ & EW & 1.014 &  0.0 & \\
Fe4383 & EW    & 1.110  & 0.05  & 2  \\
Ca4455 & EW    & 1.240  & 0.05  & 3  \\
Fe4531 & EW    & 1.078  & 0.05  & 3  \\
C4668  & EW    & 1.048  & 0.08  & 3  \\
H$\beta$ & EW & 1.031  & -0.005 & 3 \\
H$\beta_{\rm G}$ & EW & 1.051  & -0.005 & 3 \\
Mg$b$  & EW    & 0.002  & 0.05  & 3  \\
Mg$_1$ & mag   & 0.002  & 0.04  & 3  \\
Mg$_2$ & mag   & 1.112  & 0.04  & 3  \\
Fe5270 & EW    & 1.134  & 0.05  & 3  \\
Fe5335 & EW    & 1.289  & 0.05  & 3  \\
NaD    & EW    & 1.064  & 0.09  & 3  \\
\enddata
\tablenotetext{a}{Correction to zero velocity dispersion for a galaxy with a velocity dispersion
of 230 $\rm km\,s^{-1}$ and a template mix of $0.5 {\rm G1V} + 0.5 {\rm K0III}$.}
\tablenotetext{b}{Main source for radial gradient data, or the reference for a previously used
value of the aperture correction coefficient $\alpha$.}
\tablerefs{ (1) -- Davidge \& Clark (1994); (2) - Vazdekis et al.\ (1997);
(3) -- J\o rgensen (1997); (4) -- Cardiel et al.\ (1998).}
\end{deluxetable}

Galaxies have radial gradients in the line strengths. Thus, the line indices 
need to be aperture corrected. However, the aperture correction is not know for most
of the line indices derived. For line indices also measured by J\o rgensen (1997) we
have adopted the aperture correction described in that paper. For the remaining
line indices, we have estimated the aperture corrections in the same way as done
by J\o rgensen (1997), by comparing measurements of radial gradients from various
previous studies. We used data from Vazdekis et al.\ (1997), Cardiel et al.\ (1998),
and Davidge \& Clark (1994). For indices measured as equivalent width in \AA ngstrom or as a ratio (D4000 and UVcont),
we assume that the aperture correction can be written as 
\begin{equation}
{\rm index}_{\rm norm} = {\rm index}_{\rm ap} \cdot ( {\frac{r_{\rm ap}}{r_{\rm norm}} ) ^{\alpha}}
\end{equation}
For indices measured in magnitudes we use
\begin{equation}
{\rm index}_{\rm norm} = {\rm index}_{\rm ap} + \alpha \log \frac{r_{\rm ap}}{r_{\rm norm}}
\end{equation}
The standard size aperture has a radius, $r_{\rm norm}$, equivalent to 1.7 arcsec
at the distance of the Coma cluster.
The assumption about the aperture corrections and the standard size aperture are consistent 
with the method used in J\o rgensen (1997). 
For the indices Ca4227 and Ca4455 we adopted the same aperture correction
as used for the Fe-indices. 
In Table \ref{tab-apcor} we list the adopted values for 
the coefficient $\alpha$ and the main source of radial gradient data used
to estimate the coefficient.  For completeness, the corrections are also
listed for weaker indices in the blue, for H$\beta _{\rm G}$, and for 
the Mg, Fe and Na indices in the visible region, though these indices 
cannot be measured for the RXJ0152.7--1357 galaxies.
The higher order Balmer line indices (H$\gamma_{\rm A}$, H$\gamma_{\rm F}$, H$\delta_{\rm A}$,
H$\delta_{\rm F}$) have not been aperture corrected. The correction coefficient
for H$\beta$ determined by J\o rgensen (1997) is $-0.005$, giving a correction for
H$\beta$ of the order 0.5 per cent for galaxies in RXJ0152.7--1357. 
Therefore the aperture corrections for the higher order Balmer line indices 
may be small enough that they can be safely ignored. We note that the
continuum bands for the H$\gamma_{\rm A}$ index overlap the G4300 feature
and the Fe4383 line. Thus, H$\gamma_{\rm A}$ may have an aperture correction of
a similar size as the G4300 and Fe4383 indices and with the opposite sign.
Using the average of the aperture correction for those two indices, we estimate
that ignoring the aperture correction for H$\gamma_{\rm A}$, on average
may make the corrected indices 0.2 too large. This is not significant for the 
current analysis.

The Lick/IDS system was established from spectra that were not flux calibrated.
Therefore, offsets between indices measured from flux calibrated spectra and the
Lick/IDS system are expected. Several authors have derived such offsets, in particular, 
Worthey \& Ottaviani (1997) and J\o rgensen (1997).
For indices measured from narrow passbands, J\o rgensen (1997) finds no significant
offsets.  Worthey \& Ottaviani find offsets of similar size, though they 
do not comment on whether the offsets are significant.
For Mg$_1$ and Mg$_2$, both studies find significant offsets and the results
are consistent. J\o rgensen lists offsets for Mg$_1$ and Mg$_2$ of 0.007 and 0.011, 
respectively, with the Lick/IDS indices being larger.
For CN$_1$ and CN$_2$, Worthey \& Ottaviani find very small offsets, 0.003 and 0.004,
respectively.
Given the small size of the offsets, we choose in this project to only offset 
measurements of Mg$_1$ and Mg$_2$ to the Lick/IDS system. We adopt the 
offsets given by J\o rgensen (1997); these offsets are added to the measured
indices.

Tables \ref{tab-specline} lists the fully corrected values of the measured line indices
for galaxies in RXJ0152.7--1357.

\begin{deluxetable*}{rrrrrrrrrrrrr}
\tablecaption{Line Indices and EW[\ion{O}{2}] for Cluster Members\label{tab-specline} }
\tabletypesize{\scriptsize}
\tablewidth{0pc}
\tablehead{
\colhead{ID} & \colhead{CN3883}& \colhead{CaHK}& \colhead{D4000}
& \colhead{H$\delta_A$}& \colhead{CN$_1$}& \colhead{CN$_2$} 
& \colhead{G4300}& \colhead{H$\gamma_A$}& \colhead{Fe4383}& \colhead{C4668} & \colhead{EW [\ion{O}{2}]}
} 
\startdata
 338& 0.180& 23.94& 2.019& 1.39& 0.027& 0.027& 5.48& -4.32& 1.81& 4.34 & \nodata\\
 338& 0.005& 0.36& 0.005& 0.21& 0.000& 0.000& 0.31& 0.36& 0.30& 0.31 & \nodata\\
 346& 0.183& 23.26& 1.961& 1.52& 0.027& 0.040& 3.33& -3.56& 2.49& 6.89 & \nodata\\
 346& 0.003& 0.25& 0.003& 0.13& 0.000& 0.003& 0.18& 0.17& 0.12& 0.19 & \nodata\\
 422& 0.238& 21.69& 2.057& -0.50& 0.103& 0.128& 5.68& -4.50& 2.76& 8.21 & \nodata\\
 422& 0.005& 0.36& 0.005& 0.19& 0.005& 0.005& 0.25& 0.26& 0.19& 0.26 & \nodata\\
 523& 0.282& 21.25& 2.048& 0.40& 0.087& 0.121& 4.52& -3.77& 4.16& 5.37 & \nodata\\
 523& 0.003& 0.18& 0.002& 0.11& 0.003& 0.003& 0.14& 0.16& 0.14& 0.14 & \nodata\\
 566& 0.173& 24.42& 2.094& 2.97& 0.027& 0.027& 6.56& -2.80& 3.30& 4.78 & 8.4\\
 566& 0.005& 0.42& 0.005& 0.21& 0.000& 0.000& 0.29& 0.30& 0.20& 0.33 & 1.6\\
 627& 0.318& 23.26& 2.116& 0.01& 0.051& 0.088& 3.75& -4.18& 4.67& 6.42 & \nodata\\
 627& 0.007& 0.49& 0.006& 0.26& 0.006& 0.006& 0.36& 0.38& 0.26& 0.34 & \nodata\\
 643& 0.194& 23.84& 2.039& 2.61& 0.027& 0.027& 8.57& -8.64& 5.51& 6.92 & 8.2\\
 643& 0.005& 0.44& 0.005& 0.19& 0.000& 0.000& 0.27& 0.30& 0.17& 0.33 & 1.4\\
 737& 0.206& 17.73& 1.995& 1.83& 0.043& 0.068& 3.42& -3.54& 2.30& 3.24 & \nodata\\
 737& 0.005& 0.46& 0.005& 0.24& 0.006& 0.006& 0.30& 0.32& 0.22& 0.39 & \nodata\\
 766& 0.278& 22.36& 2.042& -0.78& 0.120& 0.150& 4.81& -4.77& 2.95& 8.00 & \nodata\\
 766& 0.003& 0.25& 0.003& 0.14& 0.004& 0.004& 0.18& 0.19& 0.14& 0.23 & \nodata\\
 776& 0.172& 19.88& 1.942& 1.84& 0.049& 0.072& 3.11& -1.93& 2.47& 4.39 & \nodata\\
 776& 0.003& 0.21& 0.002& 0.11& 0.003& 0.003& 0.15& 0.16& 0.12& 0.17 & \nodata\\
 813& 0.266& 20.63& 2.129& 1.31& 0.071& 0.100& 5.50& -5.47& 2.59& 8.99 & \nodata\\
 813& 0.002& 0.20& 0.002& 0.10& 0.002& 0.002& 0.13& 0.14& 0.10& 0.14 & \nodata\\
 896& 0.074& 6.83& 1.400& 4.09& 0.027& 0.027& 2.32& 2.88& -3.53& -13.52 & 21.6\\
 896& 0.006& 0.66& 0.006& 0.31& 0.000& 0.000& 0.55& 0.51& 0.35& 0.88 & 2.3\\
 908& 0.227& 20.83& 2.005& -0.02& 0.043& 0.072& 5.51& -4.65& 3.28& 8.30 & \nodata\\
 908& 0.003& 0.27& 0.003& 0.15& 0.003& 0.003& 0.18& 0.21& 0.13& 0.21 & \nodata\\
 1027& 0.323& 22.35& 2.179& -1.19& 0.073& 0.098& 4.33& -3.90& 5.49& 7.71 & \nodata\\
 1027& 0.006& 0.44& 0.006& 0.23& 0.005& 0.005& 0.30& 0.30& 0.21& 0.30 & \nodata\\
 1085& 0.269& 22.63& 2.160& -0.33& 0.097& 0.127& 5.72& -5.01& 3.42& 8.22 & \nodata\\
 1085& 0.004& 0.26& 0.003& 0.13& 0.003& 0.003& 0.18& 0.20& 0.14& 0.18 & \nodata\\
 1110& 0.274& 22.44& 2.245& 0.34& 0.120& 0.140& 3.07& -4.06& 5.04& 6.32 & \nodata\\
 1110& 0.005& 0.33& 0.004& 0.17& 0.004& 0.004& 0.22& 0.24& 0.17& 0.23 & \nodata\\
 1159& 0.145& 19.32& 1.794& 3.08& 0.027& 0.035& 1.90& 0.17& -0.49& 8.33 & 4.5\\
 1159& 0.004& 0.40& 0.004& 0.20& 0.000& 0.005& 0.27& 0.26& 0.21& 0.29 & 0.4\\
 1210& 0.230& 13.69& 1.937& -0.08& 0.068& 0.092& 4.00& -3.95& 0.84& -1.52 & \nodata\\ 
 1210& 0.006& 0.50& 0.006& 0.26& 0.006& 0.006& 0.37& 0.37& 0.26& 0.40 & \nodata\\
 1299& 0.181& 23.54& 2.079& 2.32& 0.027& 0.027& 7.55& -7.40& 5.68& 5.35 & 6.1\\
 1299& 0.006& 0.49& 0.006& 0.26& 0.000& 0.000& 0.37& 0.38& 0.23& 0.41 & 0.7\\
 1385& 0.073& 10.70& 1.459& 4.96& 0.027& 0.027& 1.40& 0.02& 0.01& -3.07 & 16.4\\ 
 1385& 0.004& 0.40& 0.004& 0.21& 0.000& 0.000& 0.32& 0.33& 0.26& 0.46 & 0.5\\
 1458& 0.238& 21.21& 2.081& 0.30& 0.119& 0.144& 5.78& -4.04& 3.77& 7.62 & \nodata\\
 1458& 0.005& 0.41& 0.005& 0.21& 0.005& 0.005& 0.28& 0.32& 0.23& 0.30 & \nodata\\
 1507& 0.163& 17.73& 1.625& 0.98& 0.064& 0.096& 4.07& -0.35& 0.15& 3.93 & \nodata\\
 1507& 0.006& 0.51& 0.005& 0.31& 0.007& 0.007& 0.41& 0.43& 0.33& 0.35 & \nodata\\
 1567& 0.344& 24.36& 2.209& -0.91& 0.067& 0.097& 3.79& -3.05& 2.77& 8.11 & \nodata\\
 1567& 0.006& 0.38& 0.005& 0.22& 0.005& 0.005& 0.28& 0.30& 0.22& 0.26 & \nodata\\
 1590& 0.214& 17.34& 2.234& -0.65& 0.145& 0.177& 5.50& -4.11& -0.55& 7.76 & \nodata\\
 1590& 0.007& 0.50& 0.006& 0.25& 0.006& 0.006& 0.30& 0.35& 0.25& 0.30 & \nodata\\
 1614& 0.308& 22.95& 1.951& -0.72& 0.084& 0.112& 3.23& -3.00& 3.64& 4.55 & \nodata\\
 1614& 0.005& 0.40& 0.005& 0.20& 0.005& 0.005& 0.27& 0.30& 0.19& 0.31 & \nodata\\
 1682& 0.186& 19.97& 1.886& 1.35& 0.027& 0.027& 3.47& -3.20& 2.95& 5.16 & \nodata\\
 1682& 0.003& 0.28& 0.003& 0.12& 0.000& 0.000& 0.19& 0.19& 0.13& 0.21 & \nodata\\
 1811& 0.116& 19.05& 2.075& -1.84& 0.085& 0.100& 6.89& -8.30& 1.96& 6.88 & \nodata\\
 1811& 0.009& 0.83& 0.009& 0.37& 0.009& 0.009& 0.50& 0.52& 0.37& 0.47 & \nodata\\
 1920& 0.044& 33.16& 2.516& -3.46& 0.027& 0.056& 5.57& -4.45& 5.39& -13.58 & 38.\\
 1920& 0.016& 1.41& 0.019& 0.62& 0.000& 0.015& 0.82& 0.89& 0.54& 1.05 & 10.\\
 1935& 0.264& 19.47& 2.393& -2.04& 0.109& 0.141& 2.67& -2.95& 2.35& 4.04 & \nodata\\
 1935& 0.006& 0.37& 0.005& 0.21& 0.005& 0.005& 0.27& 0.28& 0.24& 0.27 & \nodata\\
\enddata
\tablecomments{The second line for each galaxy lists the uncertainties.}
\end{deluxetable*}

\subsection{Presentation of the spectra and imaging}

\begin{figure*}
\epsfxsize 16.0cm
\epsfbox{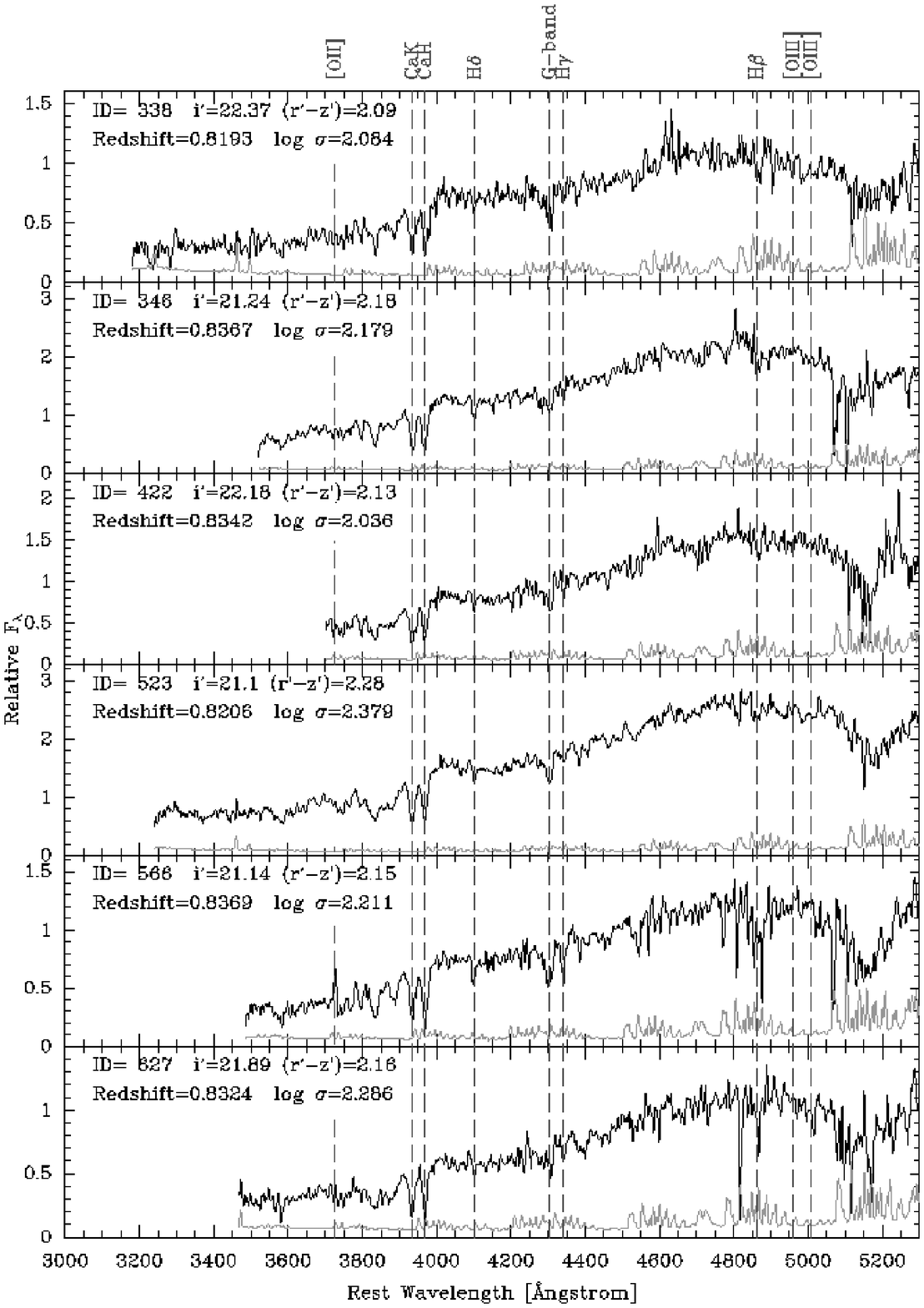}
\caption[]{Spectra of the galaxies that are considered members of the cluster.
Black lines -- the galaxy spectra; green (grey) lines -- the random noise multiplied with four.
At the strong skylines, the random noise underestimates the real noise due to systematic
errors in the sky subtraction.
Some of the absorption lines are marked. The location of the emission lines [\ion{O}{2}],
[\ion{O}{3}]$\lambda$4959, and [\ion{O}{3}]$\lambda$5007 are also marked, though these lines are only present in some of
the galaxies. The spectra shown in this figure have been processed as described
in the text, including resampling to just better than critical sampling. \label{fig-spectra} }
\end{figure*}

\begin{figure*}
\epsfxsize 17.5cm
\epsfbox{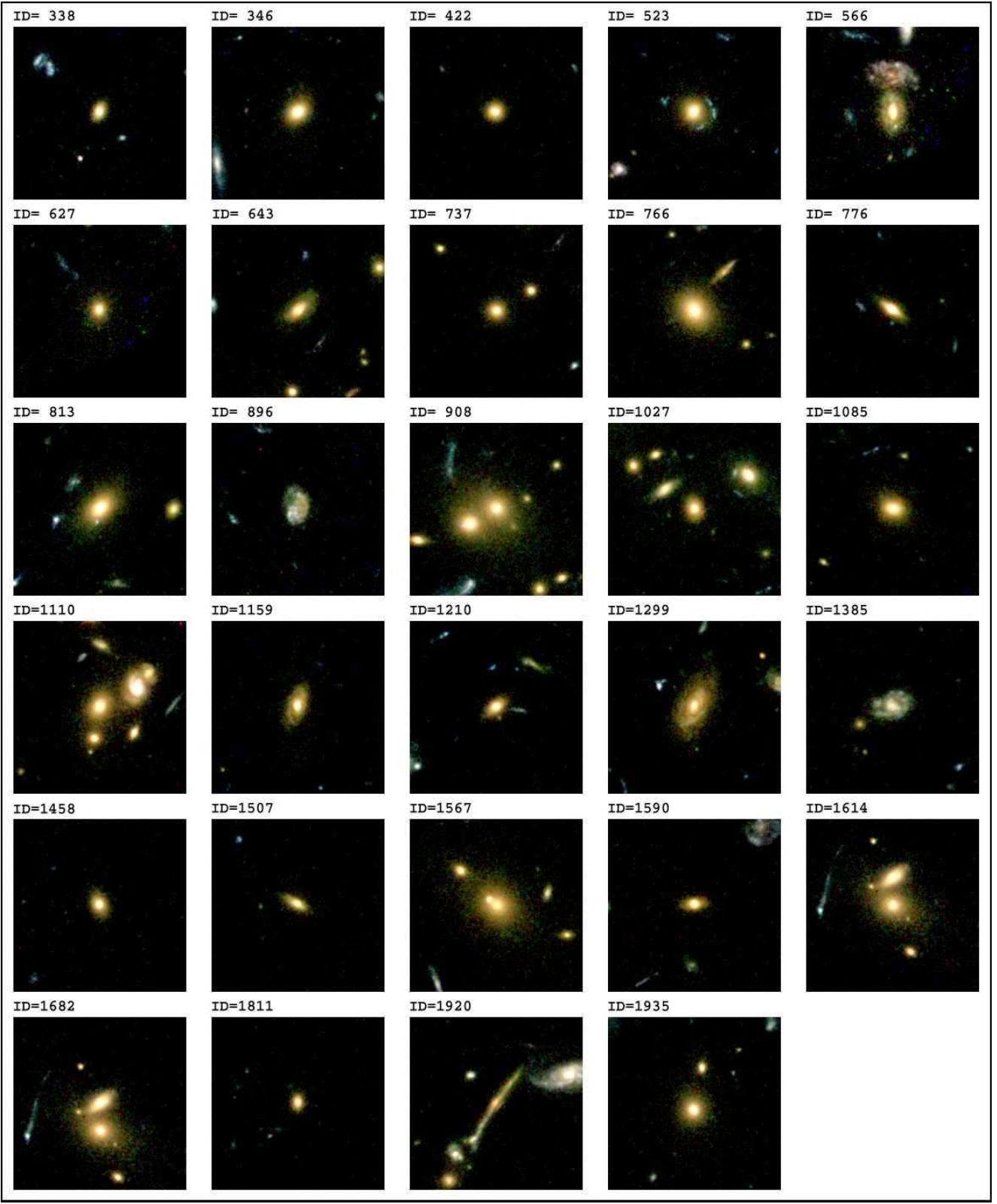}
\caption[]{
Color images of the cluster members made from archive HST imaging obtained with the ACS (program ID 9290, PI Ford).  
The HST/ACS images are reproduced as follows: F625W ($r'$) is blue, F775W ($i'$) is green, and F850LP ($z'$) is red. 
Each image covers 10 arcsec $\times$ 10 arcsec. At the distance of RXJ0152.7--1357
this corresponds to 75 kpc $\times$ 75 kpc for our adopted cosmology. North is up, East to the left.
In each panel, the galaxy with the ID listed above the image is located in the center of that image.
\label{fig-stamps} }
\end{figure*}

The spectra of the cluster members are shown in Figure \ref{fig-spectra}.
Stamp-sized color images of the galaxies are shown in Figure \ref{fig-stamps}.
We have chosen to show the color images based on the HST imaging obtained with the ACS,
rather than the GMOS-N imaging,
such that these images show the morphologies of the galaxies.

\begin{figure*}
\epsfxsize 16.0cm
\epsfbox{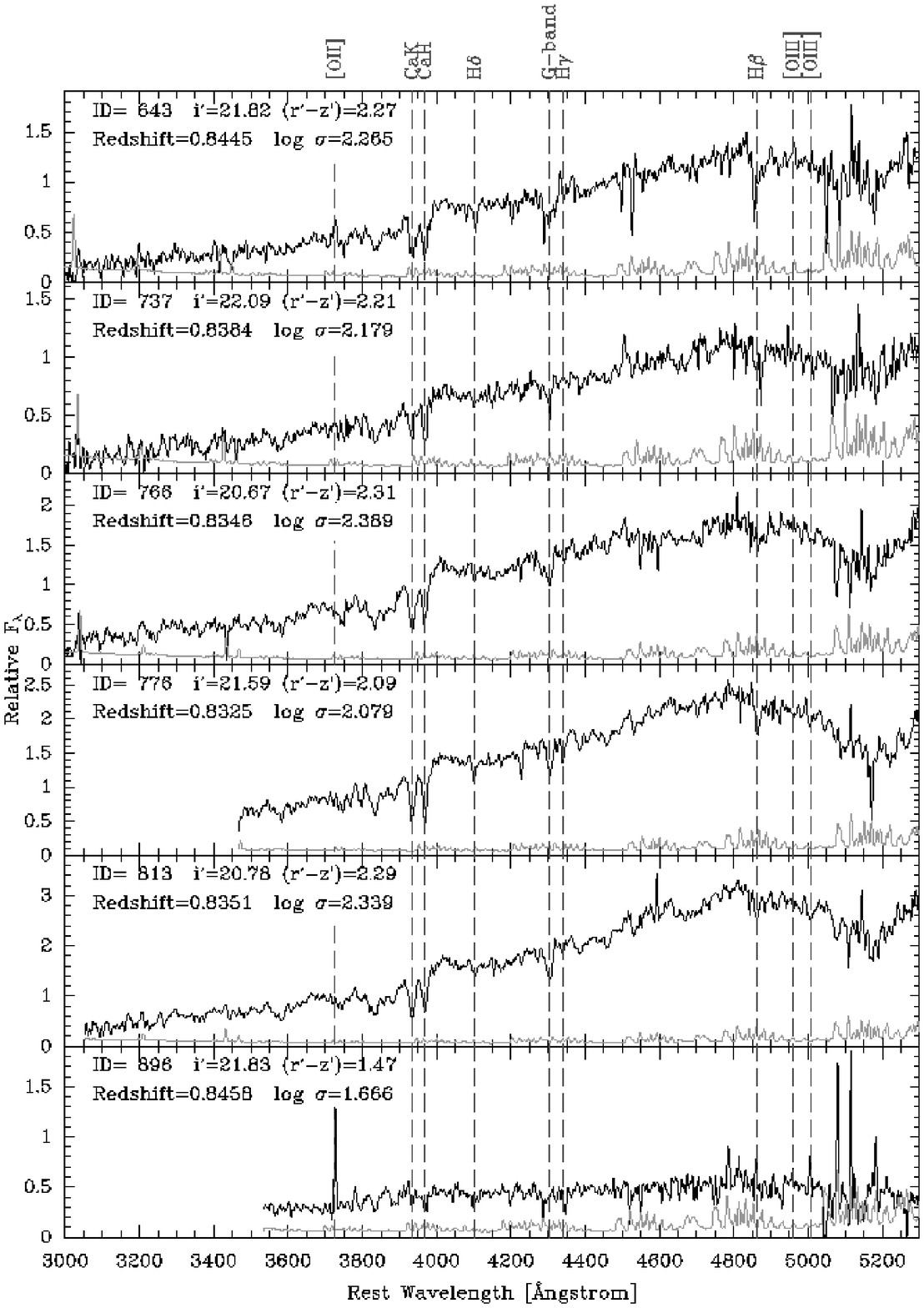}

\center{Fig.\ \ref{fig-spectra} -- {\em Continued.}}
\end{figure*}
\begin{figure*}
\epsfxsize 16.0cm
\epsfbox{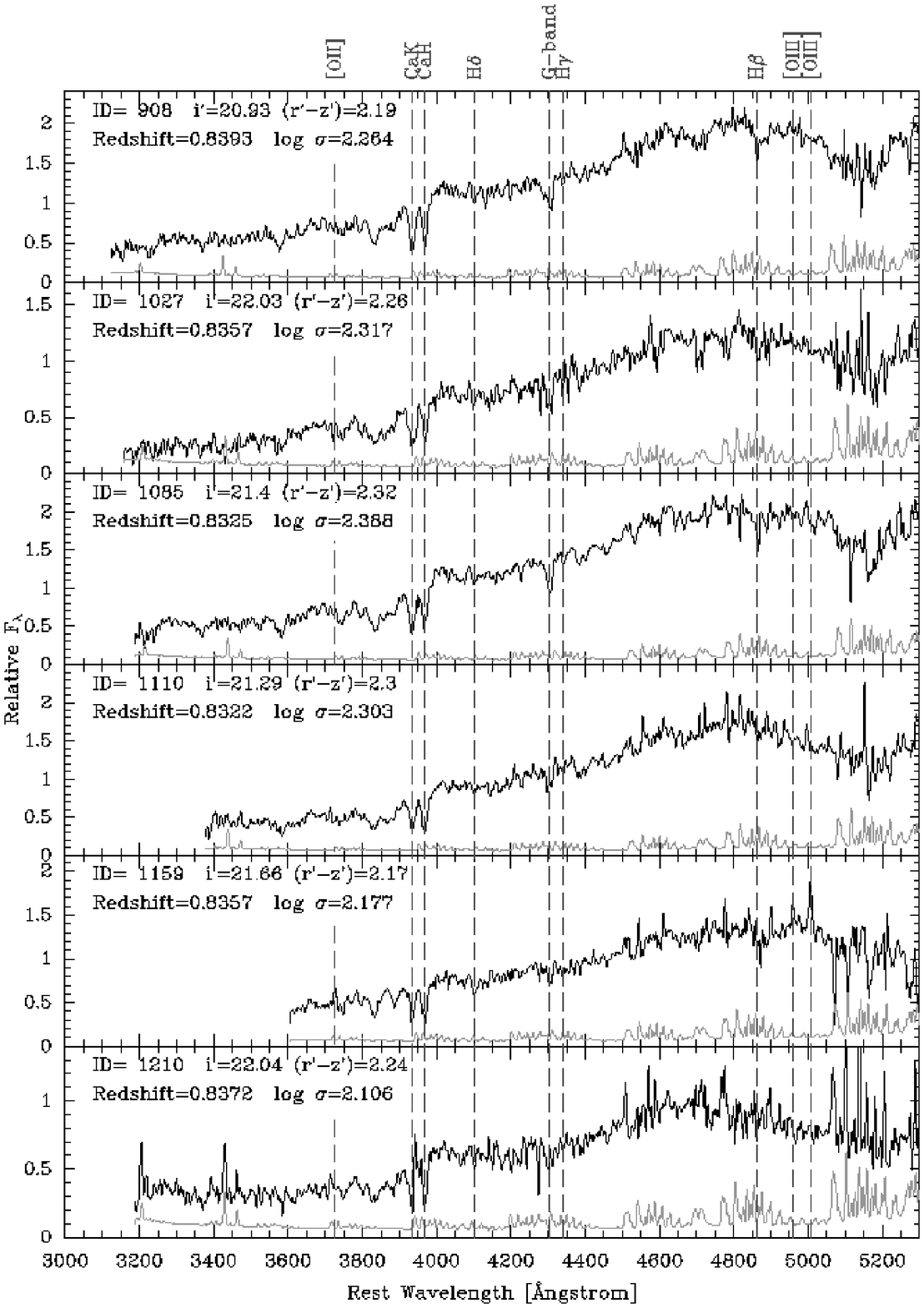}

\center{Fig.\ \ref{fig-spectra} -- {\em Continued.}}
\end{figure*}
\begin{figure*}
\epsfxsize 16.0cm
\epsfbox{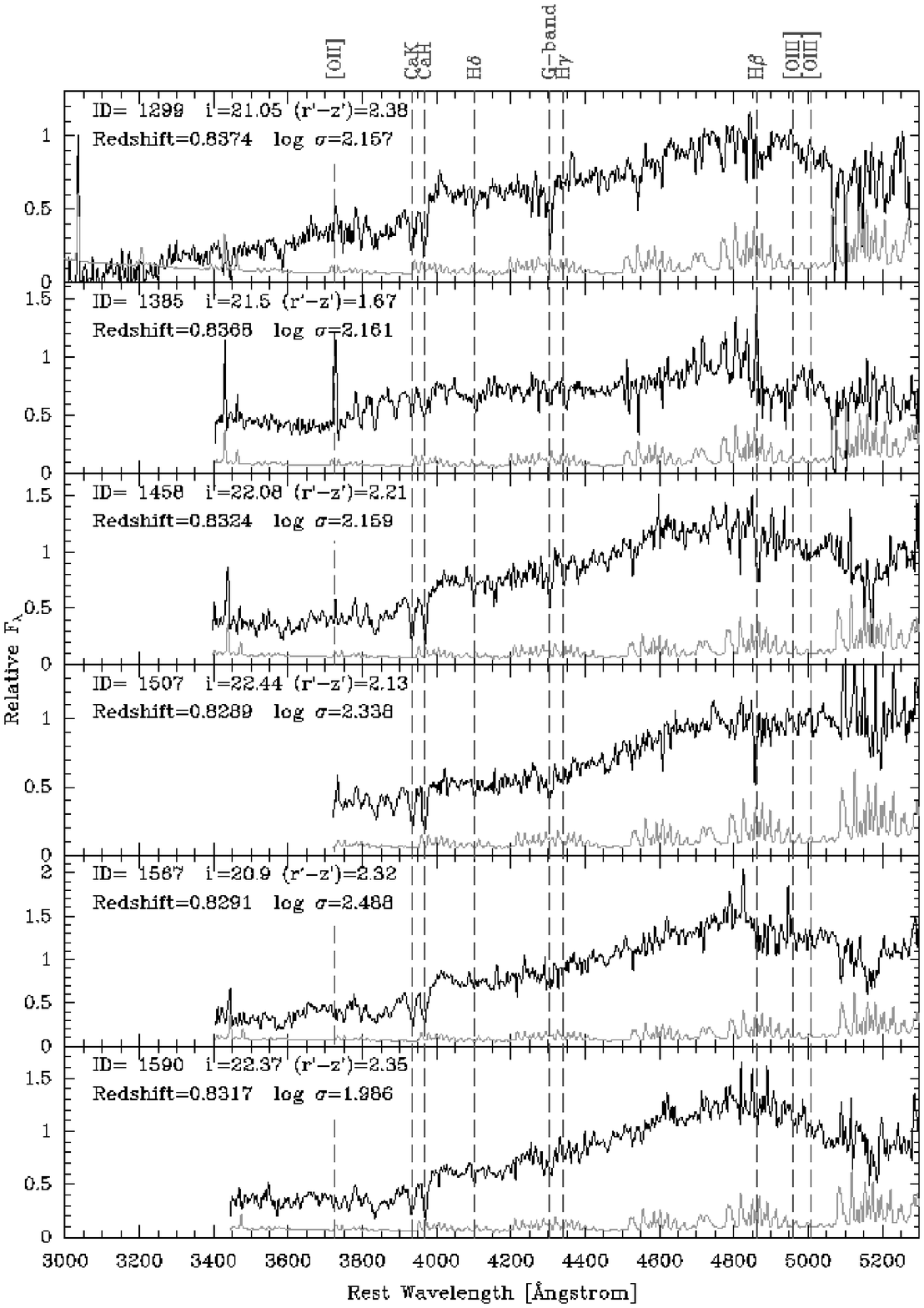}

\center{Fig.\ \ref{fig-spectra} -- {\em Continued.}}
\end{figure*}
\begin{figure*}
\epsfxsize 16.0cm
\epsfbox{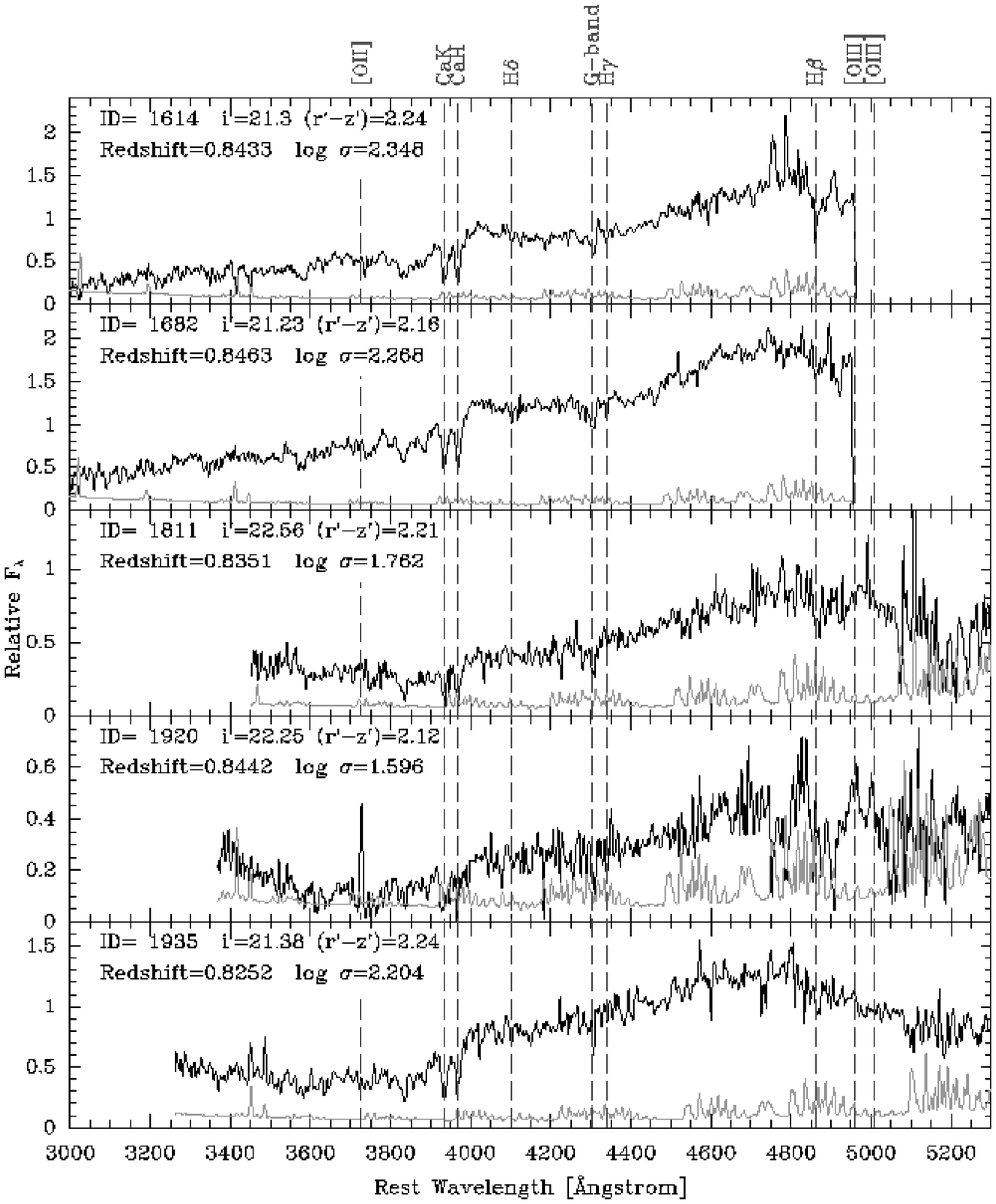}

\center{Fig.\ \ref{fig-spectra} -- {\em Continued.}}
\end{figure*}

\end{document}